# Generating function method and its applications to Quantum, Nuclear and the Classical Groups

## M. Hage-Hassan
**Université Libanaise, Faculté des Sciences Section (1)**
**Hadath-Beyrouth**

## Abstract

The generating function method that we had developing has various applications in physics and not only interress undergraduate students but also physicists. We solve simply difficult problems or unsolved commonly used in quantum, nuclear and group theory textbooks. We find simply: the generating function of the harmonic oscillator, the Feynman propagators of the oscillator and the oscillator in uniform magnetic field. We derive the invariants of SU(2) and the expressions of 3-j ,6-j symbols. We find also the octonions or Hurwitz quadratic transformations. We show that the cross-product exist only in $E_3$ and $E_7$. We determine the {p} representation of hydrogen atom in three and n-dimensions. We generalize the Cramer's rule for the calculation of the rotational spectrum of the nucleus. We find the expression of the Hamiltonian in terms of quasi-bosons for study the collective vibration. We determine the basis and the expressions of 3-j symbols of SU (3) and SU(n).We find the Schrödinger equation from Hamilton-Jacobi formalism. We present these applications in independent chapters.

## Résumé

La méthode de la fonction génératrice que nous développons a beaucoup des applications en physique et qui intéressent non seulement les étudiants de la licence mais aussi les physiciens. Nous résolvons simplement des problèmes difficiles ou non résolus en physique quantiques , nucléaire et la théorie des groups. Nous trouvons simplement : la fonction génératrice l'oscillateur harmonique, les propagateurs de Feynman de l'oscillateur et l'oscillateur dans un champ uniforme. Nous dérivons facilement les invariants et l'expression des symboles 3-j , 6-j de SU(2). Nous trouvons aussi les transformations de Hurwitz. Nous montrons que le produit vectoriel existe seulement dans $E_3$ et $E_7$. Nous déterminons la représentation {p} de l'atome d'hydrogène dans trois et n-dimensions. Nous généralisons la règle de Cramer pour l'étude de mouvement rotationnel du noyau. Nous trouvons l'Hamiltonien en fonction des quasibosons pour l'étude de vibration du noyau. Nous déterminons la représentation et les expressions des symboles 3-j des groupes SU(3) et SU(n). Nous dérivons aussi l'équation de Schrödinger du formalisme de Hamilton-Jacobi. Nous présentons ces applications dans des chapitres indépendants.

# Introduction

Quantum, nuclear and groups theories constitutes the background of any physician but there are still many problems unresolved or some of them resolved by difficult methods for undergraduates [1-9]. The generating function method, (GFM), that we develop solves some of these problems in a simple and fits naturally in these courses. But we find to make our presentation more clear and useful for students in mathematics and physics is to make a quick historical review of analytical and quantum mechanics.

## *1. A brief review of analytical mechanics and quantum mechanics*

We can represent the study of mechanics by an astronaut in space who looks at a pedestrian moving on a road. From a distance, he sees the movement is linear or curvilinear, but when he approaches he finds that the motion is zigzag and see a random stochastic, and if it approaches more he sees only its form (eyes: spin, ..), and so on.

*1.1 The first case:* In the first case the equation of the trajectory is determined by Newton formula. But we know that classical mechanics has the starting point the meeting of a very rich Tycho Brahe passion for astrophysics and a priest out of the church because he doing mathematics during the confessions. They worked together until the death of Tycho Brahe and the results were Kepler's laws and the death of Kepler poverty.

Then Galileo, a mathematics professor, loved dancing and had success. An evening in a dancing room a breeze has led Chandeliers to a pendulum motion for a very long time. This was the origin of his discovery of inertial frame and the beginning of his research.

Finally Newton introduced the acceleration and found the equation of the trajectory $\vec{F} = m\vec{\gamma}$. Undoubtedly Newton breakthrough the icy ocean of knowledge and allowed other scientists to swim in it.

The beginning of analytic mechanic [10] was with Lagrange. In France Lagrange (well respected by Napoleon) generalized Newton's formula on a variety by introducing the Lagrangian L = T - V = (kinetic energy - potential). And in the same time in Russia Euler [*] developed the calculus of variations [11] generalizing Fermat's principle of least time. After a time, Hamilton starting from Euler's variation calculus and Lagrange function, he introduced the action (S), $S = \int_{t1}^{t2} L dt$, the principle of least action and simply deduced the Lagrange's equations. He also introduced a new function, the Hamiltonian H (H = T + V) and found the canonical equations. He also discovered the quaternion and then generalized by Cayley and Grave to octonions, non associative algebra [12-13], which is very useful in mathematical physics. Hamilton's work was complemented by Jacobi and finds a new equation, the Hamilton-Jacobi equation.

But this analytical mechanics has led to the fundamental concepts of the physical system are the concepts of state and the dynamic variables: coordinates of the particles $(\vec{r})$, momentum $(\vec{p})$, the components of orbital angular momentum $(\vec{L})$ and energy (E).

Among the most important applications of mechanics are the harmonic oscillator, hydrogen atom and gravitation. But it is also important to note that the equation of the hydrogen atom or gravitation transformed into an equation of the harmonic oscillator to be resolved.

***1.2 The second case:*** The second case represents the latest approach to quantum mechanics or "the Lagrangian quantum theory" and it is the path integral introduced by Dirac and developed by Feynman [14-16]. But the methods of calculations of Feynman's propagators are beyond the undergraduate level.

It is also very important to summarize the quantum Hamiltonians [1-7] approaches and the general formalism of Dirac.

We know that Hamiltonian quantum mechanics originates from a true story:

On a sunny day in Geneva (1885) had escaped from a seller of balloons filled with hydrogen a number of them flew in the sky and in the night he found an emission of radiation picked up balloons and after that a Geneva newspaper published the wavelengths of radiation and then a secondary school teacher found the series known by his name, the Balmer series.

We must not forget that the problem of black body radiation (stove) was the basis for the introduction of light quanta by Planck, Einstein and Bohr and was also very important for the development of quantum mechanics

After much research developed two approaches are equivalent:
The matrix mechanics of Heisenberg and Schrödinger's wave mechanics. The first formulation requires appearing in any physical theory only physically observable quantities therefore the concept of the electronic orbit is unfounded at the microscopic level. But the second has its origin in the work of L. de Broglie who postulated that wave-particle duality, already predicted by Hamilton, is a general property of microscopic objects.

However, Schrödinger generalizes this notion of wave field and discovered the equation of propagation of the wave function, $H\psi = i\hbar \left(\frac{\partial \psi}{\partial t}\right)$, and a simple rule of correspondence for deriving this fundamental equation, $\vec{p} = \frac{\hbar}{i}\left(\frac{\partial}{\partial \vec{r}}\right)$.

Schrödinger also showed the equivalence of two methods, but Dirac has established the general formalism of quantum theory.

Dirac observed two weeks after reading the work of Heisenberg that the coordinates of the particles and impulses are observable but do not commute. Which implies from the mathematical point of view we need two wave functions: the first is function of coordinated and the second of momentum and are deduced from each other by means of Fourier transformation.

Because the coordinates are hermitian operators led Dirac to introduce the state function in quantum and the discovery of the delta function that bears his name (Dirac delta function) and a new formulation of quantum mechanics. In addition he introduced the ladders, or bosons, operators $(a^+, a)$ for the resolution of the harmonic oscillator, which play a fundamental role in physics.

Starting from the evolution operator $U = \exp\left[(-i(t - t_0)H/\hbar)\right]$, Heisenberg too, found that is called the Heisenberg equation of motion.

***1.3 The third case:*** The third cases are the study of elementary particles [15], but we are interested only in those lectures by the unitary groups which are very important for the study of rotational invariance and the classification of elementary particles, SU (3).

## 2. The Generating function and the unsolved problems or solved with Difficult methods

I observed that the generating function of the harmonic oscillator with complex parameter G(x, z) [8,17], can be regarded as kernel function with Gaussian measure for integration, has never been used and may be generalized to simplify the resolution or to solve many important problems in quantum , nuclear and group theory include:
1. A simple derivation of the generating function of the harmonic oscillator [6].
2. a- Schwinger [18-22] developed a method based on the Heisenberg equations of motion for calculating the Feynman propagator $\langle (x,t)|\exp[-(i/\hbar)H(t-t_0)]|x_0,t_0\rangle$.

   b- Schwinger [23-25] also start in his famous study "on angular momentum" to search operators in terms of bosons operators which generalizes the orbital angular momentum $(L) = (a^+)(S_1)(a)$ took $(J) = (a^+)(S_{1/2})(a)$, with $(S_1)$ and $(S_{1/2})$ are the Pauli matrices for spin 1 and spin ½.

Bargmann study the Fock space and use it's isomorphism with the harmonic oscillator to study the Rotation groups, following Schwinger treatment.
 But all these works are difficult to be followed by undergraduate student.
3. Connecting the equation of hydrogen atom and harmonic oscillator was performed using the quadratic transformations, (Levy-Civita, Kustaanheimo-steifel ,…)[26-29]. Kibler and (al.) have studied these transformations and have made several useful applications in physics [30-31]. But the momentum representation of hydrogen atom is not resolved using the Fourier transform except for simple cases [1-7, 32-37].
4. The development of techniques for operators of fermions and the Hamiltonian in terms of quasibosons operators proved particularly effective to study the collective Hamiltonian [38-40] and transition operators of even-even nuclei. Two development methods were used: first the Belyev and al. and the second is the Marumori and al. unfortunately these developments converge slowly when they converge, and no longer respect the Pauli principle when they are trunked.
5. The known generalization of the Euler's angles to classical groups is inconvenient [41-43] so we must seek a new parameterization.

6- Schwinger's method for the study of angular momentum, known boson method has been generalized by several authors [44-46] to study the classical groups and in particular the unitary groups. But the study of unitary groups is facing a major challenge for the explicit determination of the basis of representations of SU (n) for n> 3. In addition there are no simple formulas for 3-j symbols of SU (3) and the calculation become intractable for > 3.

## 3. The generating function method and some of it' is applications

We will present our works on these subjects in a way as simple as possible in seven chapters and appendix but prerequisites the standard graduate courses.
And we give a simple diagram for the connection between the various applications.

In the first chapter of these lectures we make a quick review of the harmonic oscillator and we find a new elementary and interesting method for the derivation of the generating function G(x, z). It has been known that the generating function with complex parameters z can be written which is the sum of the product of the basis of the oscillator $\{u_i\}$ and the basis of the analytic Hilbert space or Fock space F= $\{z^i/\sqrt{i!}\}, z \in \mathbb{C}$, with dμ (z) Is the Gaussian measure. Using the generating function and the orthogonality of Fock space we calculate simply the normalization of the delta function, the Feynman propagators of the oscillator and the charged harmonic oscillator in uniform magnetic field [16-22].

    In chapter two [47-53] we start from Schwinger's generating function of D-Wigner matrix elements and the Fock space for simple treatment of angular momentum. We have found a simple method to study and to simplifies the calculation of 3n-j symbols (n >2).
  We find also two important observations: first we find that the coordinates (x, y, z) may be written in terms of the SU(2) matrix elements as quadratic form and thereby calculate the representation {p} of hydrogen.
  Second we note that the invariants of 3-j symbols can be calculated in terms of the space of parameter of the generating functions which will allow us to find the analytic expression of 3-j symbols of SU (3) and SU (n) groups.

    We present in the third chapter the well known problem of sums squares [12-13, 28] which has-been the origin of the non-associative algebras: the quaternion, the Octonions. We show a recurrence method which gives all Octonions quadratic transformations (OQT), or Hurwitz transformations [54], and we show the connection of hydrogen and oscillators in the general case. We find the relationship between the Pauli, Dirac matrices and its generalization and the generating functions of Gegenbaeur polynomials.This generalization leads to a new algebra different from Cayley-Dixon algebra.
    The relationship between the inertia tensor and the octonions algebra was emphasized for the first time in our paper [55]. And we also show by means of the tensor of inertia and Hurwitz's theorem that the cross-products can bee defined only in Euclidian spaces of three and seven dimensions [56-57].
    The quadratic transformations that we have derived from the theory of angular momentum are related to OQT allowed us to find the momentum representation of the hydrogen atom [58-65] in the case of two and three dimensions. The general case N≥3 may be done using generating function and the Hankel's integral of Bessel functions and we determine the wave function in momentum space with the exact phase factor.

    In chapter forth we study collective motions using the Hartree-Fock variation method. In this method we approximate the ground state of the system by a Slater determinant $|\Phi_{HF}\rangle$ constructed from the states of nucleons. This wave function is not function of angular momentum, and the calculation of rotational energy can be done by using the projection operator [66-78]. But the calculation of rotations spectrum is very long. We have generalized the Cramer's rule and so the calculations can be carried out simply by the Gauss-Jacobi method we derive also the Thouless function [78].
    To study collective motions it is important to consider the residual interactions [38]. And the introduction of random phase approximation theory and more generally the quasibosons developments aim the study of these interactions. We find the generating function as expansion as product of a Fock space and Hartree-Fock basis. Using the generating function method we determine the expression of the Hamiltonian $H_b$ in terms of quasibosons operators. In many important papers we find that $H_b$ was used to study the

vibrations motion of the nucleus, e.g. [79-81]. The great utility of the GFM [82-89] encourages me after retirement (2005) to expand and develop this method.

In chapter five we generalize the Euler angles for the classical groups. We find for unitary groups the measure of integration which is the measure of Fock space. To determine the 3-j symbols of SU (3) we construct the generating function using the Schwinger method of coupling then we find also a new expression of these symbols in the case of multiplicity free [43,90-107].

In chapter six we treat the difficult problems with the help of our method and we find the basis of representations of unitary groups and its 3-j symbols with multiplicity [108-120]. The objective of this chapter is the graduate students and physicists but we need in all our works only the well known Gaussian integrals.

In the annex we treat the derivation of classical relativity and Schrodinger equation using Hamilton and Hamilton-Jacobi formalisms.

The presentation given in these lectures is simple enough to be accessible to undergraduate students and can serve as a working tool for physicists. And I limit myself to simple parts of my works that do not require much calculation over each chapter can be read independently of the others.

* Euler was the son of a Swiss baker and was sent to school early. After a time he asked his teacher to write a set of numbers and he gives the product directly. The teacher was shocked and at three and half (AM) of the morning he spent at the bakery telling the father that his son is a genius and he is incapable of educated the child. He proposed to send the son to be educated by the family of the mathematicians Bernoulli.

*The following diagram describe the connection
Between the various applications*

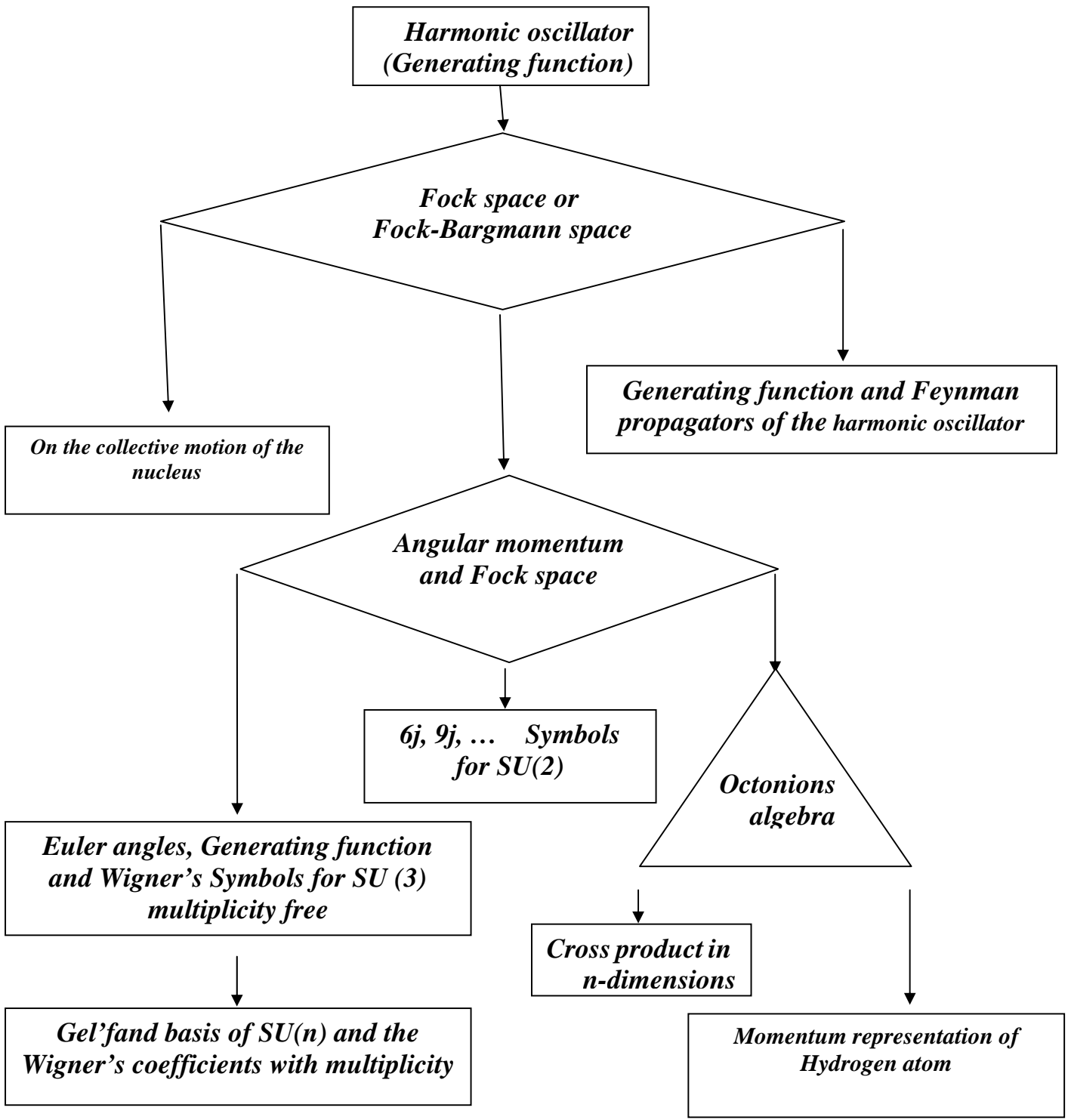

# Contents

*(Generating function method = Generating function + Fock space)

# Chapter One

## Generating function of the Harmonic oscillator, Fock space and Feynman propagators

**1. Introduction**
**2. The harmonic oscillator and Dirac transformations**
   2.1 The Schrodinger basis of harmonic oscillator
   2.2 Dirac notations in quantum mechanic
   2.3 Dirac transformation
**3. The harmonic oscillator in Dirac notations and the generating function**
   3.1 The basis of harmonic oscillator in Dirac notations
   3.2 The generating function of the harmonic oscillator
**4. The Fock space**
**5. The Dirac delta function and the normalization of the free wave**
**6. The Feynman propagator of the harmonic oscillator**
**7. The Feynman propagator of charged harmonic oscillator in uniform magnetic field**
   7.1 Isotropic charged harmonic oscillator
   7.2 The generating function of the cylindrical harmonic oscillator
   7.3 The Feynman propagator of two-dimensional isotropic charged harmonic oscillator in uniform magnetic field

# Chapter One

## Generating function of Harmonic oscillator, Fock space and Feynman propagators

## 1. Introduction

The harmonic oscillator plays a fundamental role in physics [1-7] and the solution of Schrödinger equation is well known to students. And as we have already written that Dirac determines the state of the quantum system and introduced the delta function. Furthermore, Dirac for solving the equation of harmonic oscillator start from the analogy between the form of the Hamiltonian of the harmonic oscillator and the product of a complex number and its complex conjugate and then he introduce the raising and lowering operators and found the states the oscillator with the new notations.

We know also from the standard quantum mechanics textbooks there are many important problems [17-22] with resolutions exceed the level of undergraduate's students for example: the generating function, the normalization of the delta function, the Feynman propagators etc...
In this chapter we present a new simple method, closely related to Dirac notations, for the determination of the generating function. And using this generating function and the Fock space we determine by an elementary calculation the normalization of Dirac delta function and the Feynman propagator of harmonic oscillator [18-22].

Then we review the resolution of two-dimensional isotropic charged harmonic oscillator in uniform magnetic field and after that we calculate the Feynman propagator in this case.

## 2. The Harmonic oscillator and Dirac transformations

In this part we present a brief review of the harmonic oscillator then Dirac notations of quantum mechanic and Dirac transformation [6].

### 2.1 The Schrodinger basis of harmonic oscillator
The Schrödinger equation of the harmonic oscillator in one dimension is:

$$H\psi(x) = E\psi(x) \qquad (2.1)$$

And $$H = \frac{1}{2m}(p_x^2 + m^2\omega^2 x^2) \quad [x, p_x] = i\hbar \qquad (2.2)$$

Put
$$x = \sqrt{\hbar/(m\omega)}q, \qquad (2.3)$$
we obtain
$$H = \hbar\omega(p^2 + q^2) \qquad (2.4)$$
The solution of Schrödinger equation is
$$Hu_n(q) = E_n u_n(q) \text{ and } E_n = \hbar\omega(n + (1/2)). \qquad (2.5)$$
With
$$u_n(q) = (\sqrt{\pi}\, 2^n n!)^{-\frac{1}{2}} e^{-\frac{q^2}{2}} H_n(q) \text{ And } u_n(x) = ((m\omega)/\hbar)^{1/4} u_n(q) \qquad (2.6)$$

$H_n(q)$ is the Hermite polynomial with
$$H_n(-q) = (-1)^n H_n(q) \text{ and } u_n(q) = (-1)^n u_n(-q). \qquad (2.7)$$

The generating function is given by [1]:

$$G(z,q) = \sum_{n=0}^{\infty} \frac{z^n}{\sqrt{n!}} u_n(q) \qquad (2.8)$$

*2.2 Dirac notations in quantum mechanic*

I want to do only a simplistic explanation of Dirac transformation. We know from the course of linear algebra that any Hilbert space with a basis $\{f_i\}$ has a dual space $E_n^* = \{f_i^*\}$ with $f_i^*(f_j) = (f_i, f_j)$, and $(f_i, f_j)$ is the scalar product.
We can make a change in the notations by putting:

$$|f_i\rangle = f_i, \quad \langle f_i| = f_i^* \qquad (2.9)$$

If $|q\rangle$ is the continuous eigenfunctions of the operator $\hat{q}$

We write:
$$\hat{q}|q\rangle = q|q\rangle$$

The expression of the unitary operator is

$$I = \int |q\rangle dq \langle q| \qquad (2.10)$$

Similarly, in the Hilbert space the inner product of two functions f and g is then:

$$(f,g) = \langle f\|g\rangle = \langle f|\int |q\rangle dq\langle q\|g\rangle$$
$$= \int \langle f\|q\rangle dq \langle q\|g\rangle = \int \overline{f(q)} g(q) dq \qquad (2.11)$$

For this expression is valid whatever f and g we can then deduce the famous Dirac's transformation:
$$\langle f\|q\rangle = \overline{f(q)}, \quad \langle q\|g\rangle = g(q) \qquad (2.12)$$

## 2.3 Dirac transformation

1- By definition the Dirac transformation is:

$$\langle q| : u_n \to u_n(q) = \langle q|n\rangle \tag{2.13}$$

and
$$\langle q|f\rangle = f(q),$$

2-The momentum $p = \frac{\hbar}{i}\frac{d}{dq}$ has the eigenfunctions $|p\rangle$

And
$$\int |p\rangle dp \langle p| = 1.$$

The Dirac transformation from the representation q to p is given by:

$$\langle p|\varphi\rangle = \int \langle p|q\rangle dq \langle q|\varphi\rangle$$

With
$$\langle p|q\rangle = \frac{e^{-iqp/\hbar}}{\sqrt{2\pi\hbar}}$$

And
$$\langle q|q|f\rangle = qf(q), \quad \langle q|p|f\rangle = \frac{\hbar}{i}\frac{d}{dq}f(q) \tag{2.14}$$

# 3. The Harmonic oscillator in Dirac notations and the generating function

## 3.1 The basis of harmonic oscillator in Dirac notations

Let
$$a = \frac{1}{\sqrt{2}}(q + \frac{d}{dq}), \quad a^+ = \frac{1}{\sqrt{2}}(q - \frac{d}{dq}) \tag{3.1}$$

we find
$$[a, a^+] = 1, \quad [a, a] = 0, \quad [a^+, a^+] = 0 \tag{3.2}$$

We derive from the above expressions a very useful formula:

$$af(a^+)|0\rangle = \frac{\partial f(a^+)}{\partial a^+}|0\rangle \tag{3.3}$$

And
$$H = \hbar\omega(N + \frac{1}{2}), \quad N = aa^+ \tag{3.4}$$

In Dirac notations the basis of the harmonic oscillator becomes

$$|n\rangle = \frac{a^{+n}}{\sqrt{n!}}|0\rangle, \quad \langle q|n\rangle = u_n(q) \tag{3.5}$$

With $\langle m|n\rangle = \delta_{m,n}$ and $|0\rangle$ is the vacuum state.

The energy is given by: $E_n = \hbar\omega(n + \frac{1}{2})$

And the expressions of the unitary operator and the generating functions are:

$$I = \sum |n\rangle\langle n|$$

$$G(z,q) = \sum_{n=0}^{\infty} \frac{z^n}{\sqrt{n!}} u_n(q) = \langle q|e^{za^+}|0\rangle \tag{3.6}$$

## 3.2 The generating function of the harmonic oscillator

We will find the expression of the generating function by a new simple method different from the other methods [1,6] and closely related to Dirac Ladder operators.
Using Dirac transformation and (2.14) we find:

$$\langle q|ae^{za^+}|0\rangle = \frac{1}{\sqrt{2}}(q+\frac{d}{dq})G(z,q) \tag{3.7}$$

Using also (3.3) we find: $\quad \langle q|ae^{za^+}|0\rangle = zG(z,q)$

By comparison of the above expressions we find:

$$\frac{d}{dq}G(z,q) = (\sqrt{2}z - q)G(z,q) \tag{3.8}$$

The solution of this equation is:

$$G(z,q) = c\exp\{(\sqrt{2}qz - \frac{q^2}{2}) + \varphi(z)\} \tag{3.9}$$

To determine $\varphi(z)$ we use the creation operator, we find:

$$\langle q|a^+e^{za^+}|0\rangle = \frac{1}{\sqrt{2}}(q - \frac{d}{dq})G(z,q) = \frac{\partial}{\partial z}G(z,q) \tag{3.10}$$

Using (2.13) and (3.3) we find:

$$\varphi'(z) = -z$$

After solving this equation we determine the generating function of the oscillator

$$G(z,q) = c\exp\{\sqrt{2}qz - \frac{q^2}{2} - \frac{z^2}{2}\}$$

For t = 0 we have $G(0,q) = \langle q|0\rangle = u_0(q)$ and it follows that: $c = \pi^{-\frac{1}{4}}$

So we finally get

$$G(z,q) = \sum_{n=0}^{\infty} \frac{z^n}{\sqrt{n!}} u_n(q) = \pi^{-\frac{1}{4}} \exp\{\sqrt{2}qz - \frac{q^2}{2} - \frac{z^2}{2}\} \tag{3.11}$$

# 4. The Fock space

In the expression of the generating function we note that the functions $\varphi_n(z)$

$$\varphi_n(z) = \frac{z^n}{\sqrt{n!}} \tag{4.1}$$

Constitute a basis of analytic Hilbert space that is known by Fock- Bargmann space {F}.

And $\quad \langle \varphi_i|\varphi_j\rangle = \iint \overline{\varphi_i(z)}\varphi_j(z)d\mu(z) = \iint \frac{\overline{z}^i}{\sqrt{i!}} \frac{z^j}{\sqrt{j!}} d\mu(z) = \delta_{i,j} \tag{4.2}$

$d\mu(z)$ Is the cylindrical measure or Gaussian measure:

$$d\mu(z) = e^{-(x^2+y^2)} \frac{dxdy}{\pi}, \quad z = x + iy \qquad (4.3)$$

we have also

$$f(z) = \int f(z')e^{z\bar{z}'}d\mu(z') \text{ and } e^{\alpha\beta} = \int e^{\alpha\bar{z}}e^{\beta z}d\mu(z) \qquad (4.4)$$

1-We observe also that the images of any wave function, $\Psi_f(x)$ and any operator A of the space {H}, a function f(z) and an operator $\tilde{A}(z)$ of {F} with:

$$\tilde{A}f(z) = \int \overline{(A\Psi_f(x))}\, C(x,z)\, dx \text{ and } \Psi_f(x) = \int \overline{(\tilde{A}f(z))}\, C(x,z)\, d\mu(z) \qquad (4.5)$$

We are dealing with a transformation and a problem in the oscillator basis can be transformed into a problem in the Fock space of which the resolution is simpler.

2- The generating function $|z\rangle$ is the well known coherent state with:

$$|z\rangle = e^{za^+}|0\rangle, \qquad \langle z'\|z\rangle = e^{\bar{z}'z}$$
$$a|z\rangle = z|z\rangle$$

And
$$G(z,q) = \langle q|z\rangle \qquad (4.6)$$

The unitary operator is:

$$I = \sum_n |n\rangle\langle n| = \int |z\rangle d\mu(z)\langle z| \qquad (4.7)$$

3- The transformation from the representation {q} to the representation of the Harmonic oscillator is:

$$\langle q| = \int \langle q\|z\rangle d\mu(z)\langle z| = \int G(z,q)d\mu(z)\langle z| \qquad (4.8)$$

4- The correspondence between the harmonic oscillator and Fock space may be deduced from the relation (3.3):

$$a^+ \to z, \quad a \to \delta/(\delta z) \qquad (4.9)$$

The generators of unitary group may be written in terms of raising and lowering operators of the harmonic oscillator [8-9] therefore we can write these generators in terms of the variables of Fock space.

## 5. The Dirac delta function and the normalization of the free wave

We want to determine the expression $\langle q'|q\rangle$ using the generating function of the harmonic oscillator and the orthogonality of the basis of Fock space.
We write:

$$\langle q'|q\rangle = \langle q'|I|q\rangle = \sum_{i=0} u_i(q')u_i(q) = \sum_{i,j} u_i(q') \times (\delta_{i,j}) \times u_i(q)$$

Using (4.2) and (3.11) we find:
$$\langle q'|q\rangle = \int \overline{G(z,q')} G(z,q) d\mu(z) \qquad (5.1)$$

By replacing $\overline{G(z,q')}$ and $G(z,q)$ by the expression (3.7) we obtain
$$\langle q'|q\rangle = \frac{1}{\sqrt{\pi}} \int \exp[-\frac{q'^2+q^2}{2} - \frac{\bar{z}^2+z^2}{2} + \sqrt{2}(q'\bar{z}+qz)]d\mu(z) \qquad (5.2)$$
the arrangement of this expression gives
$$\langle q'|q\rangle = \frac{1}{\pi\sqrt{\pi}} \exp(-\frac{1}{4}(q-q')^2) \int \exp([-2(x-\frac{\sqrt{2}}{4}(q+q'))^2] + \sqrt{2}iy(q-q'))dxdy$$
using the change of variables and performing the integration we find that
$$\langle q'\|q\rangle = \frac{1}{2\pi} \exp(-\frac{1}{4}(q-q')^2) \int \exp(+ik(q-q'))dk \qquad (5.3)$$
Using the Gauss integral we find that
$$\int_{-\infty}^{+\infty} \langle q'\|q\rangle dq' = \frac{1}{2\pi} \int_{-\infty}^{+\infty} \exp[-\frac{1}{4}(q-q')^2 + ik(q-q')]dq'dk = 1 \qquad (5.4)$$
But $|q\rangle$ and $|q'\rangle$ are eigenfunctions of the operator $\hat{q}$ then:

$$\langle q'|\hat{q}|q\rangle = q\langle q'\|q\rangle = q'\langle q'\|q\rangle \text{ and } (q-q')\langle q'\|q\rangle = 0$$

it follows that:

$$\exp(-\frac{1}{4}(q-q')^2)\langle q'\|q\rangle = \langle q'\|q\rangle$$

Therefore
$$\int_{-\infty}^{+\infty} \langle q'\|q\rangle dq' = \frac{1}{2\pi} \int_{-\infty}^{+\infty} \exp[+ik(q-q')]dq'dk = 1 \qquad (5.5)$$

if we use in (3.3) the free wave $Ne^{iqx}$ we find from the expression (5.5) the normalization of the wave function of free particle: $N = 1/\sqrt{2\pi}$
Finally we write
$$\langle q'\|q\rangle = \delta(q-q') = \frac{1}{2\pi} \int_{-\infty}^{\infty} e^{i(q-q')k} dk \text{ and } \int_{-\infty}^{+\infty} \delta(q-q')dq' = 1 \qquad (5.6)$$

We find in a simple and coherent way the Dirac delta function. We deduce also from (5.1) that the delta function is an even function and we obtain the normalization of the wave function of free particle without the help of distribution theory [6].

## 6. The Feynman propagator of the harmonic oscillator

The Feynman propagator of the oscillator was determined by several methods
The first one is the Feynman path integral, the second is the Schwinger method of Green function, the third is the algebraic method and finally by a direct calculation using the Mehler formula [18-19]. All these methods are difficult for undergraduates and all the text books gives only the final result. In this section we propose a simple and elementary method for the calculation of this propagator.

The Feynman propagator of the oscillator is:

$$K((x,t),(x',t_0)) = \langle x|e^{-\frac{i}{\hbar}H(t-t_0)}|x'\rangle = \langle x|e^{-\frac{i}{\hbar}H(t-t_0)}I|x'\rangle$$
$$= e^{-i\omega(t-t_0)/2}\sum_n \overline{u_n(x)}e^{-in\omega(t-t_0)}u_n(x') \qquad (6.1)$$

From the orthogonality of Fock space and (4.2) we deduce that:

$$K((x,t),(x',t_0)) = \left(\frac{m\omega}{\pi\hbar}\right)^{1/2} e^{-i\omega(t-t_0)/2}\int G(e^{-i\omega(t-t_0)/2}\overline{z},q)G(e^{-i\omega(t-t_0)/2}z,q')d\mu(z) \qquad (6.2)$$

By replacing the expressions under the integral by (2.14) and $\alpha = \omega(t-t_0)$.
We write:

$$\left(\frac{m\omega}{\pi\hbar}\right)^{1/2} e^{-i\alpha/2}\int \exp([-\frac{q^2+q'^2}{2}+e^{-i\frac{\alpha}{2}}(\overline{z}q+zq')\sqrt{2}-\frac{\overline{z}^2+z^2}{2}e^{-i\alpha}]d\mu(z)$$

After arrangement we find that:

$$K((x,t),(x',t_0)) =$$
$$= \frac{1}{\pi}\left(\frac{m\omega}{\pi\hbar}\right)^{1/2} e^{-i\alpha/2}\exp[\frac{i}{2\sin\alpha}[(q^2+q'^2)\cos\alpha - 2xx']]\times E_1 \times E_2$$

With
$$E_1 = \int \exp[-(2e^{-i\frac{\alpha}{2}}\cos\frac{\alpha}{2})[(x-\frac{\sqrt{2}}{4\cos\frac{\alpha}{2}}(q+q'))^2]]dx$$

and
$$E_2 = \int \exp[-(2ie^{-i\frac{\alpha}{2}}\sin\frac{\alpha}{2})[(y+\frac{\sqrt{2}}{4\sin\frac{\alpha}{2}}(-q+q'))^2]]dy$$

But
$$2e^{-i\frac{\alpha}{2}}\cos\frac{\alpha}{2} = (1+e^{-i\alpha}), \quad 2ie^{-i\frac{\alpha}{2}}\sin\frac{\alpha}{2} = (1-e^{-i\alpha}) \quad \text{and } e^{-i\alpha/2} = \frac{1}{\sqrt{e^{+i\alpha}}},$$

we have also

$$q = \sqrt{(m\omega)/\hbar}\, x \text{ and } \int_{-\infty}^{+\infty}e^{-az^2}dz = \sqrt{\frac{\pi}{a}}, \text{ with Re (a)>0}$$

Using the above expressions and performing the integration after change of variables we

Find that:

$$\frac{1}{\sqrt{e^{+i\alpha}}} \times E_1 \times E_2 = \pi \sqrt{\frac{1}{2i\sin\alpha}} \quad \text{With} \quad |1 \pm \cos\alpha| > 0$$

Finally we obtain the expression of Feynman propagator:

$$K((q,t),(q',t_0)) = \sqrt{\frac{m\omega}{2\pi\hbar i \sin\alpha}} \exp[\frac{i}{2\sin\alpha}[(q^2+q'^2)\cos\alpha - 2qq']] \tag{6.3}$$

Consequently we do not encounter the difficulties of the method proposed by Holstein [20] which adopted by all the authors [21-22] and especially the standard books. The same calculation can be used with the generating function to calculate the propagator of the cylindrical basis.

Our method may be applied to the calculation of the partition function and other Feynman propagators [4-5]. We can also do other calculations with the oscillator representation using the expression (4.8) and Schwinger techniques [23].

## 7. The Feynman propagator of charged harmonic oscillator in uniform magnetic field

### *7.1 Isotropic charged harmonic oscillator*

Considering an isotropic charged harmonic oscillator with electric charge q and mass µ moves in a two-dimensional plane under a uniform magnetic field B perpendiculars to the plane and the vector potentials have the following form [1-7]:

$$A_1 = -By/2, \quad A_2 = Bx/2 \tag{7.1}$$

The Hamiltonian of the system is

$$H = \frac{1}{2\mu}\left[\left(p_1 + \frac{qB}{2c}y\right)^2 + \left(p_2 + \frac{qB}{2c}x\right)^2\right] +$$

$$\frac{1}{2}\mu\omega_0^2(x^2+y^2) = H_0 - \frac{qB}{2\mu c}L_z \tag{7.2}$$

With

$$H_0 = \frac{1}{2\mu}(p_1^2 + p_2^2) + \frac{\mu\omega^2}{2}(x^2+y^2) \tag{7.3}$$

And

$$\omega = \sqrt{\omega_0^2 + \omega_c^2}, \quad \omega_c = \left(\frac{qB}{2\mu c}\right). \tag{7.4}$$

We have also $[H_0, L_z] = 0$.

To determine the eigenfunctions of $H_0, L_z$ we use the polar coordinates of two dimensions harmonic oscillators [10].
We put

$$x = \rho\cos\varphi, \quad y = \rho\sin\varphi \tag{7.5}$$
$$0 \le \rho \le \infty, \ 0 \le \varphi \le 2\pi$$

Using the method of separation of variables we find the solution of Schrödinger Equation which is the cylindrical basis:

$$\Phi_{jm}(\lambda\rho,\varphi) = \left(\frac{\mu\omega}{\pi\hbar}\right) f_{jm}(\lambda\rho) e^{-2im\varphi} =$$

$$\sqrt{\frac{\lambda^2}{\pi}} \sqrt{\frac{(j-|m|)!}{(j+|m|)!}} \exp\left(-\frac{\lambda^2\rho^2}{2}\right) L_{j+|m|}^{2|m|}(\lambda^2\rho^2)(\lambda\rho)^{2|m|} e^{-2im\varphi} \tag{7.6}$$

With $\quad j = n + |m| \quad$ and $\quad \lambda = \sqrt{\dfrac{m\omega}{\hbar}}$

The energy of the system can be given as follows:

$$E_{nm} = \hbar\omega(2n + 2|m| + 1) - m\hbar\frac{qB}{\mu c} \tag{7.7}$$

With $\quad n = 0,1,2,...,\quad$ And $|2m| = 0, \pm 1, \pm 2,.....$.

We emphasize that we can build the generating function of the cylindrical basis from the generating function of Laguerre polynomials but the calculation is more simpler with the generating function that we will build in part three.

## 7.2 The generating function of the cylindrical harmonic oscillator

In this part we review the construction of the generating function of the cylindrical basis of the harmonic oscillator which is the eigenfunctions of $(H_0, L_z)$.

We know that the Cartesian basis of harmonic oscillator in Dirac notations is

$$|n_x, n_y\rangle = a_x^{+n_x} a_y^{+n_y} |0,0\rangle \tag{7.8}$$

This ket is not eigenfunctions of $L_z$ so to obtain the basis which has this property
We must take the transformation [10]

$$A_1^+ = \frac{\sqrt{2}}{2}(a_x^+ - ia_y^+), \quad A_2^+ = \frac{\sqrt{2}}{2}(a_x^+ + ia_y^+)$$
$$L_z = (N_1 - N_2), \quad N = N_1 + N_2,$$
$$N_1 = A_1^+ A_1, \ N_2 = A_2^+ A_2$$

The new basis $|N_1, N_2\rangle$ can be written in the form

$$|N_1, N_2\rangle = |j+m, j-m\rangle = A_1^{+j+m} A_2^{+j-m} |0,0\rangle.$$

This basis is function of $L_z$ and $N$ with the values 2m and 2j.

With $\qquad H = \hbar\omega(N+1) - \dfrac{qB}{2\mu c} L_z$

And $\qquad E_{nm} = \hbar\omega(2j+1) - m\hbar\dfrac{qB}{\mu c}$

The new generating function may be written in the form:

$$|G(z_1, z_2)\rangle = \exp[z_1 A_1^+ + z_2 A_2^+]|0,0\rangle$$
$$= \exp[a_x^+ \sqrt{2}(z_1+z_2)/2 + i a_x^+ \sqrt{2}(-z_1+z_2)/2]|0,0\rangle$$

In term of Cartesian coordinates we write the generating function as:

$$G(t_1,t_2,\vec{r}) = \left(\frac{\lambda^2}{\pi}\right)^{1/2} \exp[-\lambda^2 \frac{x^2+y^2}{2} + \lambda[z_1(x+iy) + z_2(x-iy)] - z_1 z_2] =$$
$$\sum_{jm} (-1)^{j-m} \frac{z_1^{(j+m)} z_2^{(j-m)}}{\sqrt{(j+m)!(j-m)!}} \Phi_{jm}(\lambda\rho,\varphi) \qquad (7.9)$$

And 
$$\varphi_{jm}(z) = \frac{z_1^{(j+m)} z_2^{(j-m)}}{\sqrt{(j+m)!(j-m)!}}$$

Is a two dimensional Fock-Bargmann spaces [11].

## 7.3 The Feynman propagator of two-dimensional isotropic charged harmonic oscillator in uniform magnetic field

In this section, we propose a simple and elementary method for the calculation of Feynman propagator of two-dimensional charged harmonic oscillator in uniform magnetic field.
We have:

$$K((\vec{r},t),(\vec{r}',t_0)) = \langle r|e^{-\frac{i}{\hbar}H(t-t_0)}|r'\rangle = \langle r|e^{-\frac{i}{\hbar}H(t-t_0)} I|r'\rangle$$
$$= e^{-i\omega(t-t_0)} \sum_n \overline{\Phi_{jm}(\vec{r})} e^{-i(2j\alpha-2m\beta)} \Phi jm(\vec{r}') \qquad (7.10)$$

With $\quad \alpha = \omega\tau, \beta = \omega_c \tau$ and $\tau = (t-t_0)$

From the orthogonality of the basis $\varphi_{jm}(z)$ and (7.10) we deduce that:

$$K((\vec{r},t),(\vec{r}',t_0)) = \left(\frac{\mu\omega}{\pi\hbar}\right) e^{-i\omega(t-t_0)} \int [\overline{G}(\bar{a}z_1,\bar{b}z_2,\vec{r}') \times$$
$$G(az_1,bz_2,\vec{r})] d\mu(z_1) d\mu(z_2) \qquad (7.11)$$

With $\quad a = e^{-i(\alpha-\beta)/2}$ and $b = e^{-i(\alpha+\beta)/2}$

By substituting the expressions under the integral by (7.11) we write:

$$K((\vec{r},t),(\vec{r}',t_0)) = \frac{\lambda^2}{\pi} \int \exp\left[-\lambda^2 \frac{\vec{r}_1^2 + \vec{r}_2^2}{2} + \lambda[a\bar{z}_1(x_2-iy_2) + b\bar{z}_2(x_1-iy_1)] + \right.$$
$$\left. \lambda[az_1(x_1+iy_1) + bz_2(x_2+iy_2)] - ab(z_1\bar{z}_2 + \bar{z}_1 z_2)\right] d\mu(z_1) d\mu(z_2)$$

This integral is invariant by changing $z_2 \leftrightarrow \bar{z}_2$, and then we can use the well

known formula:

$$\left(\frac{1}{\pi}\right)^n \int \prod_{i=1}^{n} dx_i dy_i \exp(-\bar{z}^t X z + A^t z + \bar{z}^t \bar{B}) = (\det(X))^{-1} \exp(A^t X^{-1} \bar{B})$$

We find that
$$X = \begin{pmatrix} 1 & ab \\ ab & 1 \end{pmatrix}$$

And
$$\det(X) = 1 - a^2 b^2 = 1 - e^{-i2\alpha}$$

By an elementary calculation we find that:
$$A^t X^{-1} \bar{B} = \lambda^2 \left\{ -a^2 b^2 (\vec{r}_1^2 + \vec{r}_2^2) + a^2 (x_1 + iy_1)(x_2 + iy_2) + b^2 (x_2 - iy_2)(x_1 - iy_1) \right\} / (1 - a^2 b^2) \quad (7.12)$$

Then the propagator may be written:

$$K((\vec{r},t),(\vec{r}',t_0)) = \left(\frac{\mu\omega}{\pi\hbar}\right) \frac{e^{-i\omega(t-t_0)}}{1-e^{-i2\alpha}} \exp\left\{\lambda^2 \left[ -(\vec{r}_1^2 + \vec{r}_2^2)(1 + a^2 b^2)/2 + (a^2 + b^2)(x_1 x_2 + y_1 y_2) - i(a^2 - b^2)(x_1 y_2 - y_1 x_2) \right] / (1 - a^2 b^2) \right\} \quad (7.13)$$

It is easy to verify the following identities

$$\frac{e^{-i\alpha}}{(1-a^2 b^2)} = \frac{1}{2i \sin\alpha}, \qquad \frac{(1+a^2 b^2)}{(1-a^2 b^2)} = -i\frac{\cos\alpha}{2\sin\alpha}.$$

And
$$\frac{(a^2 + b^2)}{(1-a^2 b^2)} = -i\frac{\cos\beta}{\sin\alpha}, \qquad \frac{(a^2 - b^2)}{(1-a^2 b^2)} = \frac{\sin\beta}{\sin\alpha}$$

Substitute these relations in (7.13) we obtain the exact expression of Feynman Propagator of a charged harmonic oscillator in constant magnetic field:

$$K((\vec{r},t),(\vec{r}',t_0)) = \left(\frac{\mu\omega}{2i\pi\hbar}\right) \frac{1}{\sin(\omega\tau)} \exp\left\{\frac{i\mu\omega}{\hbar}\left[\frac{\cos(\omega\tau)}{2\sin(\omega\tau)}(\vec{r}_1^2 + \vec{r}_2^2) + -\frac{\cos(\omega_c \tau)}{\sin(\omega\tau)}(x_1 x_2 + y_1 y_2) - \frac{\sin(\omega_c \tau)}{\sin(\omega\tau)}(x_1 y_2 - y_1 x_2)\right]\right\} \quad (7.14)$$

I leave the reader to compare between our method and Schwinger's method, Reference [18] part B.

It is important to emphasize that the expression (7.14) may be obtained by the application of the transformation from the coordinates representations to the harmonic oscillator basis.

# Chapter two

# Angular momentum and Fock –Bargmann space

**1. Introduction**
**2. Schwinger approach for angular momentum**
   2.1 Preliminary
   2.2 The Boson polynomial basis of SU(2)
   2.3 Finite rotations and its generating function
   2.4 Orthogonality and normalization of D-Wigner matrix elements
   2.5 projection operator and the matrix elements of rotation
   2.6 Expression of the matrix elements
   2.7 Particulars cases of the matrix elements of rotation
**3. The spherical harmonic and the quadratic transformation $R^4 \to R^3$**
   3.1 The generating function of the spherical harmonic
   3.2 The quadratic transformation $R^4 \to R^3$
   3.3 The connection between hydrogen atom and the harmonic oscillator
   3.4 Some applications of the generating function of spherical harmonics
**4. The addition of angular momentum**
   4.1 expression of the integral over the product of three D
   4.2 expressions of 3j symbols of SU (2)
   4.3 Euler's identity and Regge symmetry of SU(2)
**5. The invariant polynomials of the 3-j symbols**
   5.1 The invariant polynomials and the D-Wigner matrix elements
   5.2 The invariant polynomials and the generating function
        of spherical harmonics
   5.3 Generating function of the invariants of SU(2)
   5.4 The Van der Wearden formula for 3j symbols
**6. Schwinger's Approach for the coupling and 6j symbols of SU(2)**
   6.1 Schwinger's Approach for the coupling
   6.2 The generating function of the 6j symbols of SU (2)
   6.3 Symmetry of 6j symbols
**7. Appendices**

# Chapter two

# Angular momentum and Fock –Bargmann space

## 1. Introduction

It is well known that the orbital angular momentum $\vec{L} = \vec{r} \times \vec{p}$ plays a central role in classical and Quantum mechanics. But in quantum mechanics we can write the components $\{L_x, L_y, L_z\}$ as quadratic forms in terms of creation and annihilation operators, and the matrix of spin -1, of the three-dimensional harmonic oscillators. Schwinger was observed that the use of spin half leads simply for the study of angular momentum with the help of two dimensions harmonic oscillator basis [23]. But Bargmann used the isomorphism between the Fock space and the basis of the oscillator to redo the famous Schwinger's work "on angular momentum "in the Fock basis [24-25] but the calculation of the 3n-j symbols became difficult for n>2.

In this work we use the Schwinger's generating function of elements of the matrix of rotations to determine the generating function of the spherical harmonics and we find a quadratic transformation $R^4 \to R^3$ very useful in physics [47]. We find using the Gaussian integral the generating functions: of Gegenbauer, Legendre polynomials and the characters of SU(2) [54].

In many physical problems we find that the wave functions is not eigenfunctions of angular momentum therefore we need the finite projection operator to obtain the good wave function [66-73]. The infinitesimal projection operators is given in the appendix.

We simply deduct the invariant of SU (2), or Van der Warden invariant, which can be generalized for the determination of Wigner's symbols of unitary groups [108-113].

We also show that the calculation of generating functions symbols 3n-j can be done simply by using symbolic computation programs.

## 2. Schwinger approach for angular momentum

We review the properties of spherical harmonics and the Schwinger basis of SU (2) and the Wigner's D- matrix elements.

### *2.1* Preliminary

***Problem***: If f and g are two quadratics forms

$$f = \sum_{ij}^{n} f_{ij} a_i^+ a_j = (a^+)^t (f)(a), \quad g = \sum_{ij}^{n} g_{ij} a_i^+ a_j = (a^+)^t (g)(a)$$

Using the commentator: $[a_i, a_j] = 0, [a_i^+, a_i^+] = 0, [a_i^+, a_j] = \delta_{ij}$

It's simple to prove that: $[f, g] = (a^+)[(f)(g) - (g)(f)](a)$

**Application**: if $(f) = (g) = \vec{S} = \vec{\sigma}/2$, $\vec{\sigma}$ Are the Pauli matrices with spin 1
With

$$S_x = \begin{pmatrix} 0 & -i & 0 \\ i & 0 & 0 \\ 0 & 0 & 0 \end{pmatrix}, S_y = \begin{pmatrix} 0 & 0 & i \\ 0 & 0 & 0 \\ -i & 0 & 0 \end{pmatrix}, S_z = \begin{pmatrix} 0 & 0 & 0 \\ 0 & 0 & -i \\ 0 & i & 0 \end{pmatrix},$$

We write the orbital angular momentum:

$$\vec{L} = \vec{r} \times \vec{p} = (a^+)^t (\vec{\sigma})(a),$$

With: $(a^+)^t = (a_x^+, a_y^+, a_z^+)$.

It's simple to verify that the solid harmonics $Y_{\ell m}(\vec{r}) \, \vec{r} = (x, y, z)$ are eigenfunctions of $\vec{L}^2$ and $L_z$ with:

$$\vec{L}^2 Y_{\ell m}(\vec{r}) = \ell(\ell + 1) Y_{\ell m}(\vec{r}), \text{ and } L_z Y_{\ell m}(\vec{r}) = m Y_{\ell m}(\vec{r})$$

### *2.2 The boson polynomial basis of SU(2)*

Schwinger in his work [1] "on angular momentum" use the spin 1/2 instead of spin 1 and find the generators of SU(2) or the infinitesimal operators in terms of creation and destruction operators of two-dimensional harmonic oscillator:
we write:

$$\vec{J} = \begin{pmatrix} a_1^+ & a_2^+ \end{pmatrix} \left(\frac{\vec{\sigma}}{2}\right) \begin{pmatrix} a_1 \\ a_2 \end{pmatrix}$$

$\vec{\sigma}$ Are the Pauli matrices with spin 1/2:

$$\vec{S}_{1/2} = \vec{\sigma}/2, \sigma_1 = \begin{pmatrix} 0 & 1 \\ 1 & 0 \end{pmatrix}, \sigma_2 = \begin{pmatrix} 0 & -i \\ i & 0 \end{pmatrix}, \sigma_3 = \begin{pmatrix} 1 & 0 \\ 0 & -1 \end{pmatrix},$$

We find

$$J_+ = J_1 + iJ_2 = a_1^+ a_2, \, J_- = J_1 - iJ_2 = a_2^+ a_1, \, J_3 = [a_1^+ a_1 - a_2^+ a_2]/2$$

And $\quad N = [a_1^+ a_1 + a_2^+ a_2]/2 \Rightarrow \vec{J}^2 = N(N+1)$

We formally write

$$N = [a_1^+ \partial/\partial a_1 + a_2^+ \partial/\partial a_2]/2, \, J_3 = [a_1^+ \partial/\partial a_1 - a_2^+ \partial/\partial a_2]/2$$

After the Euler' theorem the eigenfunctions of $J_3$, $N$ and $\vec{J}^2$ is the homogeneous functions:

$$|jm\rangle = \frac{a_1^{+(j+m)} a_2^{+(j-m)}}{\sqrt{(j+m)!(j-m)!}} |0,0\rangle$$

With $\quad j \geq |m|,\ j = 0,\ 1/2,\ 1,\ 3/2,\ 2,\ldots$

We find the Fock Bargmann basis by applying the transformations:

$$a_1^+ \to \xi,\ a_2^+ \to \eta,\ \text{and}\ a_1 \to \partial/\partial\xi,\ a_2 \to \partial/\partial\eta$$

$$\varphi_{lm}(u) = \frac{\xi^{l+m}\eta^{l-m}}{\sqrt{(l+m)!(l-m)!}},\ u = (\xi,\eta) \tag{2.1}$$

we denote $d\mu(u)$ by the measure of integration

$$d\mu(u) = d\mu(\xi)d\mu(\eta) \tag{2.2}$$

The conjugate representation of $\varphi_{jm}(z)$ in the space of Fock-Bargmann is

$$(\varphi_{jm}(z))_c = (-1)^{j-m}\varphi_{j-m}(z)$$

## 2.3 Finite rotations and its generating function

The matrix of rotation [1-7] can be deduced simply from the property $RR^* = 1$.
Using Euler angles we write:

$$D^j_{(m',m)}(\Omega) = \langle jm'|R^J|jm\rangle = \langle jm'|e^{-i\psi J_3}e^{-i\theta J_2}e^{-i\varphi J_3}|jm\rangle$$
$$= e^{-i\psi m'}d^j_{(m',m)}(\theta)e^{-i\varphi m} \tag{2.3}$$

And $\quad R_s(\Omega) = e^{-i\psi S_3}e^{-i\theta S_3}e^{-i\varphi S_3},\ \Omega = (\psi,\theta,\varphi)$

Multiplying by $\varphi_{jm'}(u)\varphi_{jm}(v)r^{2j}$ and after the summation we find the
Generating function of the matrix elements of rotation

$$\sum_{jmm'}\varphi_{jm'}(u)D^j_{(m',m)}(z)\varphi_{jm}(v) = \exp[{}^t(u)(R_s(z))(v)]$$

$$\phi(u,v,z) = \exp[(\varsigma\ \eta)\begin{pmatrix}z_1 & -\bar{z}_2 \\ z_2 & \bar{z}_1\end{pmatrix}\begin{pmatrix}\chi \\ \lambda\end{pmatrix}] \tag{2.4}$$

with

$$z_1 = \rho\exp(\varphi_1)\cos(\frac{\theta}{2}),\quad z_2 = \rho\exp(\varphi_2)\sin(\frac{\theta}{2})$$

$$\Omega = (\psi\theta\varphi),\ \varphi_1 = (\frac{\psi+\varphi}{2}),\ \varphi_2 = (\frac{\psi-\varphi}{2})$$

and $\quad 0 \leq \rho \leq \infty,\ 0 \leq \varphi \leq 2\pi,\ 0 \leq \theta \leq \pi,\ 0 \leq \psi \leq 2\pi$

$$R_s(z) = \rho R_s(\Omega) = \begin{pmatrix}z_1 & -\bar{z}_2 \\ z_2 & \bar{z}_1\end{pmatrix},\ D^j_{(m',m)}(z) = \rho^{2j}D^j_{(m',m)}(\Omega)$$

It is clear that:

$$(\frac{\partial^2}{\partial z_1 \partial\bar{z}_1} + \frac{\partial^2}{\partial z_2 \partial\bar{z}_2})D^j_{(m',m)}(z) = 0 \tag{2.5}$$

This expression may be generalized to SU(n) groups.

## 2.4 Orthogonality and normalization of D-Wigner matrix elements

We have

$$\int (\int (D^{j_1}_{(m_1',m_1)})^*(z) D^{j_2}_{(m_2',m_2)}(z) e^{-\bar{z}z} d\mu(z) = \int \rho^{2(j_1+j_2)} e^{-\bar{z}z} d\rho \int (D^{j}_{(m',m)})^*(\Omega) D^{j}_{(m',m)}(\Omega) d(\Omega)$$

With $z = (z^1, z^2)$ and $d\mu(z) = d\mu(z^1) d\mu(z^2)$.
Using the generating (2.4) we find

$$[\sum_{j_1 m_1 m_1'} \varphi_{j_1 m_1'}(u) \varphi_{j_1 m_1}(v)] \times [\sum_{j_1 m_1 m_1'} \varphi_{j_1 m_1'}(x) \varphi_{j_1 m_1}(y)] \int (D^{j}_{(m_1',m_1)})^*(z) D^{j}_{(m_1',m_1)}(z) d\mu = \quad (2.6)$$

$$\int \exp[(u_1 \ u_2) \begin{pmatrix} z_1 & -\bar{z}_2 \\ z_2 & \bar{z}_1 \end{pmatrix} \begin{pmatrix} v_1 \\ v_2 \end{pmatrix}] \exp[(x_1 \ x_2) \begin{pmatrix} z_1 & -\bar{z}_2 \\ z_2 & \bar{z}_1 \end{pmatrix} \begin{pmatrix} y_1 \\ y_2 \end{pmatrix}] d\mu(z) =$$

$$\int \exp[{}^t(u)(R_s(z))(v) + {}^t(x)(R_s(z))(y)] d\mu(z)$$

After integration we find that:

$$\int \exp[{}^t(u)(R_s(z))(v) + {}^t(x)(R_s(z))(y)] d\mu(z) = \quad (2.7)$$
$$\exp[u_1 v_1 x_2 y_2 + u_2 v_2 x_1 y_1 - v_2 u_1 x_2 y_1 - u_2 v_1 y_2 x_1]$$

After expansion of (2.7) and it's identification with (2.6) we find:

$$\frac{1}{8\pi^2} \int (D^{j_1}_{(m_1',m_1)})^*(\Omega) D^{j_2}_{(m_2',m_2)}(\Omega) d(\Omega) = \frac{1}{2j_1+1} \delta_{m_1'm_2'} \delta_{m_1 m_2} \delta_{j_1 j_2} \quad (2.8)$$

## 2.5 The finite projection operator

With the help of (2.8) we find that the angular momentum projection operator is:

$$P^{j}_{(m,k)} = \frac{2j+1}{8\pi^2} \int (D^{j}_{(m,k_1)})^*(\Omega) R(\Omega) d(\Omega) \quad (2.9)$$

And the projection of the wave function $|\Psi_k\rangle$ is $|\Psi^{j}_{(m,k)}\rangle$:

$$|\Psi^{j}_{(m,k)}\rangle = \frac{2j+1}{\sqrt{N_{j,k}} 8\pi^2} \int (D^{j}_{(m,k_1)})^*(\Omega) R(\Omega) |\Psi_k\rangle d(\Omega)$$

With

$$N_{j,k} = \frac{2j+1}{8\pi^2} \int (D^{j}_{(m,k_1)})^*(\Omega) \langle \Psi_k | R(\Omega) | \Psi_k \rangle d(\Omega)$$

We write the projection operator in the general case by:

$$P^{j}_{(m,k)} = \frac{2j+1}{8\pi^2} \int (D^{j}_{(m,k_1)})^*(\Omega) R(\Omega) d(\Omega) = \sum_{\alpha} |\alpha j m\rangle \langle \alpha j m| \quad (2.10)$$

{α} Is a set of quantum numbers.

This projection operator is called sometimes Hill-Wheeler integral.

*2.6 Expression of the matrix elements*
From the expansion of the generating function we derive the expression of the matrix elements of finite rotations in terms of the Jacobi polynomial.

$$d^j_{(m',m)}(\theta) = \left[\frac{(j+m')!(j-m')!}{(j+m)!(j-m)!}\right]^{-1/2} (\cos(\theta/2))^{m'+m} (\sin(\theta/2))^{m'-m} P^{(m'-m,m'+m)}_{j-m'}(\cos\theta) \quad (2.11)$$

*2.7 Particulars cases of the matrix elements of rotation*
We can derive the expressions of the following particulars cases:

$$c - D^j_{(j,m)}(\Omega) = \sqrt{\frac{(2j)!}{\rho^{4j}}} \varphi_{jm}(z_1,-\bar{z}_2) \quad d - D^j_{(-j,m)}(\Omega) = \sqrt{\frac{(2j)!}{\rho^{4j}}} \varphi_{jm}(z_2,\bar{z}_1)$$

$$a - D^j_{(m',j)}(\Omega) = \sqrt{\frac{(2j)!}{\rho^{4j}}} \varphi_{jm'}(z_1,z_2) \quad b - D^j_{(m',-j)}(\Omega) = \sqrt{\frac{(2j)!}{\rho^{4j}}} \varphi_{jm'}(-\bar{z}_2,\bar{z}_1)$$

$$e - D^l_{(m,0)}(\Omega) = \left(\frac{4\pi}{2l+1}\right)^{1/2} Y^*_{lm}(\theta,\varphi) \quad (2.12)$$

**Remarque's:**
1- We observe that the expression (e) is a transformation from the four dimensions to the three dimensions
2- We shall use later this important property to derive the momentum representation of Hydrogen atom.

# 3. The spherical harmonic and the quadratic transformation $R^4 \to R^3$

The relation between the Wigner's D matrix and the spherical harmonics is given by:

$$D^l_{(m,0)}(z) = \sqrt{\frac{4}{(2l+1)}} Y_{lm}(\vec{r}) \quad (3.1)$$

With: $\rho^2 = r$

We observe that the expression is a transformation from the four dimensions to the three dimensions but here we shall use the generating function for the derivation of the transformation.

*3.1 The generating function of the spherical harmonic*
We find from (2.4) that the generating function of spherical harmonics is:

$$\int e^{\lambda(\bar{v}_1\bar{v}_2)} \exp[(u_1 \ \ u_2)\begin{pmatrix} z_1 & -\bar{z}_2 \\ z_2 & \bar{z}_1 \end{pmatrix}\begin{pmatrix} v_1 \\ v_2 \end{pmatrix}]d\mu(v_1,v_2) = \exp[\frac{\lambda(\vec{a}\cdot\vec{r})}{2}] \quad (3.2)$$

$$= \sum_{lm}[\frac{4\pi}{2l+1}]^{\frac{1}{2}}\lambda^l \varphi_{lm}(u)Y_{lm}(\vec{r})$$

$\vec{a}$ is a vector of length zero, $\vec{a}\cdot\vec{a} = 0$ and has the components

$$a_1 = -u_1^2 + u_2^2, \ a_2 = -i(u_1^2 + u_2^2), \ a_3 = 2u_1 u_2 \quad (3.3)$$

### 3.2 The quadratic transformation $R^4 \to R^3$

We obtain the quadratic transformation $R^4 \to R^3$ from (3.2) or terms of coordinates $\vec{r} = (x,y,z)$ we find:

$$x = z_1\bar{z}_2 + z_2\bar{z}_1, \ y = i(z_1\bar{z}_2 - z_2\bar{z}_1), \ z = z_1\bar{z}_1 - z_2\bar{z}_2 \quad (3.4)$$

We can write these expressions in term of spin half as

$$x = (\bar{z}_1 \ \ \bar{z}_2)\begin{pmatrix} 0 & 1 \\ 1 & 0 \end{pmatrix}\begin{pmatrix} z_1 \\ z_2 \end{pmatrix}, \quad y = (\bar{z}_1 \ \ \bar{z}_2)\begin{pmatrix} 0 & -i \\ i & 0 \end{pmatrix}\begin{pmatrix} z_1 \\ z_2 \end{pmatrix},$$

$$z = (\bar{z}_1 \ \ \bar{z}_2)\begin{pmatrix} 1 & 0 \\ 0 & -1 \end{pmatrix}\begin{pmatrix} z_1 \\ z_2 \end{pmatrix} \quad (3.5)$$

If we put $z_1 = u_1 + u_2, \ z_2 = u_3 + u_4$ we find

$$x = 2(u_1 u_3 + u_2 u_4), \ y = 2(-u_1 u_4 + u_2 u_3), \ z = u_1^2 + u_2^2 - u_3^2 - u_4^2 \quad (3.6)$$

$$\rho = \sqrt{r}, \ r = z_1\bar{z}_1 + z_2\bar{z}_2, \ r^2 = x^2 + y^2 + z^2$$

The quadratic transformation $R^4 \to R^3$ [8-11] or Octonions quadratics transformation (or Hurwitz transformation) corresponds to the transformation introduced by Kustaanheimo and Steifel up to permutation on x, y, z and the $u_i's$. Recently we used it for the derivation of the momentum representation of hydrogen atom [ch.5].

We also observe that expression (3.5) can be extended to the transformation $R^8 \to R^5$ using the Dirac matrices that we will then make a generalization.

### 3.3 The connection between hydrogen atom and the harmonic oscillator

A quick calculation shows that the equation of the hydrogen atom

$$(-\frac{\hbar^2}{2\mu}\Delta - \frac{Ze^2}{r})\Psi = E\Psi . \ (\mu \text{ is the reduced mass}). \quad (3.7)$$

That may be written on the basis of harmonic oscillator in the form

$$(-\frac{\hbar^2}{2\mu}\sum_{i=1}^{4}\frac{\partial^2}{\partial u_i^2} - 4E\vec{u}^2)\Psi = -4Ee^2\Psi \quad (3.8)$$

With a constraint on the eigenfunctions: $\dfrac{\partial}{\partial \psi}\Psi = 0$

And $\qquad \omega = \sqrt{-8E/\mu}, \quad 4Ze^2 = \hbar\omega(n+2)$

And the energy is given by:

$$E = -2\mu\left(\frac{Ze^2}{\hbar(n+2)}\right)^2 \quad (3.9)$$

### 3.4 Some applications of the generating function of spherical harmonics

All calculations that we perform in the Fock Bargmann space can be solved with the Gaussian integrals in finite dimensions:

$$(1/\pi^n)\int\prod_{i=1}^{n}dx_i dy_i \exp\left(-\bar{z}^t Xz + A^t z + \bar{z}^t \overline{B}\right) = (\det(X))^{-1}\exp(A^t X^{-1}\overline{B}) \quad (3.10)$$

With $z = (z_1, z_2, ..., z_n)$

### 3.4.1 Generating function for Legendre polynomials

We put $u_1 = \bar{u}_2$ in the formula (3.10) and using (3.13) we find the generating function of Legendre polynomials.

$$\int e^{\vec{r}\cdot\vec{a}/2}d\mu(\eta) = \sum_{l=0}^{\infty} r^l P_l(\cos\theta) = \frac{1}{\sqrt{1-2r\cos\theta+r^2}} \quad (3.11)$$

### 3.4.2 Generating function of the characters $\chi(R)$ of SU (2).

In the generating function of matrix-D we replace (u) by $(\bar{v})$ we get after integration:

$$\int\exp[^t(\bar{v})(A)(v)]d\mu(v) = \sum_j r^{2j}\sum_m D^j_{(m,m)}(R) = \sum_j r^{2j}\chi(R)$$
$$= 1/(1-2r\cos\frac{\theta}{2}\cos(\varphi+\psi)+r^2) \quad (3.12)$$

## 4. The addition of angular momentum

If we consider a system of two particles the conservation of angular momentum imply that the state of the system is decomposition on the product of the states of one particle.
We have $\qquad \vec{J}_3 = \vec{J}_1 + \vec{J}_2$

$$J_1^2|j_1 m_1\rangle = j_1(j_1+1)|j_1 m_1\rangle, \quad J_2^2|j_2 m_2\rangle = j_2(j_2+1)|j_2 m_2\rangle$$
$$J_3^2|(j_1 j_2)j_3 m\rangle = j_3(j_3+1)|(j_1 j_2)j_3 m\rangle \quad (4.1)$$

And

$$|(j_1 j_2)j_3 m\rangle = \sum_{m_1 m_2}\langle j_1 m_1, j_2 m_2 \| (j_1 j_2)j_3 m_3\rangle|j_1 m_1\rangle|j_2 m_2\rangle$$
$$|j_1 m_1\rangle|j_2 m_2\rangle = \sum_j \langle j_1 m_1, j_2 m_2 \| (j_1 j_2)j_3 m_3\rangle|(j_1 j_2)j_3 m\rangle \quad (4.2)$$

$$D^{j_3}_{(m_3',m_3)}(\Omega) = \langle (j_1 j_2) j_3 m' | R^{J_3} | (j_1 j_2) j_3 m \rangle \tag{4.3}$$

The Wigner 3j symbols for angular momentum may be written in terms of the Clebsh-Gardan coefficients $\langle j_1 m_1, j_2 m_2 \| (j_1 j_2) j_3 m_3 \rangle$ like that:

$$\begin{pmatrix} j_1 & j_2 & j_3 \\ m_1 & m_2 & m_3 \end{pmatrix} = (-1)^{j_1-j_2-m_2}(2j_3+1)^{-1/2} \langle j_1 m_1, j_2 m_2 \| (j_1 j_2) j_3 - m_3 \rangle \tag{4.4}$$

With $m = -m_3 = m_1 + m_2$ and $m' = -m_3' = m_1' + m_2'$

### 4.1 expression of the integral over the product of three D's:
Using the above expressions we write:

$$\langle j_1 m_1 | \langle j_2 m_2 | R^{J_3} | j_1 m_1' \rangle | j_2 m_2' \rangle = D^{j_1}_{(m_1,m_1')}(\Omega) D^{j_2}_{(m_2,m_2')}(\Omega) =$$

$$\sum_{jm} \begin{pmatrix} j_1 & j_2 & j_3 \\ m_1 & m_2 & m_3 \end{pmatrix} \begin{pmatrix} j_1 & j_2 & j_3 \\ m_1 & m_2 & m_3 \end{pmatrix} D^{*j_3}_{(m_3,m_3')}(\Omega)$$

In multiply by $D^{*j_3}_{(m_3,m_3')}(\Omega)$ and perform the integration we find:

$$\int d\Omega D^{j_1}_{(m_1,m'_1)}(\Omega) D^{j_2}_{(m_2,m'_2)}(\Omega) D^{j_3}_{(m_3,m'_3)}(\Omega) = \begin{pmatrix} j_1 & j_2 & j_3 \\ m'_1 & m'_2 & m'_3 \end{pmatrix} \begin{pmatrix} j_1 & j_2 & j_3 \\ m_1 & m_2 & m_3 \end{pmatrix} \tag{4.5}$$

This expression is known by Gaunt formula or the integral of the product of three D.

### 4.2 expressions of 3j symbols of SU (2)
we find first the integral representation of SU (2) and the sets of generalized hypergeometric functions for 3j symbols. Then we deduce from Euler identity the Regge [48] symmetry of these symbols.

#### 4.2.1 Integral representation of 3j symbols
Using the Gaunt formula we can calculate the particular case using (4.5) or much simpler from (3.6):

$$\begin{pmatrix} j_1 & j_2 & j_3 \\ j_1 & -j_2 & j_2-j_1 \end{pmatrix} = (-1)^{-2j_1+2j_2} \sqrt{\frac{(2j_1)!(2j_2)!}{(j_1+j_2+j_3+1)!(j_1+j_2+j_3)!}} \tag{4.6}$$

Put $m'_1 = j_1$, $m'_2 = -j_2$ and $m'_3 = j_2 - j_1$.
We deduce the integral representation of 3j symbols:

$$\begin{pmatrix} j_1 & j_2 & j_3 \\ m_1 & m_2 & m_3 \end{pmatrix} = \frac{(-1)^{2(j_2-j_1)+j_2+m_2} \Gamma_3}{\Gamma_2 \Gamma_1 \sqrt{(J-2j_2)!(J-2j_1)!}} \int_0^\pi [(\cos\frac{\theta}{2})^{2(j_2-m_2)+1}(\sin\frac{\theta}{2})^{2(j_1-m_1)+1}$$

$$\times P^{((m_3+j_1-j_2,\,m_3-j_1+j_2)}_{j_3-m_3}(\cos\theta)] d\theta \tag{4.7}$$

with $\Gamma_i = \sqrt{(j_i - m_i)!(j_i + m_i)!}$

### *4.2.2 Wigner's expression for 3j symbols*

We write the Jacobi polynomials in terms of $\cot^2(\theta/2)$ in the expression (4.7) and after integration we get the Wigner's expression for 3j symbols:

$$\begin{pmatrix} j_1 & j_2 & j_3 \\ m_1 & m_2 & m_3 \end{pmatrix} = (-1)^{j_2 - j_1 + m_3} \Delta(j,m) \frac{(j_1 + j_2 + m_1)!}{(j_1 - j_2 + m_3)!} \times$$
$$_3F_2(-j_3 + m_3, -j_3 + j_1 - j_2, j_1 - m_1 - 1; j_1 - j_2 + m_3 + 1, -j_3 - j_2 - m_1; 1) \quad (4.8)$$

### *4.3 Euler's identity and Regge symmetry of SU(2)*

We determine the symmetries of the 3j symbols by new method may be generalized to other problems. We write the expression of Jacobi polynomials (4.7) in terms of hypergeometric functions [60] and then we use the Euler identity

$$F(\alpha, \beta; \gamma; z) = (1-z)^{\gamma - \alpha - \beta} F(\gamma - \alpha, \gamma - \beta; \gamma; z) \quad (4.9)$$

we find

$$\int_0^\pi d\theta (\sin^2(\theta/2))^\rho (\cos^2(\theta/2))^\sigma {}_2F_1(n + \alpha + \beta + 1, -n; 1 + \beta; \cos^2(\theta/2)) =$$
$$\int_0^\pi d\theta (\sin^2(\theta/2))^{\rho - \alpha} (\cos^2(\theta/2))^\sigma {}_2F_1(-n - \alpha, 1 + \beta + n; 1 + \beta; \cos^2(\theta/2)) \quad (4.10)$$

To find the symmetries we assume that after transformation we obtain the same expression but with the new indices.
We find the new indices $n', \alpha', \beta', \rho', \sigma'$ in terms of the old one:

$$n' = n + \alpha, \alpha' = -\alpha, \beta' = \beta, \rho' = \rho - \alpha, \sigma' = \sigma$$

In our case we find the Regge symmetry.

$$\begin{pmatrix} j_1 & j_2 & j_3 \\ m_1 & m_2 & m_3 \end{pmatrix} = \begin{pmatrix} (j_1 + j_2 - m_3)/2 & (j_1 + j_2 + m_3)/2 & j_3 \\ (j_1 - j_2 + m_1 - m_2)/2 & (j_1 - j_2 + m_1 - m_2)/2 & (-j_1 + j_2) \end{pmatrix} \quad (4.11)$$

## 5. The invariant polynomials of the 3-j symbols

Van der Wearden [49] determined the invariant of SU(2), method known to Weyl [9], and deduces the Wigner 3j symbols of this group. We will determine first the Van der Wearden invariant using the D-Wigner matrix elements and then the generating function of spherical harmonics.

### *5.1 The invariant polynomials and the D-Wigner matrix elements*

The integral of the product of three generating functions of D-Wigner matrix elements

$$G^3 = \int \prod_{i=1}^3 [\phi(x^i, y^i, z^i)] d\mu(z_1) d\mu(z_2) = \sum_{j_1 j_2 j_3} H_{(j_1 j_2 j_3)}(x) H_{(j_1 j_2 j_3)}(y) \quad (5.1)$$

Using the expression $\int \exp[\alpha z + \beta \bar{z}]d\mu(z) = e^{\alpha\beta}$ for integration we find:

$$G^3 = \exp[[x^1,x^2][y^1,y^2] + [x^1,x^3][y^1,y^3] + [x^2,x^3][y^2,y^3]]$$

With $x^i = (\xi_i, \eta_i), y^i = (u_i, v_i)$ and $J = j_1 + j_2 + j_3$
After the identification of two sides of (5.1) we find the invariant:

And $$H_{(j_1j_2j_3)}(x) = \frac{[x^2x^3]^{(J-2j_1)}[x^3x^1]^{(J-2j_2)}[x^1x^2]^{(J-2j_3)}}{\sqrt{(J+1)!(J-2j_1)!(J-2j_2)!(J-2j_3)!}} \tag{5.2}$$

We denote $H_{(j_1j_2j_3)}(x)$ by the invariant polynomials of SU (2)
And $[x^2x^3]$, $[x^3x^1]$, $[x^1x^2]$ are the elementary invariants of SU (2).

## 5.2 The invariant polynomials and the generating function of spherical harmonics

$$I = \int \exp[-\vec{r}^2 + 2(\alpha_1\vec{a}_1\cdot\vec{r} + \alpha_2\vec{a}_2\cdot\vec{r} + \alpha_3\vec{a}_3\cdot\vec{r})]dxdydz \tag{5.3}$$

With $u^1 = (\xi_1, \eta_1), u^2 = (\xi_2, \eta_2), u^3 = (\xi_3, \eta_3),$

And $\vec{a}_i \cdot \vec{a}_j = -2[u^i u^j]^2$ avec $[u^i u^j] = [\xi_i\eta_j - \eta_i\xi_j]$.

After integration we obtain:

$$I = \exp[-2\alpha_1\alpha_2\vec{a}_1\cdot\vec{a}_2 - 2\alpha_1\alpha_3\vec{a}_1\cdot\vec{a}_3 - 2\alpha_2\alpha_3\vec{a}_2\cdot\vec{a}_3] \tag{5.4}$$

the development of the integral (5.3) gives

$$I = \sum_{l_i} 2^{4L} [\prod_{i=1}^{3}(\alpha_i^{l_i}\sqrt{\frac{4\pi}{2l_i+1}})] \times [\int_0^\infty \exp\{-r^2\}r^{L+2}dr]$$
$$\times \left[\sum_{m_i}\prod_{i=1}^{3}\varphi_{l_im_i}(u^i)\int\prod_{i=1}^{3}Y_{l_im_i}(\theta\varphi)\right]\sin\theta d\theta d\varphi \tag{5.5}$$

With $L = l_1 + l_2 + l_3$
We use the well-known result of the theory of angular momentum

$$\int\prod_{i=1}^{3}Y_{l_im_i}(\theta\varphi)\sin\theta d\theta d\varphi = \prod_{i=1}^{3}(\frac{(2l_i+1)}{4\pi})]^{\frac{1}{2}}\begin{pmatrix}l_1 & l_2 & l_3\\0 & 0 & 0\end{pmatrix}\begin{pmatrix}l_1 & l_2 & l_3\\m_1 & m_2 & m_3\end{pmatrix}. \tag{5.6}$$

After the integration of (5.5) and the identification with the second member of the integral we obtain the Van der Wearden invariant $H_{(j_1 j_2 j_3)}(u)$ of 3-j symbols:

$$H_{(l_1 l_2 l_3)}(u) = \sum_{m_i} \left[\prod_{i=1}^{3} \varphi_{l_i,m_i}(u^i)\right] \begin{pmatrix} l_1 & l_2 & l_3 \\ m_1 & m_2 & m_3 \end{pmatrix}, \; m_1 + m_2 + m_3 = 0 \qquad (5.7)$$

This is the same expression as above.

### 5.3 Generating function of the invariants of SU(2)

We deduce the generating function of 3j symbols of SU (2) from (5.2) by multiplying it by:

$$\varphi_{l_3(l_1 l_2)}(\tau) = [(L+1)]^{\frac{1}{2}} \frac{[\tau_3]^{(L-2l_3)} [\tau_2]^{(L-2l_2)} [\tau_1]^{(L-2l_1)}}{\sqrt{(L-2l_3)(L-2l_2)(L-2l_1)}}, \qquad (5.8)$$

After summing with respect to $j_i = l_i = 0, 1/2, 1, ...$ we obtain the Schwinger formula:

$$\sum_{j_i,m_i} \varphi_{j_3(j_1 j_2)}(\tau) \left[\prod_{i=1}^{3} \varphi_{j_i,m_i}(u^i)\right] \begin{pmatrix} j_1 & j_2 & j_3 \\ m_1 & m_2 & m_3 \end{pmatrix} = \exp \begin{pmatrix} \begin{vmatrix} \tau_1 & \tau_2 & \tau_3 \\ \xi_1 & \xi_2 & \xi_3 \\ \eta_1 & \eta_2 & \eta_3 \end{vmatrix} \end{pmatrix} \qquad (5.9)$$

the symmetries of the 3j symbols can be deduced from the invariance of the determinant: permutation of columns, permutation of rows and transposition.

### 5.4 The Van der Wearden formula for 3j symbols

The Van der Wearden formula for 3j symbols can be derived simply form (5.2):

$$\begin{pmatrix} l_1 & l_2 & l \\ m_1 & m_2 & m_3 \end{pmatrix} = (-1)^{2(l_2 - l_1)} \Delta(l,m) \frac{(l-l_1+l_2)!(l-m_3)!(l_2-m_2)!}{(-l+l_1+l_2)!(l-l_2+m_1)!(l-l_1-m_2)!} \times$$

$$_3F_2(-l_2-m_2, -l_1+m_1, l-l_1-m_2; l-l_1-m_2+1, l-l_2+m_1+1; 1) \qquad (5.10)$$

With

$$\Delta(l,m) = (-1)^{l_2+m_2} \sqrt{\frac{(l+l_1-l_2)!(-l+l_1+l_2)!(l+m_3)!(l_1-m_1)!}{(l-l_1+l_2)!(l+l_1+l_2+1)!(l-m_3)!(l_1+m_1)!(l_2+m_2)!(l_2-m_2)!}}$$

The method of invariants has been the subject of many studies [24, 49-51] but the generalization of this method for SU (n) for n> 3 is very difficult.

## 6. Schwinger's Approach for the coupling and 6j symbols of SU(2)

We will present the Schwinger's method very interesting for the determination of the the many couplings states and we use it later for the determination of SU (3) basis. For the calculation of 6j and 9j symbols, it is more convenient to follow Bargeman's method with a simple change of variables which makes the calculations simpler.

## 6.1 Schwinger's Approach for the coupling

The polynomials invariant of SU (2) has the generating function:

$$\exp[\tau_3[z^1,z^2] + \tau_2[z^1,z^3] + \tau_1[z^2,z^3]] = \sum \Phi_{j_1 j_2 j_3}(\tau) H_{[j_1 j_2 j_3]}(z) \qquad (6.1)$$

To determine the generating function of the coupling of two angular momentums we change $z_3^1$ by $-z_3^2$ and $z_3^2$ by $z_3^1$ in the above expression. We obtain the Schwinger's formula:

$$G(\alpha, z) = \exp[\{\alpha_3[x,y] + \alpha_1(zy) + \alpha_2(xz)\}] =$$

$$\exp[\{\alpha_3[\frac{\partial}{\partial u}, \frac{\partial}{\partial v}] + \alpha_2[z_1 \frac{\partial}{\partial u_1} + z_2 \frac{\partial}{\partial u_2}] + \alpha_1[z_1 \frac{\partial}{\partial v_1} + z_2 \frac{\partial}{\partial v_2}]]\} \exp[(uy) + (vx)]$$

With $[x,y] = x_1 y_2 - x_2 y_1$.

This formula can be applied to the calculation of several coupling of angular momentum where the great interest of this method.

## 6.2 The generating function of the 6j symbols of SU (2)

The expression of the 6j symbols [6] is given by:

$$\begin{Bmatrix} j_1 & j_2 & j_3 \\ l_1 & l_2 & l_3 \end{Bmatrix} =$$

$$\sum_{\mu_1 \mu_2 \mu_3} (-1)^{l_1+l_2+l_3+\mu_1+\mu_2+\mu_3} \begin{pmatrix} l_1 & l_2 & j_3 \\ \mu_1 & -\mu_2 & m_3 \end{pmatrix} \begin{pmatrix} l_2 & l_3 & j_1 \\ \mu_2 & -\mu_3 & m_1 \end{pmatrix} \begin{pmatrix} l_3 & l_1 & j_2 \\ \mu_3 & -\mu_1 & m_2 \end{pmatrix} \begin{pmatrix} j_1 & j_2 & j_3 \\ m_1 & m_2 & m_3 \end{pmatrix}$$

With help of Fock-Bargmann space and the expression (5.2) we write

$$\begin{pmatrix} j_1 & j_2 & j_3 \\ m_1 & m_2 & m_3 \end{pmatrix} = \int \left[\prod_{i=1}^{3} \overline{\varphi}_{j_i m_i}(x^i)\right] \frac{[x^2 x^3]^{(j-2j_1)} [x^3 x^1]^{(j-2j_2)} [x^1 x^2]^{(j-2j_3)}}{\sqrt{(j+1)!(j-2j_1)!(j-2j_2)!(j-2j_3)!}} d\mu(x)$$

We need to replace the four 3j symbols by expressions of this form to calculate the generating function of 6j symbols.

$$G(\tau) = \int \exp[D_0 + D_1 + D_2 + D_3] d\mu(\xi, \xi', \eta, \eta') \qquad (6.2)$$

We obtain the generating function for the 6j symbols
To avoid the Bargmann complex calculations we make a change of variables $\eta_i, \eta_i'$ by $\overline{\eta}_i, \overline{\eta}_i'$ we obtain:

$$D_0 = \begin{vmatrix} \tau_{01} & \tau_{02} & \tau_{03} \\ \xi_1' & \xi_2' & \xi_3' \\ \eta_1' & \eta_2' & \eta_3' \end{vmatrix}, D_1 = \begin{vmatrix} \tau_{10} & \tau_{12} & \tau_{13} \\ \overline{\eta}_1' & \xi_2 & \overline{\eta}_3 \\ -\overline{\xi}_1' & \eta_2 & -\overline{\xi}_3 \end{vmatrix}, D_2 = \begin{vmatrix} \tau_{23} & \tau_{20} & \tau_{21} \\ \overline{\eta}_1 & \overline{\eta}_2' & \xi_3 \\ -\overline{\xi}_1 & -\overline{\xi}_2' & \eta_3 \end{vmatrix}, D_3 = \begin{vmatrix} \tau_{32} & \tau_{31} & \tau_{30} \\ \xi_1 & \overline{\eta}_2 & \eta_3' \\ \eta_1 & -\overline{\xi}_2 & -\xi_3' \end{vmatrix},$$

$$D_0 = \begin{vmatrix} \tau_{01} & \tau_{02} & \tau_{03} \\ \xi_1' & \xi_2' & \xi_3' \\ \overline{\eta}_1' & \overline{\eta}_2' & \overline{\eta}_3' \end{vmatrix}, D_1 = \begin{vmatrix} \tau_{10} & \tau_{12} & \tau_{13} \\ \eta_1' & \xi_2 & \eta_3 \\ -\overline{\xi}_1' & \overline{\eta}_2 & -\overline{\xi}_3 \end{vmatrix}, D_2 = \begin{vmatrix} \tau_{23} & \tau_{20} & \tau_{21} \\ \eta_1 & \eta_2' & \xi_3 \\ -\overline{\xi}_1 & -\overline{\xi}_2' & \overline{\eta}_3 \end{vmatrix}, D_3 = \begin{vmatrix} \tau_{32} & \tau_{31} & \tau_{30} \\ \xi_1 & \eta_2 & \eta_3' \\ \overline{\eta}_1 & -\overline{\xi}_2 & -\overline{\xi}_3' \end{vmatrix},$$

Then the generating function for the 6j symbols may be written in the form:

$$G(\tau) = \left(\frac{1}{\pi}\right)^n \int \prod_{i=1}^{n} dx_i dy_i \exp(-\bar{z}^t X z) = (\det(X))^{-1}$$

This expression can be executed simply with a symbolic program (maple, Scientific Work,) and we find an expression for the 6j with one summation analogue to Racah formula:

$$G(\tau) = (\det(X))^{-1}$$
$$\det(X) = (g(\tau))^2, \quad g(\tau) = 1 + \sum_{i=0}^{3} a_i + \sum_{i=1}^{3} b_i$$
$$a_0 = \tau_{10}\tau_{20}\tau_{30}, \quad a_1 = \tau_{01}\tau_{31}\tau_{21}, \quad a_2 = \tau_{32}\tau_{02}\tau_{12}, \quad a_3 = \tau_{23}\tau_{13}\tau_{03},$$
$$b_1 = \tau_{01}\tau_{10}\tau_{23}\tau_{32}, \quad b_2 = \tau_{02}\tau_{20}\tau_{13}\tau_{31}, \quad b_3 = \tau_{03}\tau_{30}\tau_{13}\tau_{31},$$

### *6.3 Symmetry of 6j symbols*

If we denote as Bargmann done the power of $(\tau_{ij})$ by $(k_{ij})$ it is simple to observe that $g(\tau)$ is invariant by the permutation of rows and columns of:

$$\begin{vmatrix} \tau_{10} & \tau_{20} & \tau_{30} \\ \tau_{01} & \tau_{31} & \tau_{21} \\ \tau_{32} & \tau_{02} & \tau_{12} \\ \tau_{23} & \tau_{13} & \tau_{03} \end{vmatrix} \Leftrightarrow \begin{vmatrix} k_{10} & k_{20} & k_{30} \\ k_{01} & k_{31} & k_{21} \\ k_{32} & k_{02} & k_{12} \\ k_{23} & k_{13} & k_{03} \end{vmatrix} \quad (6.4)$$

# 7. Appendices

### *Appendix 1*
The matrix (X) for the 6j symbols

$$\begin{pmatrix}
0 & 0 & 0 & 0 & 0 & 0 & 0 & \tau_{12} & 0 & 0 & 0 & -\tau_{13} \\
0 & 0 & 0 & 0 & 0 & 0 & 0 & 0 & \tau_{23} & -\tau_{21} & 0 & 0 \\
0 & 0 & 0 & 0 & 0 & 0 & \tau_{31} & 0 & 0 & 0 & -\tau_{32} & 0 \\
0 & -\tau_{03} & \tau_{02} & 0 & 0 & 0 & 0 & 0 & 0 & 0 & 0 & 0 \\
\tau_{03} & 0 & \tau_{01} & 0 & 0 & 0 & 0 & 0 & 0 & 0 & 0 & 0 \\
-\tau_{02} & \tau_{01} & 0 & 0 & 0 & 0 & 0 & 0 & 0 & 0 & 0 & 0 \\
0 & 0 & 0 & 0 & \tau_{21} & 0 & 0 & 0 & -\tau_{20} & 0 & 0 & 0 \\
0 & 0 & 0 & 0 & 0 & \tau_{32} & -\tau_{10} & 0 & 0 & 0 & 0 & 0 \\
0 & 0 & 0 & \tau_{13} & 0 & 0 & 0 & -\tau_{10} & 0 & 0 & 0 & 0 \\
0 & 0 & 0 & 0 & 0 & \tau_{31} & 0 & 0 & 0 & 0 & -\tau_{30} & 0 \\
0 & 0 & 0 & \tau_{12} & 0 & 0 & 0 & 0 & 0 & 0 & 0 & -\tau_{10} \\
0 & 0 & 0 & 0 & \tau_{23} & 0 & 0 & 0 & 0 & -\tau_{20} & 0 & 0
\end{pmatrix}$$

*Appendix 2*
*The infinitesimal projection operator*

$$J_{\pm}|jm\rangle = \sqrt{(j \mp m)(j \pm m + 1)}|j(m \pm 1)\rangle, \quad \text{And } J_{\pm}|j \pm j\rangle = 0.$$

$$|\Phi\rangle = \sum_{j=|m|}^{\infty} C_j |jm\rangle, \quad \text{And } P_{jm}|\Phi\rangle = C_j |jm\rangle$$

We find the triangular system as in the projection of oscillator

$$\gamma_k | \quad J_-^k J_+^k |\Phi\rangle = \sum_{l=j+k}^{\infty} C_j \frac{(l-m)!(l+m+k)!}{(l-m-k)!(l+m)!} |jm\rangle, \quad \text{For } k = 1, 2, \text{ etc.}$$

$$\gamma_r = (-1)^r \frac{(2j+1)!}{(r)!(2j+r+1)!}$$

$$P_{jm} = \frac{(2j+1)(j+m)!}{(j-m)!} \sum_{r=0}^{\infty} (-1)^i \frac{J_-^{j-m+r} J_+^{j-m+r}}{r!(2j+r+1)!}$$

One can show that CGC can be presented in the form:

$$\langle j_1 m_1 |\langle j_2 m_2 \|(j_1 j_2) j_3 m_3\rangle = \frac{\langle j_1 m_1 |\langle j_2 m_2 | P_{m_3;j_3}^{j_3} | j_1 j_1\rangle | j_2 (j_3 - j_1)\rangle}{\langle j_1 j_1 |\langle j_2 (j_3 - j_1) | P_{j_3;j_3}^{j_3} | j_1 j_1\rangle | j_2 (j_3 - j_1)\rangle}$$

Using the explicit formulas (3.1) and (3.2) one can easily obtain the final formula of su(2)-CGC

*Appendix 3*

The calculus of 9j coefficients is very useful in physics and we present the calculations as a problem. The 9-j coefficients are:

$$\begin{Bmatrix} j_{11} & j_{12} & j_{13} \\ j_{21} & j_{22} & j_{23} \\ j_{31} & j_{32} & j_{33} \end{Bmatrix} = \sum_{\text{all } m's} \begin{pmatrix} j_{11} & j_{12} & j_{13} \\ m_{11} & m_{12} & m_{13} \end{pmatrix} \begin{pmatrix} j_{21} & j_{22} & j_{23} \\ m_{21} & m_{22} & m_{23} \end{pmatrix} \begin{pmatrix} j_{31} & j_{32} & j_{33} \\ m_{31} & m_{32} & m_{33} \end{pmatrix}$$

$$\times \begin{pmatrix} j_{11} & j_{21} & j_{31} \\ m_{11} & m_{21} & m_{31} \end{pmatrix} \begin{pmatrix} j_{12} & j_{22} & j_{32} \\ m_{12} & m_{22} & m_{32} \end{pmatrix} \begin{pmatrix} j_{13} & j_{23} & j_{33} \\ m_{13} & m_{23} & m_{33} \end{pmatrix}$$

1. Prove that the generating function is:

$$\int \exp\{\sum_{i=1}^{6} D(\xi, \xi', \tau)\} d\mu(\xi) d\mu(\xi')$$

2. Use the change of variables $\eta_i \to \bar{\eta}_i$ to find the above expression in the form $Exp\{-Z^t X Z\}$

Use a symbolic program to find the expression of the generating function.

Note: It's very important to note that many physical applications [52] have as a starting point the cylindrical basis therefore we treat these cases in the paper [53].

# Chapter three

# Octonions algebra and the cross product in n-dimensions
(Why do we love Octonions?)

**1. Introduction**
**2. The problem of sums squares**
   2.1 The Hurwitz theorem on sums of squares
   2.2 Solution of Hurwitz Problem
**3. The Octonions quadratic transformations**
   3.1 Levi-Civita Transformation
   3.2 Octonions Transformations
   3.3 Cayley- Hamilton transformation and the Hurwitz's matrices
**4. Pauli, Dirac matrices and Gegenbauer polynomials**
   4.1 Pauli, Dirac matrices and quadratic transformations
   4.2 Gegenbauer polynomials and the quadratic transformations
   4.3 New generalization of quadratic transformations and
      Generalized Dirac algebra
   4.4 the connection of Hydrogen atom and Harmonic oscillator
**5. The cross product in n-dimensions and Hurwitz theorem**
   5.1 Inertia Tensor
   5.2 Inertia tensor and the quaternion
   5.3 The cross product in n-dimensions
**6. The generating matrices and Cartan-Weyl basis**

# Chapter three

# Octonions algebra and the cross product in n-dimensions

## 1. Introduction

It is well known that the old problem of Sums squares [12-13] has been the source of the well known division algebra R, C, Q = H, O which are very important in mathematics and physics. The particular quadratic transformations: Levi-Civita and Kutaanheimo - Steifel can be deduced simply from these algebras [26-29]. Kibler and al. [30-31] have studied these transformations and made several useful applications. We present a recurrence method for the determination of all quadratic transformations, Hurwitz or octonions quadratic transformations, which are derived as a transformation using the Cayley-Hamilton. In addition we found similar transformations from the theory of angular momentum [22] and the connection hydrogen atom and oscillator can be generalized to all these transformations [58-59].

We note that these quadratic transformations are related to the Pauli, Dirac matrices and generating functions of Gegenbauer polynomials and generalizations of Dirac algebra.

The relationship between the inertia tensor and the octonions algebra was emphasized for the first time in the paper [55]. And we also show by means of the tensor of inertia and Hurwitz's theorem that there are only cross-products in the Euclidian space of seven and three dimensions [55-57].

## 2. The problem of sums squares

### 2.1 The Hurwitz theorem on sums of squares

The problem of sums squares is an old problem and Hurwitz find the final solution.

**<u>The general question we ask is</u>**: For which r=s=n is there an identity

$$(u_1^2 + u_2^2 + ..u_r^2)(v_1^2 + v_2^2 + ...v_s^2) = x_1^2 + x_2^2 + ..x_n^2 \tag{2.1}$$

Where the x's are algebraically determined by:

$$x_i = \sum_{i,j}^{n} a_{ij} u_i v_j \tag{2.2}$$

From the historical point of view we know that:
1- n=2 is known to Indian Brahmagupta(605?) and to Fermat.
2- n= 4 is the Euler's identity

3- In 1843, Euler's identity was rediscovered by Hamilton in his work on quaternion.
4- Graves and Cayley independently found an 8-square Identity.
5- In 1898, Hurwitz proved a theorem that killed this subject:

Hurwitz proved only the dimension constraints n = 1, 2, 4, and 8, it is also the case that, up to a linear change of variables, the only sum of squares identities in these dimensions are the ones associated to multiplication in the four classical real division algebras of dimensions 1, 2, 4 and 8: the real numbers, complex numbers, quaternion's and octonions.

R, C, Q, O are the division algebra and play very important role in physics and math.

This problem is very important in numbers theory and it has been the basis of the theory of spinners, the Cayley-Dixon algebra and Clifford algebra for r≠s=n.

*2.2 Solution of Hurwitz Problem*
For n=2:
$$U = H_2 = \begin{pmatrix} u_1 & -u_2 \\ u_2 & u_1 \end{pmatrix} \tag{2.3}$$

And Z=UV we find:
$$Z^t Z = (x_1^2 + x_2^2) = (u_1^2 + u_2^2)(v_1^2 + v_2^2)$$
$$U = u_1 I + u_2 J, \quad J^2 = -1$$
U belong to the complex field or the Clifford algebra C(0,1)

For n=4
$$U = H_4 = \begin{pmatrix} u_1 & -u_2 & -u_3 & -u_4 \\ u_2 & u_1 & -u_4 & u_3 \\ u_3 & u_4 & u_1 & -u_2 \\ u_4 & -u_3 & u_2 & u_1 \end{pmatrix} \tag{2.4}$$

$$\Rightarrow U = u_1 1 + u_2 I + u_3 J + u_4 K$$
U belong to the H-field =C (0, 2) or the quaternion's of Hamilton

For n=8
$$H_8 = \begin{pmatrix} u_1 & u_2 & u_3 & u_4 & u_5 & u_6 & u_7 & u_8 \\ -u_2 & u_1 & u_4 & -u_3 & u_6 & -u_5 & -u_8 & u_7 \\ -u_3 & -u_4 & u_1 & u_2 & u_7 & u_8 & -u_5 & -u_6 \\ -u_4 & u_3 & -u_2 & u_1 & u_8 & -u_7 & u_6 & -u_5 \\ -u_5 & -u_6 & -u_7 & -u_8 & u_1 & u_2 & u_3 & u_4 \\ -u_6 & u_5 & -u_8 & u_7 & -u_2 & u_1 & -u_4 & u_3 \\ -u_7 & u_8 & u_5 & -u_6 & -u_3 & u_4 & u_1 & -u_2 \\ -u_8 & -u_7 & u_6 & u_5 & -u_4 & -u_3 & u_2 & u_1 \end{pmatrix} \tag{2.5}$$

It can be developed as a linear combination of Clifford matrices.

$$H_8 = u_1 I - \sum_{i=2}^{8} u_i \Gamma_i^t, \quad \Gamma_i \Gamma_j + \Gamma_j \Gamma_i = -2I, \quad (i,j = 2,8)$$

We notice that matrices $H_n$ are anti-symmetric, orthogonal and the rows and columns are the components of a vector V (u).

## 3. The Octonions quadratic transformations

In what follows, we will expose a recurrence method for the determination of the matrices (H). It's starting point the transformation of Levi-Civita and the orthogonality of the matrices (H). We deduce also these transformations from Cayley-Hamilton transformation.

### *3.1 Levi-Civita Transformation*
For n=2 Levi-Civita introduced the conformal transformation which is an application of $R^2 \to R^2$.

$$z_1 = u_1^2 - u_2^2, \quad z_2 = 2u_1 u_2$$

That is written 
$$\begin{pmatrix} z_1 \\ -z_2 \end{pmatrix} = \begin{pmatrix} u_1 & u_2 \\ -u_2 & u_1 \end{pmatrix} \begin{pmatrix} u_1 \\ -u_2 \end{pmatrix} = (H_2)(U_2) \qquad (3.1)$$

### *3.2 Octonions Transformations*
For the generalization of the transformations of Levi-civita we put

$$\begin{pmatrix} z_1 \\ z_2 \end{pmatrix} = 2 \begin{pmatrix} u_1 & u_2 \\ -u_2 & u_1 \end{pmatrix} \begin{pmatrix} u_3 \\ u_4 \end{pmatrix} = 2(H_2)(U_2')$$

Using the orthogonality of $(H_2)$ we find

$$z_1^2 + z_2^2 = 2(u_1^2 + u_2^2)(u_3^2 + u_4^2)$$

And if we put $\quad z_3 = (u_1^2 + u_2^2) - (u_3^2 + u_4^2)$

We write

$$\begin{pmatrix} z_1 \\ -z_2 \\ -z_3 \\ 0 \end{pmatrix} = \begin{pmatrix} u_3 & u_4 & u_1 & u_2 \\ -u_4 & u_3 & u_2 & -u_1 \\ -u_1 & -u_2 & u_3 & u_4 \\ -u_2 & u_1 & -u_4 & u_3 \end{pmatrix} \begin{pmatrix} u_1 \\ u_2 \\ u_3 \\ u_4 \end{pmatrix} = (H_4)(U_4) \qquad (3.2)$$

Thus we find the transformation of $R^4 \to R^3$ known by Kustaanheimo-steifel transformation.

To obtain $(H_8)$ and $(H_{16})$ we repeat the same process while replacing $(H_2), (U_2')$ and $z_3$ by $(H_4), (U'_4)^t = (u_5,\ldots,u_8)$ and $z_5$ we deduce $(H_8)$ then we adopt the same way for $(H_{16})$.

## 3.3 Cayley- Hamilton transformation and the Hurwitz's matrices

The Cayley transformation [121-124] for the orthogonal groups $O_n$ is:

$$O_n = \frac{I - S_n}{I + S_n} \quad (3.3)$$

$S_n$ Is a skew symmetric matrix of order n.

In order to obtain $O_n$ in terms of the variables {u}: we multiply by $u_1$ the numerator
And the denominator of (3.3).

We multiply also $O_n$ by $r_n$. $r_n = \vec{u}^2 = \sum_{i=1}^{N} u_i^2$

To simplify the notation we replace $u_1 S_n$ by $S_n$ in the expression of $O_n(u)$ we obtain

$$O_n(u) = \vec{u}^2 \frac{u_1 I - S_n}{u_1 I + S_n} \quad (3.4)$$

### 3.3.1 The transformation $R^2 \to R^2$

For n=2 we have

$$S_2 = \begin{pmatrix} 0 & u_2 \\ -u_2 & 0 \end{pmatrix}$$

And simple calculation gives

$$O_2 = \frac{1}{r_2} \begin{pmatrix} u_1^2 - u_2^2 & -2u_1 u_2 \\ 2u_1 u_2 & u_1^2 - u_2^2 \end{pmatrix} \quad (3.5)$$

### 3.3.2 The transformation $R^4 \to R^3$

Using a computer symbolic program we find the Weyl's expression

$$O_3(u) = (r_3 I - 2u_1 S_3 + 2S_3^2)$$
$$= \begin{pmatrix} u_1^2 - u_2^2 - u_3^2 + u_4^2 & -2(u_2 u_1 + u_3 u_4) & -2(u_3 u_1 - u_2 u_4) \\ 2(u_2 u_1 - u_4 u_3) & u_1^2 - u_2^2 + u_3^2 - u_4^2 & -2(u_1 u_4 + u_2 u_3) \\ 2(u_3 u_1 + u_2 u_4) & 2(u_1 u_4 - u_2 u_3) & u_1^2 + u_2^2 - u_3^2 - u_4^2 \end{pmatrix} \quad (3.6)$$

In the space of 4-dimensions we derive also the expression

$$\begin{pmatrix} (H_3^t) & V_3^t \\ -V_3 & u_1 \end{pmatrix} \begin{pmatrix} (H_3^t) & -V_3^t \\ V_3 & u_1 \end{pmatrix} = \begin{pmatrix} O_3(u) & 0_3^t \\ 0_3 & \vec{u}^2 \end{pmatrix} \quad (3.7)$$

With $\quad V_3 = (u_4 \quad -u_3 \quad u_2)$ and $0_3 = (0 \quad 0 \quad 0)$.

### 3.3.3 The transformation $R^8 \to R^7$

We also find, using the symbolic program, an analogue expression as above-mentioned:

$$(O_7) = \frac{1}{r_5}(r_5 I - 2u_1 S_7 + 2S_7^2) \tag{3.8}$$

And in the 8-dimensions space we derive the expression

$$\begin{pmatrix} (H_7^t) & V_7^t \\ -V_7 & u_1 \end{pmatrix} \begin{pmatrix} (H_7^t) & -V_7^t \\ V_7 & u_1 \end{pmatrix} = \begin{pmatrix} O_7(u) & 0_7^t \\ 0_7 & \vec{u}^2 \end{pmatrix} \tag{3.9}$$

With $\quad V_7 = (u_8 \ u_7 \ -u_6 \ -u_5 \ u_4 \ u_3 \ -u_2)$

And $\quad 0_7^t = (0 \ 0 \ 0 \ 0 \ 0 \ 0 \ 0)$.

## 4. Pauli, Dirac matrices and Gegenbauer polynomials

There is a close link between the Octonions Quadratics transformations and spinor theory [54-55]. In this regard, we put $z_i$ in quadratic form in terms of $(\bar{v})^t$ and $(v)$

### 4.1 Pauli, Dirac matrices and the Octonions transformations
**a- Pauli matrices and the Transformation $R^8 \to R^2$.**

$$z_i = (\bar{v})^t (\sigma_i)(v) \tag{4.1}$$

With $(\bar{v})^t = (v_1 v_2)$ and $(\sigma_i)$ denotes the Pauli matrices (11).

$$(\sigma_1) = \begin{pmatrix} 0 & 1 \\ 1 & 0 \end{pmatrix}, \quad (\sigma_2) = \begin{pmatrix} 0 & -i \\ i & 0 \end{pmatrix}, \quad (\sigma_3) = \begin{pmatrix} 1 & 0 \\ 0 & -1 \end{pmatrix}$$

**b- Dirac matrices and the Transformation $R^8 \to R^5$.**

Put $(v)^t = (v_1 \ v_2 \ v_3 \ v_4)$, then by explicit calculation we find:

$$\begin{aligned} z_1 &= (\bar{v})^t \gamma^5(v), & z_2 &= i(\bar{v})^t \gamma^2(v) \\ z_3 &= i(\bar{v})^t \gamma^2(v), & z_4 &= i(\bar{v})^t \gamma^1(v) \\ z_4 &= (\bar{v})^t \gamma^0(v) \end{aligned} \tag{4.2}$$

It is clear that γ-matrices are the famous Dirac representation.

$$\gamma^0 = \begin{pmatrix} I & 0 \\ 0 & I \end{pmatrix}, \quad \gamma^i = \begin{pmatrix} 0 & \sigma_i \\ -\sigma_i & 0 \end{pmatrix}, \quad \gamma^5 = \begin{pmatrix} 0 & I \\ I & 0 \end{pmatrix} \tag{4.3}$$

Finally we can change the Euclidean by a pseudo-Euclidean space (11) which doesn't affect our treatment.

## 4.2 Gegenbauer polynomials and the quadratic transformations

We noticed a relationship between the generating function of Gegenbauer polynomials and the Octonions algebra and this part aims to present this relationship.

### a- The Gaussian integral and Levi-Civita transformation

The quadratic transformation $R^2 \to R^2$ is

$$y' = u_1^2 - u_2^2, \quad x' = 2u_1 u_2, \quad r' = u_1^2 + u_2^2 \tag{4.4}$$

Put $(x, y, z) = (x_1, x_2, x_3)$

We write:

$$zr' + ixx' + iyy = (u_1 \quad u_2) A_1 \begin{pmatrix} u_1 \\ u_2 \end{pmatrix} = (u_1 \quad u_2) \begin{pmatrix} x_3 + ix_2 & ix_1 \\ ix_1 & x_3 - ix_2 \end{pmatrix} \begin{pmatrix} u_1 \\ u_2 \end{pmatrix} \tag{4.5}$$

The Gaussian integral in this case is

$$\int \exp[\alpha(x_3 r' + ix_1 x' + ix_2 y')] d\mu(u) = \frac{1}{\sqrt{1 - 2x_3 \alpha + \alpha^2 r^2}} \tag{4.6}$$

The second part is the generating function of Gegenbauer polynomials

We also

$$A_1 = \begin{pmatrix} x_3 + ix_2 & ix_1 \\ ix_1 & x_3 - ix_2 \end{pmatrix} = x_3 I + x_2 \Gamma_2 + x_1 \Gamma_1 \tag{4.7}$$

With $\quad \Gamma_1^2 = \Gamma_2^2 = -1$

### b- The Gaussian integral and the quaternion

It is well known that

$$A_2 = \begin{pmatrix} x_4 + ix_3 & x_2 + ix_1 \\ -x_2 + ix_1 & x_4 - ix_3 \end{pmatrix} = \sum_{i=1}^{4} x_i \Gamma_i, \tag{4.8}$$

With $\quad \Gamma_4 = 1, \Gamma_1^2 = \Gamma_2^2 = \Gamma_3^2 = -1$

Where $(I, \Gamma_3, \Gamma_2, \Gamma_1)$ are representations matrices of the quaternion. We find by a direct calculation the Gaussian integral

$$\int \exp\left[\alpha(\bar{z}_1 \quad \bar{z}_2) A_2 \begin{pmatrix} z_1 \\ z_2 \end{pmatrix}\right] d\mu(u) = \frac{1}{\sqrt{1 - 2x_4 \alpha + \alpha^2 r^2}} \tag{4.9}$$

with $\quad z_1 = u_1 + iu_2, \quad z_2 = u_3 + iu_4$

$$(\bar{z}_1 \quad \bar{z}_2) A_2 \begin{pmatrix} z_1 \\ z_2 \end{pmatrix} = x_4 r' + x_3 x_3' + x_2 x_2' + x_1 x_1'$$

The second part is the generating function of Gegenbauer polynomials

*c- The Gaussian integral and the Octonions transformations*

The $R^8 \rightarrow R^5$ is given by

$$(x_1 + ix_2) = 2(z_1 z_3 + z_2 z_4), \ (x_3 + ix_4) = 2(z_1 z_4 - z_2 z_3)$$
$$r_1 = u_1^2 + u_2^2 + u_3^2 + u_4^2, \ r_2 = u_5^2 + u_6^2 + u_7^2 + u_8^2 \tag{4.10}$$
$$x_5 = r_1 - r_2, \ r = r_1 + r_2$$

We consider as previously:

$$(\bar{z}^t A_3 z) = x_6 r + i(x_1 x_1' + x_2 x_2' + x_3 x_3' + x_4 x_4' + x_5 x_5'),$$

$$A_3 = \begin{pmatrix} x_6 + ix_5 & 0 & -x_1 + ix_2 & -x_4 + ix_3 \\ 0 & x_6 + ix_5 & -x_4 - ix_3 & x_1 + ix_2 \\ x_1 + ix_2 & x_4 - ix_3 & x_6 - ix_5 & 0 \\ x_4 + ix_3 & -x_1 + ix_2 & 0 & x_6 - ix_5 \end{pmatrix} = \sum_{i=1}^{6} x_i \Gamma_i \tag{4.11}$$

t is the transpose and $(z)^t = (z_1, z_2, z_3, z_4)$

We also write $\Gamma_6 = 1$, $\Gamma_i^2 = -1$, $i \leq 5$, $\Gamma_i \Gamma_j + \Gamma_j \Gamma_i = 0$

The Gaussian integral in this case is:

$$\int \exp[\alpha(\bar{z})^t A_3(z)] d\mu(u) = \frac{1}{\sqrt{1 - 2x_6 \alpha + \alpha^2 r^2}} \tag{4.12}$$

with $\quad z_i = \sum_{i=0}^{3} u_{2i+1} + iu_{2i+2}$

We find again that the second member is the generating function of Gegenbauer Polynomials.

## *4.3 New generalization of quadratic transformations generalized Dirac algebra*

we can generalize $A_1, A_2$ and $A_3$ by writing:

$$A_n = \begin{pmatrix} (x_{2n} + ix_{2n-1})I_{2^{n-2}} & A_{n-1} \\ -\bar{A}_{n-1}^t & (x_{2n} + ix_{2n-1})I_{2^{n-2}} \end{pmatrix} = \sum x_i \Gamma_i$$

We note that the generalization of the quadratic transformations (5.12) can be written as:

$$z^t A_n z = x_{2n} u + i(\sum_{i=1}^{2n-1} x_i x_i') \tag{4.13}$$

With $\quad r^2 = \sum_{i=1}^{2n} x_i^2, \ u^2 = \sum_{i=1}^{2n-1} x_i'^2$

$A_n$ Are the Pauli matrices for n = 2 and the Dirac matrices for n = 3.

It is also important to note that we deduce a new quadratics transformations and new algebra different from the Cayley-Dixon algebra for n> 3.
Using a symbolic program we find also:
For n=4

$$\int \exp[\alpha(\bar{z})^t A_4(z)] d\mu(z) = \frac{1}{(1 - 2x_{2n}\alpha + \alpha^2 u^2)^4}, \qquad (4.14)$$

And $\qquad (\bar{z})^t A_n(z) = x_{2n} r' + i(\sum_{i=1}^{2n-1} x_i x_i')$

For n=5
$$\int \exp[\alpha(\bar{z})^t A_5(z)] d\mu(z) = \frac{1}{(1 - 2x_{2n}\alpha + \alpha^2 u^2)^8}, \qquad (4.15)$$

We find also after integration the generating functions of Gegenbauer polynomials:

$$\int d\mu(z) \exp\{\alpha \bar{z}^t A_n z\} = 1/(1 - 2\alpha x_{2n} - \alpha^2 r^2)^m \qquad (4.16)$$

we find also that:
$$A_n = \sum x_i \Gamma_i \text{ and } \Gamma_i^2 = -1, \Gamma_i \Gamma_j + \Gamma_j \Gamma_i = \delta_{ij} \qquad (4.17)$$

$\Gamma_i$ are elements of Clifford algebra.
We deduce from the above mentioned that there is a close relationship between the Clifford algebra and the generating functions of Gegenbauer polynomials
  It is important to note that the integration with Grassmann variables of the formula (4.12) becomes $(1 - 2\alpha x_{2n} - \alpha^2 r^2)^m$. This result can be considered as the extension of the generating function of Gegenbauer polynomials.
Our variables $\{x_i'\}$ are in the form: $x_i' = \sum_{ij} a_{ij} \bar{z}_i z_j$ this is not the case of Cayley –Dixon algebra for n>3. Then our algebra is a new algebra which generalized Dirac algebra.

### *4.4 The connection of Hydrogen atom and Harmonic oscillator*

  *4.3.1 Introduction:* We want to prove the important formula

$$\Delta_{u,N}(u) f(x) = 4\vec{u}^2 \Delta_{x,n} f(x) \qquad (4.18)$$
$$\text{With } (n, N) = (2,2), (3,4), (5,8), (9,16).$$
$\Delta_{x,n}$ is the Laplacian of SO(n), n=2, 3, 5 and 9.

**Solution**
  We derive the solution With the help of the relations
$$\frac{\partial^2 x_l}{\partial u_i^2} = 0, (l \leq n) \text{ and } \frac{\partial x_{n+1}}{\partial u_i} = \begin{cases} 2u_i \text{ if } i \leq n \\ -2u_i \text{ if } i > n \end{cases} \qquad (4.19)$$
We have also $x_i$ is a homogenous function in terms of $u_i$ and the matrix $H_n$ is orthogonal

*4.3.2 The connection between hydrogen atom and the harmonic oscillator*

A quick calculation shows that the equation of the hydrogen atom

$$(-\frac{\hbar^2}{2m}\Delta_{(x,n)}(x) - \frac{\alpha}{r})f(x) = Ef(x) \quad (4.20)$$

That may be written on the basis of harmonic oscillator in the form

$$(\Delta_{u,N}(u) - 4E\vec{u}^2)f(u) = 4\alpha f(u)$$

With $\quad \dfrac{m\omega^2}{2} = -4E, \ 4\alpha = \hbar\omega(2n+v+1), v = L-1+D/2$

And the energy is given by: $\quad E_{n,v} = -\dfrac{m\alpha^2}{4\hbar^2}(n+v+1/2)^{-2} \quad (4.21)$

# 5. The cross product in n-dimensions and Hurwitz theorem

We demonstrated using an elementary method that the tensor of inertia of a material point and the cross product of two vectors were only possible in a three or seven dimensional space [55-57]. The representation matrix of the cross product in the seven dimensional space and its properties were given.

*5.1 Inertia Tensor*

The kinetic energy of a particle of mass m=1 which moves in a system in rotation with angular velocity $(\vec{\omega})$ is:

$$T = \frac{1}{2}(\vec{\dot{x}})\cdot(\vec{\dot{x}}) = \frac{1}{2}(\vec{\omega}\times\vec{r})\cdot(\vec{\omega}\times\vec{r})$$

$$\begin{pmatrix}\dot{x}_1\\\dot{x}_2\\\dot{x}_3\end{pmatrix} = \begin{pmatrix}0 & z & -y\\-z & 0 & x\\y & -x & 0\end{pmatrix}\begin{pmatrix}\omega_1\\\omega_2\\\omega_3\end{pmatrix} \quad (5.1)$$

$$\Rightarrow T = \frac{1}{2}(\omega)^t(V_3)^t(V_3)(\omega) \text{ and } (M) = (V_3)^t(V_3)$$

We write the inertia matrix as:

$$(M) = \begin{pmatrix}m_{11} & m_{12} & m_{13}\\m_{12} & m_{22} & m_{23}\\m_{13} & m_{23} & m_{33}\end{pmatrix} = \begin{pmatrix}\vec{r}^2-x^2 & -xy & -xz\\-xy & \vec{r}^2-y^2 & -yz\\-xz & -yz & \vec{r}^2-z^2\end{pmatrix} \quad (5.2)$$

*5.2 Inertia tensor and the quaternion*

The identification of two sides of the equation (5.2) may be written as:

$$\begin{aligned}
m_{11} + x^2 &= \vec{r}^2 & m_{12} + xy &= 0 & m_{13} + xz &= 0 \\
m_{12} + xy &= 0 & m_{22} + y^2 &= \vec{r}^2 & m_{23} + yz &= 0 \\
m_{13} + xz &= 0 & m_{23} + yz &= 0 & m_{33} + z^2 &= \vec{r}^2 \\
\text{with} & & x^2 + y^2 + z^2 &= \vec{r}^2.
\end{aligned} \quad (5.3)$$

We can express these systems in matrix form as $(H_4)^t (H_4) = \vec{r}^2 I$

$$\begin{pmatrix} & & & -x \\ & (V_3)^t & & -y \\ & & & -z \\ x & y & z & 0 \end{pmatrix} \begin{pmatrix} & & & x \\ & (V_3) & & y \\ & & & z \\ -x & -y & -z & 0 \end{pmatrix} = \begin{pmatrix} \vec{r}^2 & 0 & 0 & 0 \\ 0 & \vec{r}^2 & 0 & 0 \\ 0 & 0 & \vec{r}^2 & 0 \\ 0 & 0 & 0 & \vec{r}^2 \end{pmatrix} \quad (5.4)$$

We replace the matrix $(V_3)$ by its expression in (5.1); we deduce the orthogonal and anti-symmetric matrix:

$$(H_4) = \begin{pmatrix} 0 & z & -y & x \\ -z & 0 & x & y \\ y & -x & 0 & z \\ -x & -y & -z & 0 \end{pmatrix} \quad (5.5)$$

With $(H_4)$ is the matrix representation of the quaternion

$$\begin{aligned}
h &= xe_1 - ye_2 + ze_3 \\
e_1^2 &= -1, \ e_2^2 = -1, \ e_3^2 = -1 \\
e_1 e_2 &= e_3, \ e_2 e_3 = e_1, \ e_3 e_1 = e_2
\end{aligned} \quad (5.6)$$

$(H_4)$ Is the Hurwitz matrix and $e_1, e_2$ and $e_3$ are the generators of the quaternion algebra.

### 5.3 The cross product in n-dimensions

The generalization of the tensor of inertia in an intuitive way is:

$$(M) = \begin{pmatrix} m_{11} & m_{12} & \dots & m_{1n} \\ m_{21} & m_{22} & \dots & m_{2n} \\ \vdots & \dots & \dots & \vdots \\ m_{1n} & m_{2n} & \dots & m_{nn} \end{pmatrix} = \begin{pmatrix} \vec{r}^2 - x_1^2 & -x_1 x_2 & \dots & -x_1 x_n \\ -x_1 x_2 & \vec{r}^2 - x_2^2 & \dots & -x_2 x_n \\ \vdots & \dots & \dots & \vdots \\ -x_1 x_n & -x_2 x_n & \dots & \vec{r}^2 - x_n^2 \end{pmatrix} \quad (5.7)$$

With $\vec{r} = \sum_{i=1}^{n} x_i \vec{e}$ and

$$\begin{aligned}
m_{ii} + x_i^2 &= \vec{r}^2, i = 1,\dots,n \\
m_{ij} + x_i x_j &= 0, i \neq j \ et \ i,j = 1,\dots,n
\end{aligned}$$

And the matrix system $(H_n)^t(H_n) = \vec{r}^2 I$ takes the form:

$$\begin{pmatrix} & & -x_1 \\ (V_n)^t & & \vdots \\ & & -x_n \\ x_1 & \ldots & x_n & 0 \end{pmatrix} \begin{pmatrix} & & & x_1 \\ & (V_n) & & \vdots \\ & & & x_n \\ -x_1 & \ldots & -x_n & 0 \end{pmatrix} = \begin{pmatrix} \vec{r}^2 & 0 & \ldots & 0 \\ 0 & \vec{r}^2 & \ldots & 0 \\ \vdots & \ldots & \ldots & \vdots \\ 0 & \ldots & \ldots & \vec{r}^2 \end{pmatrix} \qquad (5.8)$$

$$(H_n)^t(H_n) = \vec{r}^2 I, \text{ and } h = \sum_{i=1}^{7} x_i e_i \qquad (5.9)$$

The generators of the Octonions algebra satisfy:

$$e_i^2 = -1, \ e_i e_j = -e_j e_i, \ i,j = 1,\ldots 7$$

Hurwitz showed that we can only build orthogonal and anti-symmetric matrix which lines are a linear combination of components of a vector only if n=1, 2,4 or 8. Consequently the matrix (M) is orthogonal if n+1=8, it results that dim($R^n$)=1,3 or 7.
After simple calculus we find the matrix:

$$(V_7) = \begin{pmatrix} 0 & x_7 & -x_6 & -x_5 & x_4 & x_3 & -x_2 \\ -x_7 & 0 & -x_5 & x_6 & x_3 & -x_4 & x_1 \\ x_6 & x_5 & 0 & x_7 & -x_2 & -x_1 & -x_4 \\ x_5 & -x_6 & -x_7 & 0 & -x_1 & x_2 & x_3 \\ -x_4 & -x_3 & x_2 & x_1 & 0 & x_7 & -x_6 \\ -x_3 & x_4 & x_1 & -x_2 & -x_7 & 0 & x_5 \\ x_2 & -x_1 & x_4 & -x_3 & x_6 & -x_5 & 0 \end{pmatrix} \qquad (5.10)$$

***Properties of the matrix (V)***

$$\begin{aligned} a - \ & (V_3)^3 = -(\vec{r}^2)(V_3), \ (V_7)^3 = -(\vec{r}^2)(V_7) \\ b - \ & Exp[-i\theta(V_n)] = 1 - i\sin\theta(V_n) - (1-\cos\theta)(V_n)^2 \\ c - \ & (H_4)^2 = -I, \ (V_3)^3 = -(V_3), \ (H_8)^2 = -I, \ (V_7)^3 = -(V_7) \end{aligned} \qquad (5.11)$$

## 6. The generating matrices and the Cartan-Weyl basis

The adjoint representations of the orthogonal groups are anti-symmetric and the number of elements is n (n-1)/2. The matrix Hn is anti-symmetric and function of (n-1) parameters, {u}, and develops in terms of the adjoint representation of SO (n). To generate the Cartan-Weyl basis we need consequently n/2 matrices, this number is in agreement with the number of the simple roots of the orthogonal groups.
By analogy with the generating functions we call these matrices by the generating matrices of the Cartan-Weyl basis and we build it for the cases.

We start with the link of the Cartan-Weyl basis for the group SO(3), SO(4) and SO(5) with Hurwitz matrices.

*A-Generating matrices of SO(3)*

$$\text{For } n = 2 \quad\quad H_2 = u_1 I + u_2 \widehat{S}_3$$
$$\text{For } n = 3 \quad\quad H_3 = u_1 I + u_2 \widehat{S}_3 + u_3 \widehat{S}_2 + u_4 \widehat{S}_1 \tag{6.1}$$

*B-Generating matrices of SO(4)*

For n= 4 we obtain by Cayley transformation two orthogonal matrices

$$H_4^1 = u_1 I + u_2 \widehat{S}_3 + u_3 \widehat{S}_2 + u_4 \widehat{S}_1$$
$$H_4^2 = u_1 I + u_2 \widehat{T}_3 + u_3 \widehat{T}_2 + u_4 T_1 \tag{6.2}$$

*C-Generating matrices of SO(5)*

For n= 5 we must add only $\widehat{U}_1, \widehat{U}_2, \widehat{V}_1, \widehat{V}_2$ to the above-mentioned matrices and we write

$$H_5^1 = u_1 I + u_2 \widehat{S}_3 + u_3 \widehat{S}_2 + u_4 \widehat{S}_1 + u_5 \widehat{U}_1 + u_6 \widehat{U}_2 + u_7 \widehat{V}_1 + u_8 \widehat{V}_2$$

$$= \begin{pmatrix} u_1 & u_2 & u_3 & u_4 & u_5 \\ -u_2 & u_1 & u_4 & -u_3 & u_6 \\ -u_3 & -u_4 & u_1 & u_2 & u_7 \\ -u_4 & u_3 & -u_2 & u_1 & u_8 \\ -u_5 & -u_6 & -u_7 & -u_8 & u_1 \end{pmatrix} \tag{6.3}$$

We must change S by T to obtain the other matrix $H_5^2$

**D-The generators of SO (5)**
We put
$$L_{ij} = -i(x_i (\partial / \partial x_j) - x_j (\partial / \partial x_i)) \tag{6.4}$$

The generators of SO (5) groups are:

$$S_1 = L_{23} + L_{14}, \quad S_2 = L_{31} + L_{24}, \quad S_3 = L_{12} + L_{34}$$
$$T_1 = L_{23} - L_{14}, \quad T_2 = L_{31} - L_{24}, \quad T_3 = L_{12} - L_{34}$$
$$U_1 = L_{15}, \quad U_2 = L_{25}, \quad U_\pm = L_{15} \pm iL_{25} \tag{6.5}$$
$$V_1 = L_{35}, \quad V_2 = L_{45}, \quad V_\pm = L_{25} \pm iL_{45}$$

# Chapter four

# Momentum representation of hydrogen atom and Octonions quadratic transformations

## Part I-On the hydrogen wave function in Momentum-space

**1. Introduction**
**2. Generating function of hydrogen atom in momentum representation**
   2.1 The generating function of Laguerre polynomial
   2.2 The generating function of spherical harmonics
   2.3 Generating function for the basis of the hydrogen atom
**3. The connection of $R^3$ hydrogen atom and $R^4$ Harmonic oscillator**
   3.1 The Octonions quadratic transformation $R^4 \to R^3$
   3.2 The volume element
**4. The wave functions of hydrogen atom in momentum space**
   4.1 The generating function in {u} representation
   4.2 The generating function of momentum-space
   4.3 The wave functions in momentum-space

## Part II-On the N-dimensional hydrogen atom in momentum representation

**1. Introduction**
**2. The wave function of hydrogen atom in representation space**
   2.1 The wave function of hydrogen atom
   2.2 The radial function
   2.3 The angular function
**3. The momentum representation of N-dimensional hydrogen atom**
   3.1 The Generating function and the momentum representation
   3.2 Derivation of the generating function using Hankel's integral
   3.3 The Wave function in the momentum representation
**4. Appendix**

# Chapter four

# Momentum representation of hydrogen atom and Octonions quadratic transformations

# Part I-On the hydrogen wave function in Momentum-space

## 1. Introduction

    The problem of the hydrogen atom has played a central role in the development of Quantum mechanics. Schrödinger solved his equation and found the wave function in terms of coordinates. The problem in momentum space has been reformulated by Fock [32] and led to an integral form of the Schrödinger equation and the eigenfunctions are then expanded in terms of spherical harmonics. Despite the importance of Fock's work and the interest of many authors [33-37] to study the wave function in momentum space it must not hide [1-7] that the direct calculation of Fourier transform of the wave function of coordinates is up till now undone and our aim in this work is to fill this gap.

  The wave function of coordinates [6-7] has the form $\psi_{nlm}(\vec{r}) = R_{nl}(\omega r)Y_{lm}(\theta\varphi), \omega = 2/n$. Where $R_{nl}(\omega r)$ is the radial part, $Y_{lm}(\theta\varphi)$ is the spherical harmonic and ($\Omega$) the solid angle. The difficulty for the determination of the wave function in momentum space comes from $\omega$ and the appearance of the term "r" in the exponential of the radial part.

  We propose to circumvent these problems by using the quadratic transformation and the generating function method where $\omega = 2/n_0$ is a constant for all the elements of the basis. After calculation of the Fourier transform we found in the expansion of order n a function and we replace $\omega$ by $2/n_0$ and then we obtain the analytic expression of the wave function of hydrogen atom in momentum representation.

## 2. Generating function of hydrogen atom in momentum representation

The wave function of hydrogen atom in momentum representation is

$$\Psi_{nlm}(\vec{p}) = \frac{1}{(2\pi)^{3/2}} \int e^{-i\vec{p}\cdot\vec{r}} \Psi_{nlm}(\vec{r}) d\vec{r}$$

$$\psi_{nlm}(\vec{r}) = e^{-(\omega r)/2} R_{nl}(x) Y_{lm}(\theta\varphi), \vec{r} = (r,\theta,\varphi), \omega = 2/n \quad (2.1)$$

With $R_{nl}(x)$ is the radial part $R_{nl}(x) = \frac{N_{nl}}{(n+l)!} x^l L_{n-l-1}^{(2l+1)}(x)$     (2.2)

And $\quad x = \omega r, N_{nl} = \dfrac{2}{n^2}\sqrt{\dfrac{(n-l-1)!}{[(n+l)!]}}, \omega = 2\delta, \delta = \dfrac{1}{n}$ (2.3)

$L_n^\alpha(r)$ is the associated Laguerre polynomial. Atomic unit are used through the text.

## 2.1 The generating function of Laguerre polynomial $L_{n-l-1}^{2l+1}(r)$

The generating function of Laguerre polynomial $L_n^\alpha(r)$ is:

$$\sum_{n=0}^{\infty}\dfrac{z^n}{(n+\alpha)!}L_n^{(\alpha)}(r) = \dfrac{1}{(1-z)^{\alpha+1}}e^{-\dfrac{z}{(1-z)}r}$$

From the property $\quad \dfrac{d}{dr}L_n^{(\alpha)}(r) = -L_{n-1}^{(\alpha+1)}(r)$

We deduce that $\quad \sum_{n=0}^{\infty}\dfrac{z^n}{(n+l+1)!}L_{n-l-1}^{2l+1}(r) = \dfrac{(z)^{l+1}}{(1-z)^{2l+2}}e^{-\dfrac{z}{(1-z)}r}$ (2.4)

## 2.2 The generating function of spherical harmonics

$$\dfrac{(\vec{a}\cdot\vec{r})^l}{2^l l!} = [\dfrac{4\pi}{2l+1}]^{\dfrac{1}{2}}\sum_m \varphi_{lm}(\xi)Y_{lm}(\vec{r})$$ (2.5)

With $\vec{a}$ is a vector of length zero, $\vec{a}\cdot\vec{a} = 0$ and its components

$$a_1 = -\xi_1^2 + \xi_2^2, \ a_2 = -i(\xi_1^2 + \xi_2^2), \ a_3 = 2\xi_1\xi_2, \vec{a}^2 = 0$$

With $\quad \varphi_{lm}(\xi) = \dfrac{\xi^{l+m}\eta^{l-m}}{\sqrt{(l+m)!(l-m)!}}$.

## 2.3 Generating function for the basis of the hydrogen atom

We multiply $\psi_{nlm}(\vec{r})$ by $[\dfrac{4\pi}{2l+1}]^{\dfrac{1}{2}}z^n\varphi_{lm}(\alpha\xi)$, and summing with respect to $n, l, m$

$$G(z,\alpha\xi,\vec{r}) = \sum_{nlm}[\dfrac{4\pi}{2l+1}]^{\dfrac{1}{2}}[\dfrac{\pi(n+l)!}{4(n-l-1)!}]^{\dfrac{1}{2}}z^n\varphi_{lm}(\alpha\xi)\psi_{nlm}(\vec{r}) =$$

$$\sum_{n=0}^{\infty}e^{-(\omega r)/2}\dfrac{z^n}{(n+l+1)!}L_{n-l-1}^{2l+1}(\omega r)\sum_m \alpha\varphi_{lm}(\alpha\xi)\psi_{nlm}(\vec{r})$$

Substituting (2.4) and (2.5) in the above expression we obtain:

$$G(z,\alpha\xi,\vec{r}) = [\dfrac{\omega}{\sqrt{\pi}}\dfrac{z}{(1-z)^2}]\exp[-\dfrac{\omega r(1+z)}{2(1-z)} - \dfrac{\omega z(\vec{a}\cdot\vec{r})}{(1-z)^2}]$$ (2.6)

# 3. The connection of $R^3$ hydrogen atom and $R^4$ Harmonic oscillator

We will make a revision of the derivation of the quadratic transformations then we determine the volume element. A summary of the connection between the wave function of hydrogen atom and harmonic oscillator is given also.

## *3.1 The quadratic transformation $R^4 \to R^3$*

Consider the relationship between the well-known Wigner's D matrix spherical harmonics polynomials

$$\rho^{2j} D^l_{(m,0)}(\Omega) = \left(\frac{4\pi}{2l+1}\right)^{1/2} Y^*_{lm}(\theta,\varphi), \quad z = (z_1, z_2) \tag{3.1}$$

$$z_1 = u_1 + iu_2, \; z_2 = u_3 + iu_4,$$
$$\rho = \sqrt{r}, \; r = z_1 \bar{z}_1 + z_2 \bar{z}_2, \; r^2 = x^2 + y^2 + z^2$$

We write in terms of Euler's angles or Cayley-Klein parameterization.

$$z_1 = u_1 + iu_2 = \sqrt{r}\cos\frac{\theta}{2} e^{-\frac{i(\phi+\psi)}{2}}, z_2 = u_3 + iu_4 = \sqrt{r}\sin\frac{\theta}{2} e^{-\frac{i(\phi-\psi)}{2}} \tag{3.2}$$

And $D^j_{(m',m)}(z_1, \bar{z}_1, z_2, \bar{z}_2) = u^{2j} D^j_{(m',m)}(\psi\theta\varphi), j = 1, 1/2,...$

It is important to emphasize that the elements of the matrix D are solution of Laplacian $\Delta_4$ .
If we put $l =1$ in (3, 1) we obtain the quadratic transformation

$$x = 2(u_3 u_1 + u_4 u_2) = z_1 \bar{z}_2 + z_2 \bar{z}_1, \; y = 2(u_4 u_1 - u_3 u_2) = i(z_1 \bar{z}_2 - z_2 \bar{z}_1),$$
$$z = u_1^2 + u_2^2 - u_3^2 - u_4^2 = z_1 \bar{z}_1 - z_2 \bar{z}_2, \tag{3.3}$$

## *3.2 The volume element*

We consider the transformation $(u_1, u_2, u_3, u_4) \to (r, \psi\theta\varphi)$

With $\quad 0 \leq \theta \leq \pi, \; 0 \leq \psi, \varphi \leq 2\pi, \; 0 \leq r \leq \infty, \; -\infty \leq u_i \leq +\infty, \; i = 1,...,4$

And $\quad d^4\vec{u} = |J| dr d\theta d\varphi d\psi$

The calculation of the Jacobian gives $|J| = (u^2/8)\sin\theta \; but \; d^3\vec{r} = r^2 dr d\theta d\varphi d\psi$

Therefore $\quad 8u^2 d\vec{u} = d\vec{r} d\psi$,
And

$$\int f(x,y,z) d^3\vec{r} = \frac{4}{\pi} \int f(x(u), y(u), z(u)) u^2 d^4\vec{u} \tag{3.4}$$

# 4. The wave functions of hydrogen atom in momentum space

We write first the Fourier transform in the representation (u) and with the help of Bargmann integral we determine the generating function in momentum representation. Finally the development of this function gives us the hydrogen atom wave functions in Momentum space.

## *4.1 The generating function in {u} representation*

We denote the generating function by $G(z, \alpha\xi, \vec{p})$ in the representation $\{u\}$. But to determine the generating function (2.6) we must multiply by $4/\pi$ to reflect the change in the measure of integration. We write

$$\Psi_{nlm}(\vec{p}) = \frac{1}{(2\pi)^{3/2}} \int e^{-i\vec{p}\cdot\vec{r}} \Psi_{nlm}(\vec{r}) d\vec{r} \tag{4.1}$$

To calculate this expression we must write (4.1) in the (u) representation using the formula (3.4):

$$\Psi_{nlm}(\vec{p}) = \frac{4}{\pi} \frac{1}{(2\pi)^{3/2}} \int e^{-i\vec{p}\cdot\vec{r}} \Psi_{nlm}(\vec{r}) u^2 d^4u \tag{4.2}$$

In the expression of $\Psi_{nlm}(\vec{p})$ there is the term $u^2$ for that we consider a new generating function:

$$G(z, \alpha\xi, \vec{p}, \beta) = \frac{1}{(2\pi)^{3/2}} \frac{4}{\pi} \frac{z}{(1-z)^2} \times$$

$$\int e^{-i\vec{p}\cdot\vec{r}} \exp\left[-\frac{\omega r(1+z)}{2(1-z)} + \frac{\alpha \omega z(\vec{a}\cdot\vec{r})}{2(1-z)^2}\right] e^{-\beta u^2} d^4\vec{u} \tag{4.3}$$

We assume that $\beta \geq 0$ therefore there is no problem of convergence.

We write then: $G(z, \alpha\xi, \vec{p}) = -\partial G(z, \alpha\xi, \vec{p}, \beta)/\partial\beta\big|_{\beta=0}$ \hfill (4.4)

With $\qquad G(z, \alpha\xi, \vec{p}) = \sum_{nlm} \left(\frac{4\pi}{2l+1}\right)^{\frac{1}{2}} \frac{z^n}{N_{nl}} \alpha^l \varphi_{lm}(\xi) \psi_{nlm}(p)$ \hfill (4.5)

## *4.2 The generating function of momentum-space*

We can do the integration of (4.3) by a direct calculation with the variables (u) we can perform the integration using the Gauss formula

$$\left(\frac{1}{\pi}\right)^n \int \prod_{i=1}^n dx_i dy_i \exp(-\bar{z}^t X z + A^t z + \bar{z}^t \bar{B}) = (\det(X))^{-1} \exp(A^t X^{-1} \bar{B}) \tag{4.6}$$

With $z = (z_1, z_2, ..., z_n)$

We have

$$-i\vec{p}\cdot\vec{r} = -ip_x(z_1\bar{z}_2 + z_2\bar{z}_1) + p_y(z_1\bar{z}_2 - z_2\bar{z}_1) - ip_z(z_1\bar{z}_2 - z_2\bar{z}_1),$$

$$\vec{a}\cdot\vec{r} = a_x(z_1\bar{z}_2 + z_2\bar{z}_1) + ia_y(z_1\bar{z}_2 - z_2\bar{z}_1) + a_z(z_1\bar{z}_2 - z_2\bar{z}_1) \tag{4.7}$$

We obtain then

$$X = \begin{pmatrix} \dfrac{\omega(1+z)}{2(1-z)} + \beta - ip_z + \dfrac{\omega z}{2(1-z)^2} a_z & -ip_x + \dfrac{\omega z}{2(1-z)^2} a_x - p_y + \dfrac{1}{i}\dfrac{\omega z}{2(1-z)^2} a_y \\ -ip_x + \dfrac{\omega z}{2(1-z)^2} a_x + p_y - \dfrac{1}{i}\dfrac{\omega z}{2(1-z)^2} a_y & \dfrac{\omega(1+z)}{(1-z)} + \beta + ip_z - \dfrac{\omega z}{2(1-z)^2} a_z \end{pmatrix}$$

Because $\vec{a}^2 = 0$ we deduce that:

$$\det(X) = \left[ \left( \frac{\delta(1+z)}{(1-z)} + \beta \right)^2 + \vec{p}^2 + i\alpha \frac{2\delta z}{(1-z)^2} \vec{a} \cdot \vec{p} \right], \quad \delta = \omega/2 \tag{4.8}$$

We find therefore the generating functions

$$G(z, \alpha\xi, \vec{p}, \beta) = \frac{2}{\sqrt{2\pi}} \frac{z}{\left[ (\delta(1+z) + \beta(1-z))^2 + (1-z)^2 \vec{p}^2 + 2i\alpha\delta \vec{a} \cdot \vec{p} \right]} \tag{4.9}$$

In applying the relation (4.4) we find the generating function $G(z, \alpha\xi, \vec{p})$

$$G(z, \alpha\xi, \vec{p}) = \frac{4\delta}{\sqrt{2\pi}} \frac{z(1-z^2)}{\left[ (\delta(1+z) + \beta(1-z))^2 + (1-z)^2 \vec{p}^2 + 2i\alpha\delta \vec{a} \cdot \vec{p} \right]^2} \tag{4.10}$$

*4.3 The wave functions in momentum-space*

We drive the basis of momentum-space using the formula

$$\left[ \varphi_{jm}(\partial/\partial\xi) \frac{1}{n!} \frac{\partial^n}{\partial z^n} \frac{1}{l!} \frac{\partial^l}{\partial \alpha^l} G(z, \alpha\xi, \vec{p}) \right]_0 = \left( \frac{4\pi}{2l+1} \right)^{1/2} \frac{1}{N_{nl}} \psi_{nlm}(\vec{p}) \tag{4.11}$$

In this case we must take $\delta = 1/n$ and to execute the calculations we proceed by step:

**1 - Derivation with respect to α**

$$\left[ \frac{1}{l!} \frac{\partial^l}{\partial \alpha^l} G(z, \alpha\xi, \vec{p}) \right]_0 = (i)^l \frac{(l+1)!}{\sqrt{2\pi}} \times (4\delta)^{l+1} \times \frac{(1-z^2)z^{l+1}}{[(\delta(1+z))^2 + (1-z)^2 \vec{p}^2]^{l+2}} \frac{(\vec{a}.\vec{p})^l}{2^l l!} \tag{4.12}$$

We have

$$(\delta(1+z))^2 + (1-z)^2 \vec{p}^2 = ((\vec{p}^2 + \delta^2) - 2z(\vec{p}^2 - \delta^2) + z^2(\vec{p}^2 + \delta^2)$$

$$= (\vec{p}^2 + \delta^2)[1 - 2zx + z^2], \quad x = \left( \frac{\vec{p}^2 - \delta^2}{\vec{p}^2 + \delta^2} \right)$$

We deduce that

$$\left[ \frac{1}{l!} \frac{\partial^l}{\partial \alpha^l} G(z, \alpha\xi, \vec{p}) \right]_0 = (i)^l \frac{(l+1)!}{\sqrt{2\pi}} \times \frac{(4\delta)^{l+1}}{(\vec{p}^2 + \delta^2)^{l+2}} \times \frac{(1-z^2)z^{l+1}}{[1 - 2zx + z^2]^{l+2}} \frac{(\vec{a}.\vec{p})^l}{2^l l!} \tag{4.13}$$

## 2- Derivation with respect to z

Using the familiar formula for the generating function of Gegenbauer polynomials

$$(1 - 2rt + r^2)^{-\alpha} = \sum_{m=0}^{\infty} r^m C_m^{\alpha}(t) \tag{4.14}$$

We write

$$\frac{(1-z^2)z^{l+1}}{[1-2zx+z^2]^{l+2}} = \sum_{m=0}^{\infty} (1-z^2)z^{l+1} C_m^{l+2}(x)$$

$$= \sum_{n=1}^{\infty} z^n \left[ C_{n-l-1}^{l+2}(x) - C_{n-l-3}^{l+2}(x) \right]$$

With $m + l + 1 = n$, $m + l + 3 = n$ and $\delta = 1/n$ therefore

$$\left[ \frac{1}{n!} \frac{\partial^n}{\partial z^n} \frac{1}{l!} \frac{\partial^l}{\partial \alpha^l} G(z, \alpha\xi, \vec{p}) \right]_0 = (i)^l \frac{(l+1)!}{\sqrt{2\pi}} \times \frac{(4\delta)^{l+1}}{(\vec{p}^2 + \delta^2)^{l+2}} \times$$
$$\left[ C_{n-l-1}^{l+2}(x) - C_{n-l-3}^{l+2}(x) \right] \frac{(\vec{a}.\vec{p})^l}{2^l l!} \tag{4.15}$$

Put $\vec{y} = (y_1, y_2, y_2, y_3, y_4) = \left( \frac{\delta p_x}{(\vec{p}^2 + \delta^2)}, \frac{\delta p_y}{(\vec{p}^2 + \delta^2)}, \frac{\delta p_y}{(\vec{p}^2 + \delta^2)}, \frac{(\vec{p}^2 - \delta^2)}{(\vec{p}^2 + \delta^2)} \right)$

We obtain $\vec{y}.\vec{y} = 0$.
Thus we find the transformation introduced by Fock.

## 3- Derivation with respect to $\varphi_{lm}(\frac{\partial}{\partial \xi})$

By using the formula (2.5) we get the following expression

$$\left[ \varphi_{jm}(\partial/\partial \xi) \frac{1}{n!} \frac{\partial^n}{\partial z^n} \frac{1}{l!} \frac{\partial^l}{\partial \alpha^l} G(z, \alpha\xi, \vec{p}) \right]_0 = (i)^l \frac{(l+1)!}{\sqrt{2\pi}} \times \frac{(4\delta)^{l+1}}{(\vec{p}^2 + \delta^2)^{l+2}} \times$$
$$\left[ C_{n-l-1}^{l+2}(x) - C_{n-l-3}^{l+2}(x) \right] Y_{lm}(\vec{p}) \tag{4.16}$$

4- The wave functions in momentum space
The comparisons of (4.16) and (4.12) give us the result:

$$\psi_{nlm}(\vec{p}) = (i)^l N_{nl} a^{-3/2} \frac{(l+1)!}{\sqrt{2\pi}} \times (4\delta)^{l+1} \frac{\left[ C_{n-l-1}^{l+2}(x) - C_{n-l-3}^{l+2}(x) \right]}{(\vec{p}^2 + \delta^2)^{l+2}} Y_{lm}(\vec{p}) \tag{4.17}$$

And with the help of the recurrences formula [10]:
$$(n + \alpha) C_{n+1}^{(\alpha-1)}(x) = (\alpha - 1)[C_{n+1}^{(\alpha)}(x) - C_{n-1}^{(\alpha)}(x)]$$

We derive finally the wave functions in momentum space:

$$\psi_{nlm}(\vec{p}) = (i)^l N_{nl} \frac{(l)!}{\sqrt{2\pi}} \times \frac{n(4\delta)^{l+1}}{(\vec{p}^2 + \delta^2)^{l+2}} C_{n-l-1}^{l+1}\left(\frac{\vec{p}^2 - \delta^2}{\vec{p}^2 + \delta^2}\right) Y_{lm}(\vec{p}) \tag{4.18}$$

It is clear that we obtain by an elementary method and direct calculus not only the wave function in momentum representation but also the phase factor.

# Part II- On the N-dimensional hydrogen atom in momentum representation

## 1. Introduction

In a previous part we presented a new and elementary method for the determination of the wave function in momentum space for two and three dimensions using the generating function and Octonions quadratic transformations for a direct integration of the Fourier transform. But Octonions transformations are valid only for N = 2, 3, 5 and 9.

In this work we present, the generalization for N≥3 by using the technique of generating function and the Hankel's integral [60] of Bessel functions and therefore we determine the wave function in momentum space with the exact phase factor for any order N.

## 2. The wave function of hydrogen atom in representation space

In this part we exhibit only the well known wave function solution of the Schrödinger equation for the hydrogen atoms in N-dimensions [62-63].

### 2.1 The wave function of hydrogen atom
The Schrödinger equation of the hydrogen atom in N-dimensions space is

$$\left(-\frac{\hbar^2}{2\mu}\Delta_N - \frac{e^2}{r}\right)\Psi(\vec{r}) = E\Psi(\vec{r}) \tag{2.1}$$

Where $\mu$ is the reduced mass.
We write $\vec{r} = (x_1, x_2, \ldots, x_N)$ in spherical coordinates as $\vec{r} = (r, \theta_1, \theta_2, \ldots, \theta_{N-2}, \varphi)$.

$$x_1 = r\sin\theta_1 \sin\theta_1 \sin\theta_2 \ldots \sin\theta_{N-2} \cos\varphi$$
$$x_2 = r\sin\theta_1 \sin\theta_2 \sin\theta_3 \ldots \sin\theta_{N-2} \sin\varphi$$
$$x_3 = r\sin\theta_1 \sin\theta_2 \sin\theta_3 \ldots \cos\theta_{N-2}$$
$$\vdots$$
$$x_N = r\cos\theta_1 \tag{2.2}$$

With $0 \leq \theta_j \leq \pi$, $j = 1, \ldots, N-2$, $0 \leq \varphi \leq 2\pi$.
And atomic unit are used through the text.
The method of separation of variables is used for the resolution of the Schrödinger equation and we write only the solution:

$$\Psi_{n,l,\{\mu\}}(\vec{r}) = R_{n,l}(r)Y_{l,\{\mu\}}(\Omega_r) = N_{n,l}\omega^{N/2}(\omega r)^l e^{-\frac{r}{2\lambda}}L_{n-l-1}^{2l+N-2}(\omega r)Y_{l,\{\mu\}}(\Omega_r) \tag{2.3}$$

$$N_{n,l} = \left\{\frac{(n-l-1)!}{2(n+(N-3)/2)(n+l+N-3)!}\right\}^{1/2}, \omega = 2\delta_n = \frac{2}{(n+(N-3)/2)}$$

## 2.2 The radial function

$L_{n-l-1}^{(\alpha)}(x)$ Are the Laguerre polynomials

$$\int_0^\infty e^{-x} x^\alpha (L_n^\alpha(x))^2 dx = \frac{\Gamma(\alpha+n+1)}{n!} \tag{2.4}$$

The generating function of Laguerre polynomials is:

$$\sum_{n=0}^\infty z^n L_n^{(\alpha)}(x) = \frac{1}{(1-z)^{\alpha+1}} e^{-\frac{z}{1-z}x} \tag{2.5}$$

$$\frac{d^k}{dx^k} L_n^{(\alpha)}(x) = (-1)^k L_{n-k}^{(\alpha+k)}(x)$$

But $\alpha + k = 2l + N - 2$, $n - k = n - l - 1$

Consequently $\alpha = l + N - 3$

We deduce also that

$$\frac{(-z)^{l+1}}{(1-z)^{2l+N}} \exp[-x\frac{z}{(1-z)}] = \sum_{n=0}^\infty z^n L_{n-l-1}^{2l+N-2}(x) \tag{2.6}$$

## 2.3 The angular function

$Y_{l,\{\mu\}}(\Omega_r)$ Is the Hyperspherical function

$$Y_{l,\{\mu\}}(\Omega_r) = \frac{1}{\sqrt{2\pi}} A_{l,\{\mu\}} e^{im\varphi} \prod_{j=1}^{N-2} C_{\mu_j-\mu_{j+1}}^{\alpha_j+\mu_{j+1}}(\cos\theta_j)(\sin\theta_j)^{\mu_{j+1}} \tag{2.7}$$

With $\quad A_{l,\{\mu\}} = \prod_{j=1}^{N-2} \Gamma(\alpha_j + \alpha_{j+1}) \sqrt{\frac{(\alpha_j + \mu_j)(\mu_j - \mu_{j+1})!}{\pi 2^{1-2\alpha_j-2\mu_{j+1}} \Gamma(2\alpha_j + \mu_j + \mu_{j+1})}}$

And $\quad 2\alpha_j = N - j - 1, l = \mu_1 \geq \mu_2 \geq \ldots \geq |\mu_{N-1}| = |m|, (l,\{\mu\}) = (l, \mu_2, \ldots, \mu_{N-1})$.

$C_n^\alpha(\cos\theta)$ Is the Gegenbauer polynomial of degree n and parameter α.

## 3. The momentum representation of N-dimensional hydrogen atom

Using the development of the free wave in space of N-dimensions, the generating functions of Laguerre polynomials and Hankel's integral we determine the generating function of momentum representation and hence the wave function in momentum space.

### 3.1 The Generating function and the momentum representation

The wave function of hydrogen atom in momentum representation is:

$$\Psi_{n,l,\{\mu\}}(\vec{p}) = \frac{1}{(2\pi)^{N/2}} \int e^{-i\vec{p}.\vec{r}} \Psi_{n,l,\{\mu\}}(\vec{r}) d\vec{r} \tag{3.1}$$

$$= \frac{N_{n,l}}{(2\pi)^{N/2}} \int e^{-i\vec{p}.\vec{r}-\frac{\omega r}{2}} (\omega r)^l L_{n-l-1}^{2l+N-2}(\omega r) Y_{l,\{\mu\}}(\Omega_r) d\vec{r} \tag{3.2}$$

$d\vec{r}$ Is defined by:

$$d\vec{r} = dx_1 dx_2 \ldots dx_N = r^{N-1} d\Omega_r . \tag{3.3}$$

We note that $i\vec{p}\cdot\vec{r}+\delta r$ can be regarded as the scalar product of two vectors in Euclidean space $E_{N+1}$. The first one is a vector of zero lengths $(x_1, x_2, \ldots, x_N, ir)$ and the second vector is defined by the angles $\vec{K}=(\theta, \theta_1^k, \ldots, \theta_{N-1}^k)$ and the length $K=\sqrt{\vec{p}^2+\delta_n^2}$

Using the development of the wave function of the free particle in N-dimensions [4]:

$$e^{i\vec{p}\cdot\vec{r}} = (2\pi)^{N/2} \sum_{[L]}^{\infty} i^l Y_{[L]}(\Omega_r) Y_{[L]}^*(\Omega_p) J_\nu(pr)/(pr)^{\frac{N}{2}-1}, \quad \nu = l + \frac{N}{2} - 1 \quad (3.4)$$

We find

$$\Psi_{n,l,\{\mu\}}(\vec{p}) = \frac{1}{(2\pi)^{N/2}} \int e^{-i\vec{p}\cdot\vec{r}} \Psi_{n,l,\{\mu\}}(\vec{r}) d\vec{r}$$

$$= N_{n,l} \left( \int_0^\infty (\omega r)^l e^{-\frac{\omega r}{2}} L_{n-l-1}^{2l+N-2}(\omega r) J_\nu(pr)/(pr)^{\frac{N}{2}-1} r^{N-1} dr \right) \left( (-i)^l Y_{[L]}(\Omega_p) \right) \quad (3.5)$$

Multiply by $1/(N_{n,l}) z^n /(\omega^{l+N/2})$ and do the summation we write first the generating function for the basis and the generating function with ω=constant which is very useful for the determination of wave function in momentum space.

$$G(p, z, \delta_n) = \sum_{n=0}^\infty \frac{(p)^{\frac{N}{2}-1}}{N_{n,l}} \frac{z^n}{\omega^{l+N/2}} \Psi_{n,l}(\vec{p}) \quad (3.6)$$

And

$$G(p, z, \delta) = \sum_{n=0}^\infty z^n \left( \int_0^\infty (r)^l e^{-\frac{r}{2\lambda}} L_{n-l-1}^{2l+N-2}(\omega r) J_\nu(pr)/(pr)^{\frac{N}{2}-1} r^{N-1} dr \right)$$

$$= \frac{(-z)^{l+1}}{(1-z)^{2l+N-1}} \int_0^\infty e^{-\gamma r} J_\nu(pr) r^{\nu+1} dr \quad \nu = l + \frac{N}{2} - 1 \quad (3.7)$$

With

$$\gamma = \frac{\omega}{2} \frac{1+z}{1-z}, \quad \omega = 2\delta \quad (3.8)$$

### *3.2 Derivation of the generating function using Hankel's integral*

Using Hankel's integral[60]:

$$\int_0^\infty e^{-\gamma r} J_\nu(pr) r^{\nu+1} dr = \frac{2\gamma(2p)^\nu (\Gamma(\nu+3/2))}{\sqrt{\pi} \times (\gamma^2+p^2)^{\nu+3/2}}, [\text{Re}(\nu) > -1] \quad (3.9)$$

We find the generating function as follows:

$$\frac{(-z)^{l+1}}{(1-z)^{2l+N-1}} \frac{2\gamma(2p)^\nu (\Gamma(\nu+3/2))}{\sqrt{\pi} \times (p^2+\gamma^2)^{\nu+3/2}} =$$

$$= \frac{\omega}{\sqrt{\pi}} \frac{1+z}{(1-z)^{2l+N}} (2p)^{l+\frac{N}{2}-1} \Gamma(l+\frac{N+1}{2}) \frac{(-z)^{l+1}(1-z)^{2l+N+1}}{(p^2(1-z)^2+\delta^2(1+z)^2)^{l+\frac{N+1}{2}}}$$

$$= \frac{2\delta(2p)^{l+\frac{N}{2}-1} \Gamma(l+\frac{N+1}{2})}{\sqrt{\pi}(p^2+\delta^2)^{l+\frac{N+1}{2}}} \frac{(-z)^{l+1}(1-z^2)}{(1-2zx+z^2)^{l+\frac{N+1}{2}}}, x = \left(\frac{p^2-\delta^2}{p^2+\delta^2}\right) \qquad (3.10)$$

This function is the generalization of the generating function (4.10) of the paper [36], by a minus sign due to the derivation of (4.9).

### *3.3 The Wave function in the momentum representation*

We have
$$\frac{1}{n!}\frac{d^n}{dz^n}G(p,z,\delta_n)\bigg|_{z=0} = \frac{1}{n!}\frac{d^n}{dz^n}[G(p,z,\delta)|\delta=\delta_n]\bigg|_{z=0} \qquad (3.11)$$

Using the development
$$\frac{(-z)^{l+1}(1-z^2)}{(1-2zx+z^2)^{l+\frac{N+1}{2}}} = (-1)^{l+1}\sum_{n=1}^{\infty} z^n [C_{n-l-1}^{l+\frac{N+1}{2}}(x) - C_{n-l-3}^{l+\frac{N+1}{2}}(x)] \qquad (3.12)$$

And the formula [16] $\qquad (n+\alpha)C_{n+1}^{(\alpha-1)}(x) = (\alpha-1)[C_{n+1}^{(\alpha)}(x) - C_{n-1}^{(\alpha)}(x)] \qquad (3.13)$

We find that the expression (3.11) may be written as

$$\frac{(p)^{\frac{N}{2}-1}}{N_{n,l}(2\delta_n)^{l+\frac{N}{2}}}\Psi_{n,l}(\vec{p}) = (-1)^{l+1}\frac{(2\delta_n)(2p)^{l+\frac{N}{2}-1}\Gamma(l+\frac{N-1}{2})(n+\frac{(N-3)}{2})}{\sqrt{\pi}(p^2+\delta_n^2)^{l+\frac{N+1}{2}}}C_{n-l-1}^{l+(N-1)/2}(x)$$

Finally we derive the wave function in the momentum representation

$$\Psi_{n,l,\{\mu\}}(\vec{p}) = -(i)^l \left\{\frac{(n-l-1)!(n+(N-3)/2)}{2\pi(n+l+N-3)!}\right\}^{1/2} \frac{2^{2l+N}(\delta_n)^{\frac{N}{2}+1}(\delta_n p)^l \Gamma(l+\frac{N-1}{2})!}{(p^2+\delta_n^2)^{l+\frac{N+1}{2}}} \times$$
$$C_{n-l-1}^{l+(N-1)/2}(x)Y_{[L]}(\Omega_p) \qquad (3.14)$$

This hyperphysical function may be written in term of the components of the vector defined by the angles $\vec{K}' = (-\frac{\pi}{2}+2\theta, \theta_1^k, \ldots, \theta_{n-1}^k)$ and the length $K' = \sqrt{P^2+\delta_n^2}$.
This vector may be derived from $\vec{K}$ by rotation about the vector perpendicular to the space $E_N$ with angle of rotation $-\frac{\pi}{2}+\theta$.
We can also determine the representation {p} by this method if the potential has an additional term $A/r^2$.

# 4. Appendix

*Problem*:

The Four dimensions of Laplacian have two solutions:
Our objective is to derive the passage formulas between these basis .
1- If we consider the coordinates

$$x_1 = \sqrt{r}\sin\frac{\theta}{2}\cos(\psi-\varphi) \qquad x_3 = \sqrt{r}\cos\frac{\theta}{2}\cos(\psi+\varphi)$$

$$x_2 = \sqrt{r}\sin\frac{\theta}{2}\sin(\psi-\varphi) \qquad x_4 = \sqrt{r}\cos\frac{\theta}{2}\sin(\psi+\varphi)$$

Prove that the solution of the Laplacian is: $r^j D^j_{(m',m)}(\psi\theta\varphi)$

2- If we consider the coordinates

$$x_1 = r\sin\chi\sin\theta\cos\psi \qquad x_3 = r\sin\chi\cos\theta$$

$$x_2 = r\sin\chi\sin\theta\sin\psi \qquad x_4 = r\cos\chi$$

Find the solution is the spherical basis $Y_{nlm}(\chi\theta\psi)$

$$Y_{nlm}(\vec{v}) = 2^{l+1} v^{n-l+1} l! \left(\frac{n(n-l-1)}{2\pi(n+l)!}\right)^{1/2} C^{l+1}_{n-l-1}(\cos\chi) Y_{lm}(\vec{r}), \ \vec{v} = (\vec{r}, q), \vec{r} = (x_1, x_2, x_3).$$

3- We consider the quadratic transformation

$$(\bar{z}_1 \ \bar{z}_2)\begin{pmatrix} q+iz & x+iy \\ -x+iy & q-iz \end{pmatrix}\begin{pmatrix} z_1 \\ z_2 \end{pmatrix} = qq' + ixx' + iyy' + izz'$$

$$= qq' + i\vec{r}\cdot\vec{r}'$$

Find the two expressions of the developments, $\exp[qq'+ixx'+iyy'+izz']$
in terms of spherical and D-matrix elements.

$$e^{z\cos\chi}(\frac{z}{2}\sin\chi)^{\frac{1}{2}-\alpha} J_{l+1/2}(z\sin\chi) = \sum_{n=0}^{\infty} \frac{\Gamma(2l+2)}{\Gamma(l+1+1/2)\Gamma(2l+2+n)} C^{(l+1)}_n(\cos\chi) z^n$$

With $e^{i\vec{r}\cdot\vec{r}'} = 4\pi \sum_{l=0}^{\infty}\sum_{m=-l}^{l} i^l j_l(kr) Y^*_{lm}(\theta'\varphi') Y_{lm}(\theta\varphi)$ and $j_l(\rho) = \sqrt{\frac{\pi}{2\rho}} J_{L+1/2}(\rho)$

4- Use the two expressions to derive the formula:

$$Y_{nlm}(\vec{v}) = (-i)^l (n/2)^{1/2} \sum_{m1m2} (-1)^{(n-1)/2} (2l+1)^{1/2} \begin{pmatrix} (n-1)/2 & (n-1)/2 & l \\ m_1 & m_2 & m \end{pmatrix} D^{(n-1)/2}_{(m_1,m_2)}(U_2)$$

# Chapter *V*

# On the collective motion of the nucleus "Microscopic theory"

**Introduction**

**Part I-Generalization of Cramer's rule and its application to the projection of Hartree-Fock wave function**

*1. Introduction*
*2. Generalization of Cramer's rule*
*3. The projection of the Hartree-Fock wave function*
   3.1 The Hartree-Fock basis
   3.2 the energy levels using the projection
      of the Hartree-Fock wave function
*4. Derivation of Löwdin formula and Thouless theorem*
   4.1 Generalization of Cramer's rule and Löwdin formula
   4.2. Generalization of Cramer's rule and Thouless theorem

**Part II-Collective vibration of the nucleus, G.F. Method and quasi-bosons approximation**

*1. Introduction*
*2. The RPA equation of motion*
   2.1 The equation of motion harmonic oscillator
   2.2 The RPA equation of motion
*3. The generating function method (revision)*
   3.1 The Generating coordinates Method
   3.2 New Interpretation of the generating function
   3.3 Generalization of the generating function
   3.4 applications to harmonic oscillator
   3.5 applications to Lipkin model
*4. The renormalized Thouless function as generating function for the many body problem*
   4.1 The Thouless function
   4.2 Renormalization of the Thouless function
*5. The quasi-boson expansion of the Hamiltonian*
   5.1. The Hamiltonian of the nucleus
   5.2-The image of the operators: particles-holes
   5.3-The image of the operators: two particles-two holes
   5.4 The expression of the Hamiltonian in terms of bosons operators

# Chapter V

# On the collective motion of the nucleus "Microscopic theory"

## Introduction

The Hartree-Fock variation method provides an approximate determination of ground states and ground state energies of quantum mechanical systems, and widely used in physics and chemistry. In Hartree-Fock method [66-73] we approximate the ground state of the system by a Slater determinant $|\Phi_{HF}\rangle$ constructed from the states of nucleons which are eigenstates of a single particle Hamiltonian called Hartree-Fock Hamiltonian. This approximation reduces the problem of many interacting particles to one of non-interacting particles in a field. This wave function is not function of angular momentum, and the calculation of rotational energy [68] can be done by using the integral representation of Hill-Wheeler operator. But the calculation of the rotations spectrum is Very long [70-72].

We have generalized the Cramer's rule and so the calculations can be carried out simply by the Gauss-Jacobi method [72].
Using this generalization of Cramer's rule we determine the Thouless function [78] and we take this function as the generating function of the Hartree-Fock basis.

It is obvious that this approximation neglects much of the interaction forces between particles. These forces are the residual interaction. To study collective motions it is important to consider these interactions. And the introduction of random phase approximation theory and more generally the quasibosons developments aim the study of these interactions. It is therefore important to express the Hamiltonian in terms of quasi-bosons and apply the Bogoliubov transformation to determine the frequencies of collective vibrations. But the methods developed by Belyaev and Zelevinsky [39] and that of Marumori et al. [40] converge slowly and don't respect Pauli principal.

For the sake of clarity we did a quick revision of the generating function method and we apply our method to the well known model in nuclear physics the Lipkin model [81] or in the SU (2) case we got the same result of Holstein-Primakoff mapping.

We know that the Thouless function is developed on the product of the Hartree-Fock basis and its image in the Fock space which is an orthogonal basis without normalization. So that this function to be useful we need to normalize the basis to get the correct generating function $|G(z)\rangle$. So using the generating function method we find the expression of the Hamiltonian in terms of quasi-bosons operators. This Hamiltonian was used by many others for studying the collective vibrations of the nucleus. We made reference only to some papers despite the importance of the other works [82-89].

We have already used also the generating function method for calculating the Moshinsky-Smirnov coefficients which are very useful for numerical calculations of the matrix elements of nuclear potential [79].

# Part I-Generalization of Cramer's rule and its application to the projection of Hartree-Fock wave function

## 1. Introduction

The great recent interest [72-77] to study the projection theory and their application in nuclear physics leads me to resume my former works on the projection of angular momentum [72].

Löwdin [69] proposed a formula for the calculation of the spectrum of energy levels, but this method requires a long calculation and does not take account the conditions of stability resulting from the minimization of the energy of the system using the Hartree-Fock theory.

We observe that the calculation of the rotational energy implies the calculation of a determinant, the overlap of rotation, and a set of determinants which differs from each other by the change of two columns [72]. This leads us to the generalization of Cramer's rule of linear algebra this allowing us to calculate all these determinants by Gauss elimination method.

This method takes into account the conditions of stability and minimizes the time of executions. Using this generalization we derive also the Löwdin formula [69] and the well known Thouless theorem [78].

## 2. Generalization of Cramer's rule

Let E be a vector space of dimension (n) with basis $\vec{e}_1, \vec{e}_2, \ldots, \vec{e}_n$. $\vec{a}_1, \vec{a}_2, \ldots, \vec{a}_n$ Is a set of linearly independent vectors, belonging to E. $\vec{b}_1, \vec{b}_2, , \vec{b}_s$, $s \leq n$ is another set of linearly independent vectors, belonging to E. We denote by (A) the matrix formed by the components of the vectors $(\vec{a}_i)$ and $\det(A) = \det(\vec{a}_1, \vec{a}_2, \ldots, \vec{a}_n) = |(A)|$ is the determinant of the matrix (A).

**Theorem**: Consider the following systems

$$\sum_{j=1}^{n} \vec{a}_j x(k,j) = \vec{b}_k, \ k = 1, 2, \ldots, s, \ s \leq n \tag{2.1}$$

We find the determinant formed from det (A) by substituting the components of some vector $(\vec{a}_j)$ by the components of the vectors $\vec{b}_i$, $(1 \leq i \leq s) \leq n$ by the formula:

$$\det(a_1,\ldots,\vec{a}_{i_1},\vec{b}_1,\ldots,\vec{a}_{i_s},\vec{b}_s,\ldots,a_n) = \det(A)\begin{vmatrix} x(1,i_1) & \ldots & x(1,i_s) \\ \vdots & \vdots & \vdots \\ x(s,i_1) & \ldots & x(s,i_s) \end{vmatrix} \quad (2.2)$$

**Prove**: We will proceed by induction, s = 1 then 2, etc.

It is well known from the multilinear algebra that the space $\overset{n}{\wedge} E$ has only one basic vector $\vec{e}_1 \wedge \vec{e}_2 \wedge \ldots \wedge \vec{e}_n$ and $\vec{a}_1 \wedge \vec{a}_2 \wedge \ldots \wedge \vec{a}_n = \det(A)\vec{e}_1 \wedge \vec{e}_2 \wedge \ldots \wedge \vec{e}$.

1-For s = 1 this is the case of Cramer's rule. We shall do a brief revision.

Multiply the two terms of the expression (2.1), the right by $\wedge \vec{a}_{i+1} \wedge \ldots \wedge \vec{a}_n$ and the left by $\vec{a}_1 \wedge \ldots \wedge \vec{a}_{i-1} \wedge$ we obtain:

$$\sum_{j=1}^{n} x(k,j)\vec{a}_1 \wedge \ldots \wedge \vec{a}_{i-1} \wedge \vec{a}_j \wedge \vec{a}_{i+1} \wedge \ldots \wedge \vec{a}_n = \vec{a}_1 \wedge \ldots \wedge \vec{a}_{i-1} \wedge \vec{b}_k \wedge \vec{a}_{i+1} \wedge \ldots \wedge \vec{a}_n$$

The summation in the first member is zero unless j = i, it follows that:

$$x(k,i)\vec{a}_1 \wedge \ldots \wedge \vec{a}_n = \vec{a}_1 \wedge \ldots \wedge \vec{a}_{i-1} \wedge \vec{b}_k \wedge \vec{a}_{i+1} \wedge \ldots \wedge \vec{a}_n \quad (2.3)$$

then we deduce that

$$\det(A)x(k,i) = \det(\vec{a}_1,\ldots,\vec{a}_{i-1},\vec{b}_k,\vec{a}_{i+1},\ldots,\vec{a}_n) \quad (2.4)$$

2-For s = 2, we multiply the two terms of (2.1), the right by
$$\wedge \vec{a}_{i+1} \wedge \ldots \wedge \vec{a}_{l-1} \wedge \vec{b}_r \wedge \ldots \wedge \vec{a}_n$$

And the left by $\quad \wedge \vec{a}_{i+1} \wedge \ldots \wedge \vec{a}_{l-1} \wedge \vec{b}_r \wedge \ldots \wedge \vec{a}_n,$ (2.5)

We obtain:

$$\sum_{j=1}^{n} x(k,j)\vec{a}_1 \wedge \ldots \wedge \vec{a}_{i-1} \wedge \vec{a}_j \wedge \vec{a}_{i+1} \wedge \ldots \wedge \vec{a}_n =$$
$$\vec{a}_1 \wedge \ldots \wedge \vec{a}_{i-1} \wedge \vec{b}_k \wedge \vec{a}_{i+1} \wedge \ldots \wedge \vec{a}_{l-1} \wedge \vec{b}_r \wedge \ldots \wedge \vec{a}_n \quad (2.6)$$

The first member is zero unless $j = i$ or $j = l$, it follows that

$$x(k,i)\det(\vec{a}_1,\ldots,\vec{a}_i,\ldots,\vec{a}_{l-1},\vec{b}_r,\vec{a}_{l+1},\ldots,\vec{a}_n) +$$
$$x(k,l)\det(\vec{a}_1,\ldots,\vec{a}_l,\vec{a}_{i+1},\ldots,\vec{a}_{l-1},\vec{b}_r,\vec{a}_{l+1},\ldots,\vec{a}_n) =$$
$$\det(\vec{a}_1,\ldots,\vec{a}_{i-1},\vec{b}_k,\vec{a}_{i+1},\ldots,\vec{a}_{l-1},\vec{b}_r,\vec{a}_{l+1}\ldots,\vec{a}_n) \quad (2.7)$$

in the second term of the first member, we can interchange the order of vectors and we change the sign, by replacing the expressions of the first member using their value of (2.4) we get the expression.

$$det(A)[x(k,i)x(r,l) - x(k,l)x(r,i)] = \det(A)\begin{vmatrix} x(k,i) & x(k,l) \\ x(r,i) & x(r,l) \end{vmatrix} \quad (2.8)$$

$$= \det(\vec{a}_1,\ldots,\vec{a}_{i-1},\vec{b}_k,\vec{a}_{i+1},\ldots,\vec{a}_{l-1},\vec{b}_r,\vec{a}_{l+1},\ldots,\vec{a}_n)$$

3-We assume that (2.2) is true for s-1, we prove that is true for the case s.
    Multiply the two terms of system (2.1), the left by

$$\vec{a}_1 \wedge \ldots \wedge \vec{a}_{i-1} \wedge \vec{b}_1 \wedge \ldots \wedge \vec{a}_{l-1} \wedge \vec{b}_{s-1}$$

and the right by $\wedge \vec{a}_{l+1} \wedge \ldots \wedge \vec{a}_n$, we obtain a similar expression of (2.7) and the summation is zero unless $j = i_1, i_2, \ldots, i_s$.

Using the result of the case s-1 and we note the minors by min, we find:
$\det(A)[x(s,i_1)\min(x(s,i_1)) - x(s,i_2)\min(x(s,i_2)) + \ldots + x(s,i_s)\min(x(s,i_s))] =$

$$= \det(A) \begin{vmatrix} x(1,i_1) & \ldots & x(1,i_s) \\ \vdots & \vdots & \vdots \\ x(s,i_1) & \ldots & x(i_s,i_s) \end{vmatrix} = \det(a_1, \ldots, \vec{a}_{i_1}, \vec{b}_1, \ldots, \vec{a}_{i_s}, \vec{b}_s, \ldots, a_n) \qquad (2.9)$$

## 3. The projection of the Hartree-Fock wave function

We present at first the basis of Hartree-Fock and then the calculation of the spectrum of rotations. For the calculation of the spectrum, we applied the projection of the Hartree-Fock wave function, and the application of Cramer's rule's generalization.

### *3.1 The Hartree-Fock basis*
the variation method leads to a Hamiltonian called Hartree-Fock Hamiltonian whose eigenfunctions are the states of particles $\{|c_i\rangle\}$.

We denote the occupied states by $a_1, a_2, \ldots, a_n$ and $b_1, b_2, \ldots, b_i, \ldots$ the unoccupied states.

In the second quantization formalism we write the wave functions of the system with the creation and destruction operators $\{a_i^+, a_j\}, \{b_k^+, b_l\}, 1 \leq i, j \leq n, 1 \leq k, l \leq \infty$ and we choose the wave function of Hartree-Fock as starting point.

$$|\Phi_{HF}\rangle = a_1^+ a_2^+ \ldots a_n^+ |0\rangle \qquad (3.1)$$

We note the states $\{|\Phi_i^j\rangle = b_j^+ a_i |\Phi_{HF}\rangle\}$ by particle-hole states $|1p - 1h\rangle$ and the states $\{|\Phi_{ij}^{lm}\rangle = b_l^+ b_m^+ a_i a_j |\Phi_{HF}\rangle\}$ by the 2particles-2holes states $|2p - 2h\rangle$, etc. All these states form a basis which we call the Hartree-Fock basis.

### *3.2 The energy levels using the projection of the Hartree-Fock wave function*
The spectrum of energy levels is given in the Peierls-Yoccoz theory [4] by

$$E_j = \frac{\int D_{(m,m)}^{*j}(\Omega)\langle\Phi_{HF}|HR(\Omega)|\Phi_{HF}\rangle d\Omega}{\int D_{(m,m)}^{*j}(\Omega)\langle\Phi_{HF}|R(\Omega)|\Phi_{HF}\rangle d\Omega} \qquad (3.2)$$

With H = T + V is the Hamiltonian, T is the kinetic energy and V is the potential energy. $\Omega = (\psi\beta\varphi)$ Is the solid angle and $D_{(m,m)}^j(\Omega) = e^{-im(\psi+\varphi)} d_{(m,m)}^j(\beta)$ is an element of rotations matrix.

In order to assure that the average value of H is minimal [1-2], the variation method imposes the condition:

$$\langle \Phi_{HF} | H | 1p-1h \rangle = \langle \Phi_{HF} | H b_j^+ a_i | \Phi_{HF} \rangle = 0, \forall i,j \qquad (3.3)$$

If we introduce the unitary operator of Hartree-Fock basis between H and R of the expression $\langle \Phi_{HF} | HR(\Omega) | \Phi_{HF} \rangle$ and taking into account the condition (3.3), we find in the case of axial symmetry:

$$E_j = E_{HF} + \frac{1}{4} \sum_{ij} \langle ij | \widetilde{V} | kl \rangle \frac{\int d_{(m,m)}^{*j}(\beta) \langle \Phi_{HF} | a_i^+ a_j^+ b_l b_k e^{-i\beta J_y} | \Phi_{HF} \rangle \sin \beta d\beta}{\int d_{(m,m)}^{*j}(\beta) \langle \Phi_{HF} | e^{-i\beta J_y} | \Phi_{HF} \rangle \sin \beta d\beta} \qquad (3.4)$$

With $\quad \langle ij | \widetilde{V} | kl \rangle = \langle ij | V | kl \rangle - \langle ij | V | lk \rangle$

And $\langle ij | V | kl \rangle$ are the elements of the potential matrix.

We prove by simple calculation that [2]:

$$\langle \Phi_{HF} | e^{-i\beta J_y} | \Phi_{HF} \rangle = \det(\vec{a}_1, \vec{a}_2, \ldots, \vec{a}_n) \text{ With } a_{ij} = \langle a_i | e^{-i\beta J_y} | a_j \rangle$$

And

$$\langle \Phi_{HF} | a_i^+ a_j^+ b_l b_k e^{-i\beta J_y} | \Phi_{HF} \rangle = \det(\vec{a}_1, \ldots, \vec{a}_{i-1}, \vec{b}_k, \vec{a}_{i+1}, \ldots, \vec{a}_{j-1}, \vec{b}_l, \vec{a}_{j+1}, \ldots, \vec{a}_n) \qquad (3.5)$$

With

$$(b_r)_m = \langle b_r | e^{-i\beta J_y} | a_m \rangle, 1 \le m \le n, r = k, l. \qquad (3.6)$$

According to the preceding theorem, we deduce the final expression of energy.

$$E_j = E_{HF} + \frac{1}{4} \sum_{ij} \langle ij | \widetilde{V} | kl \rangle \frac{\int d_{(m,m)}^{*j}(\beta) \langle \Phi_{HF} | e^{-i\beta J_y} | \Phi_{HF} \rangle \begin{vmatrix} x(k,i) & x(k,j) \\ x(l,i) & x(l,j) \end{vmatrix} \sin \beta d\beta}{\int d_{(m,m)}^{*j}(\beta) \langle \Phi_{HF} | e^{-i\beta J_y} | \Phi_{HF} \rangle \sin \beta d\beta} \qquad (3.7)$$

We performed the calculations of $x(k,i)$ using the Gauss elimination method and the integration by Gauss method or Gauss-Legendre integration.

## 4. Derivation of Löwdin formula and Thouless theorem

### 4.1 Generalization of Cramer's rule and Löwdin formula

We can extend the definition of variables $x(k,i)$ by

$$x(k,i) = \langle \Phi_{HF} | R c_k^+ c_i | \Phi_{HF} \rangle / \langle \Phi_{HF} | R | \Phi_{HF} \rangle \qquad (4.1)$$

With $\quad x(k,i) = 0, \text{ if } i > n$

We find that $x(k,i) = \sum_{j=1}^{n} \langle c_k | R | a_j \rangle A_{ji}^{-1}$ and $A_{kj}^{-1}$ are the minor of the matrix (A).

We deduce that the one body potential may be written in the formalism of second quantization:

$$\langle \Phi_{HF} | TR | \Phi_{HF} \rangle = \sum \langle a_i | T | c_k \rangle \sum_k \langle c_k | R | a_j \rangle A_{ji}^{-1} \langle \Phi_{HF} | R | \Phi_{HF} \rangle \quad (4.2)$$

But $\sum_{k=1}^{\infty} |c_k\rangle\langle c_k|$ is the unitary operator in the space of one particle state.

Finally
$$\langle \Phi_{HF} | TR | \Phi_{HF} \rangle = \sum_{ij} \langle a_i | TR | a_j \rangle A_{ji}^{-1} \langle \Phi_{HF} | R | \Phi_{HF} \rangle \quad (4.3)$$

Following the same method we find for the two body potential

$$\langle \Phi_{HF} | VR | \Phi_{HF} \rangle = \sum_{ijkl} \langle a_i a_j | \widetilde{V}R | a_k a_l \rangle \left[ \begin{pmatrix} A_{ki}^{-1} & A_{kj}^{-1} \\ A_{li}^{-1} & A_{lj}^{-1} \end{pmatrix} \right] \langle \Phi_{HF} | R | \Phi_{HF} \rangle \quad (4.4)$$

The inconvenient of Löwdin formula is the calculation of the elements $\{\langle a_i a_j | VR | a_k a_l \rangle\}$ that require long calculation.

### *4.2. Generalization of Cramer's rule and Thouless theorem*

Let $|\Psi\rangle$ and $|\Phi\rangle$ be two wave functions such that $|\Psi\rangle = U|\Phi\rangle$, U is an invertible linear transformation and I is the unit operator of Hartree-Fock basis.

We have
$$Uc_i^+ U^{-1} = \sum_j \langle c_j | U | c_i \rangle c_j^+ \quad (4.5)$$

And
$$|\Psi\rangle = U|\Phi\rangle = IU|\Phi\rangle = \langle \Phi | U | \Phi \rangle |\Phi\rangle + \sum_{ik} \langle \Phi | a_i^+ b_k U | \Phi \rangle b_k^+ a_i |\Phi\rangle + \quad (4.6)$$
$$\sum_{ijkl} \langle \Phi | a_i^+ a_j^+ b_k b_l U | \Phi \rangle b_l^+ b_j^+ a_j a_l |\Phi\rangle + \ldots$$

Applying the theorem we get:

$$|\Psi\rangle = U|\Phi\rangle = IU|\Phi\rangle = \langle \Phi|U|\Phi\rangle|\Phi\rangle + \langle \Phi|U|\Phi\rangle \left[ \sum_{ik} x(k,i) b_k^+ a_i |\Phi\rangle \right]$$
$$+ \langle \Phi|U|\Phi\rangle \left[ \sum_{ijkl} \begin{vmatrix} x(l,i) & x(l,j) \\ x(k,i) & x(k,j) \end{vmatrix} b_l^+ b_k^+ a_j a_i |\Phi\rangle + \ldots \right]$$
$$= \langle \Phi|U|\Phi\rangle \left[ 1 + (\sum_{ik} x(k,i) b_k^+ a_i) + \frac{1}{2!}(\sum_{ik} x(k,i) b_k^+ a_i)^2 + \ldots \right] |\Phi\rangle \quad (4.7)$$

This expression is written in the form

$$|\Psi\rangle = U|\Phi\rangle = \langle \Phi|U|\Phi\rangle \exp\left[ \sum_{k,i} x(k,i) b_k^+ a_i \right] |\Phi\rangle \quad (4.8)$$

In the particular case where $\langle \Phi|U|\Phi\rangle = 1$ we obtain the Thouless function [78].

# Part II- Collective vibration of the nucleus, G.F. Method and quasi-bosons approximation

## 1. Introduction

We shall be concerned with the approximations and the excited states. But in the Hartree-Fock theory the residual interaction is neglected. And if one is interested in collective states then the residual cannot be neglected.
If first we limit the approximation to the Hartree-Fock function and 1p-1h states and then we solve the Schrödinger equation. This approximation or the Random phase approximation (RPA) cannot give good results for phenomena involving two or more particle correlations [38].
The development of techniques for operators of fermions in terms of operators of creation and annihilation of quasibosons proved particularly effective to study the collective Hamiltonian and transition operator of even-even nuclei.

Two development methods were used: that Belyaev and Zelevinsky [38] and that of Marumori and al. [40]. Unfortunately these developments converge slowly when they are trunked.

We intend to show how the generating function method [80] allows the construction of developments in terms of quasibosons that respect the Pauli principle and are also more rapidly convergent.

## 2- The RPA equation of motion

The RPA derives from the well known equations of motion method solving the harmonic oscillator problem.

### 2.1 The equation of motion harmonic oscillator
The equation of motion harmonic oscillator in Quadratic form is:

$$H = \alpha a^{+2} + \beta a^+ a + \gamma a^2 \tag{2.1}$$

Put: $O = xa^+ - ya$   $[H, O] = -\omega O$   and $[H, O^+] = \omega O^+$ ,

We obtain:
$$[a,[H,O^+]] = \omega[a,O^+], \quad And \quad [a^+,[H,O^+]] = \omega[a^+,O^+] \tag{2.2}$$

### 2.2 The RPA equation of motion
In the Hartree-Fock theory the Hamiltonian is

$$H = H_0 + H_{res} \tag{2.3}$$

*If we put* $\quad O^+ = \sum_{mi} Y_{mi} a_m^+ a_i - Z_{mi} a_i^+ a_m$

$$[a_m^+ a_i, [H, O^+]] = \omega [a_m^+ a_i, O^+], \quad \text{And} \quad [a_i^+ a_m, [H, O^+]] = \omega [a_i^+ a_m, O^+] \tag{2.4}$$

We obtain the usual equations of the RPA:
$$\begin{aligned}(\varepsilon_m - \varepsilon_i) Y_{mi} + \sum_{nj} (\widetilde{V}_{mj,in} Y_{mi} + \widetilde{V}_{mn,ij} Z_{nj}) &= \omega Y_{mi} \\ (\varepsilon_m - \varepsilon_i) Z_{mi} + \sum_{nj} (\widetilde{V}_{in,mj} Z_{nj} + \widetilde{V}_{ij,mn} Z_{nj}) &= -\omega Z_{mi}\end{aligned} \tag{2.5}$$

We write these equations in the form:

$$\begin{pmatrix} A & B \\ -B^* & -A^* \end{pmatrix} \begin{pmatrix} Y \\ Z \end{pmatrix} = \omega \begin{pmatrix} Y \\ Z \end{pmatrix} \tag{2.6}$$

### 3. The generating function method (revision)

*3.1 The Generating coordinates Method*

The theory of rotational energy is done by Peierls-Yoccoz [68] using the well known Hill-Wheeler generating coordinate's method:

$$\Psi(x) = \int f(z) \Phi(x, z) dz \tag{3.1}$$

And $\Phi(x, z)$ is the trial function

To study the vibration I was proposing to change the trail function by the generating function and to use the Fock-Bargmann space for integration.

*3.2 New Interpretation of the generating function*

The generating function of the harmonic oscillator is:

$$G(z, q) = \sum_{n=0}^{n} \frac{\overline{z}^n}{\sqrt{n!}} u_n(q) \tag{3.2}$$

We denote the Fock-Bargmann Space $\{u_n(q)\}$ by $\{B\}$, and the Space of waves functions by $\{F\}$. The generating function is given by: $\{\overline{z}^n / \sqrt{n!}\}$

We have $\quad G(z, q) = trace\{\{B\} \otimes \{F\}\}$

And $\quad \hat{A} f(z) = \int f(z') \overline{G(z', q)} A G(z, q) dq \tag{3.3}$

A is an operator belong to $\{F\}$ and $\hat{A}(z)$ belong to $\{B\}$

*3.3 Generalization of the generating function*:

We denote by B the space of orthogonal polynomials $P_m(z) \in B$

We consider the transformation:
$$P_m(z) \in B \leftarrow \cdots \cdots \rightarrow |m\rangle \in F \tag{3.4}$$

And we define the generating function by:
$$|G(z)\rangle = \sum_m \overline{P}_m(z)|m\rangle$$

The image of the operator A is $\hat{A}(z)$

It is very simple to prove that:
$$\langle m|A|m'\rangle = \langle p_m|\hat{A}|p_{m'}\rangle$$

And $\quad \hat{A}^+ = \widehat{A^+}, \quad \widehat{(AB)} = \hat{A}\hat{B})$

So our method satisfies the conditions of B-Z and Marumori et al.
Using the formula
$$\hat{A}f(z) = \int \bar{f}(z')\langle G(z)|A|G(z')\rangle dq \tag{3.5}$$

We write: $\quad \hat{A}\langle G(z)| = \langle G(z)|A \tag{3.6}$

This expression is very useful for computing.

### *3.4 applications to harmonic oscillator*
a- In one dimension harmonic oscillator we have:
$$H = \hbar\omega(a^+a + 1/2) \tag{3.7}$$
Using $\quad \hat{A}\langle G(z,q)| = \langle G(z,q)|A$,

We find $\hat{a} = d/dz, \quad \hat{a}^+ = z, \quad \hat{H} = \hbar\omega(z\dfrac{d}{dz} + 1/2)$

b- In the three dimensions harmonic oscillator:
$$H = \hbar\omega(a_x^+a_x + a_y^+a_y + a_z^+a_z + 3/2)$$
$$\hat{A}\langle G(z,q)| = \langle G(z,q)|A, \Rightarrow \hat{H} = \hbar\omega(z_1\frac{\partial}{\partial z_1} + z_2\frac{\partial}{\partial z_2} + z_3\frac{\partial}{\partial z_3} + 1/2), \tag{3.8}$$
$$P_{(n)}(z) = \frac{z_1^{n_x} z_2^{n_y} z_3^{n_z}}{\sqrt{n_x!n_y!n_z!}}, \quad |n_x, n_y, n_z\rangle = \int P_{(n)}(z)|G(z,q)\rangle d\mu(z)$$

### *3.5 Application to Lipkin model*
The method that has just been developed will now be applied to the model of Lipkin [8] to compare our results and some of those previously obtained.
The Hamiltonian of the system studied is written:
$$H = eJ_0 + \frac{1}{2}(J_+^2 + J_-^2) \tag{3.9}$$

With $[J_+, J_-] = 2J_0$, $[J_0, J_\pm] = 2J_\pm$

The basis $|n\rangle$ of representation is defined by:

$$J_0|n\rangle = (n-J)|n\rangle$$
$$J_+|n\rangle = [(2J-n)(n+1)]|n+1\rangle \qquad (3.10)$$
$$J_-|n\rangle = [(2J-n+1)n]|n-1\rangle$$

And $\quad J_+^{N+1}|n\rangle = 0 \ \forall n, \ N = 2J$

The generating function is:

$$|\Phi(Z)\rangle = \sum_{i=0}^{N} \frac{\overline{Z}^i}{\sqrt{i!}}|i\rangle \qquad (3.11)$$

Will determine the images of the operators $J_0, J_+, J_+^2$

1-Image of $\quad \widehat{J}_0 \langle \Phi(Z)| = \langle \Phi(Z)|J_0 \qquad (3.12)$

We find that $\widehat{J}_0 \langle \Phi(Z)| = (Z\frac{\partial}{\partial Z} - J)\langle \Phi(Z)|$

Therefore $\widehat{J}_0 = (Z\frac{\partial}{\partial Z} - J)$

2-Image of $J_+$

$$\widehat{J}_+ \langle \Phi(Z)| = \langle \Phi(Z)|J_+ = \sum_{n=0}^{N} \frac{Z^n}{\sqrt{n!}}\langle n|J_+ = \sum_{n=1}^{N} \frac{Z^n}{\sqrt{n!}}\langle n-1|[n(2J-n+1)]^{1/2}$$

Note that $\widehat{J}_+$ must be developed in the form:

$$\widehat{J}_+ = \alpha_0 Z + \alpha_1 Z_1 \frac{\partial}{\partial Z} + \alpha_2 Z^2 (\frac{\partial}{\partial Z})^2 + ... \qquad (3.13)$$

We have

$$\widehat{J}_+ \langle \Phi(Z)| = \sum_{ij} \frac{(i-1)!Z^i}{\sqrt{(i-1)!(i-j-1)!}}\langle i-1| = \sum_{n=1}^{N} \frac{Z^n}{\sqrt{n!}}\langle n-1|[n(2J-n+1)]^{1/2}$$

The vectors $\{|n\rangle\}$ are linearly independents, we deduce that

$$\sum_{j=0}^{n-1} \alpha_j \frac{(n-1)!}{(n-j-1)!} = [2J-n+1]^{1/2}$$

Calculating the coefficients $\alpha_j$ is done by induction

3-Image of $\quad \widehat{J}_+^2 \langle \Phi(Z)| = \langle \Phi(Z)|J_+^2 \qquad (3.14)$

We put $\quad \widehat{J}_+^2 = \beta_0 Z + \beta_1 Z_1 \frac{\partial}{\partial Z} + \beta_2 Z^2 (\frac{\partial}{\partial Z})^2 + ...$

We find after calculation that:

$$\alpha_0 = \sqrt{2J}, \; \alpha_1 = \sqrt{2J-1} - \sqrt{2J}, \ldots$$
$$\beta_0 = \sqrt{2J(2J-1)}, \; \beta_1 = \sqrt{(2J-1)(2J-2)} - \sqrt{2J(2J-1)}, \ldots \quad (3.15)$$

Our method makes it possible to quickly condense the form of the development.

# 4. The renormalized Thouless function as generating function for the many body problem

*4.1 The Thouless function is:*

$$|\Psi\rangle = U|\Phi\rangle = \langle\Phi|U|\Phi\rangle \exp\left[\sum_{\alpha,i} x(\alpha,i) b_i^+ a_\alpha\right]|\Phi\rangle \quad (4.1)$$

Put 
$$A = \sum_{\alpha,i} z_{\alpha i}\, a_\alpha^+ b_i \text{ and } \langle\Phi|U|\Phi\rangle = 1$$

We write: 
$$|\Psi\rangle = \exp[A^+]|\Phi\rangle, \; z_{\alpha i} \in C$$

And
$$|\Psi\rangle = \sum_n \frac{A^{+n}}{n!}|\Phi\rangle = |\Phi\rangle + \left[\sum_{i\alpha} z_{(\alpha,i)}\, b_\alpha^+ a_i\, |\Phi\rangle\right] +$$
$$+\left[\sum_{ijkl} \begin{vmatrix} z_{(\alpha,i)} & z_{(\alpha,j)} \\ z_{(\beta,i)} & z_{(\beta,j)} \end{vmatrix} b_\alpha^+ b_\beta^+ a_j a_i |\Phi\rangle + \ldots\right] \quad (4.2)$$

-We denote the holes by $(\alpha,\beta,\ldots)$ and the particles $(i,j,k,\ldots)$

*4.2 Renormalization of the Thouless function*

The generating function may be obtained by the renormalization of Thouless function. We consider:

$$\bar{z}_{(\alpha,i)}, \begin{vmatrix} \bar{z}_{(\alpha,i)} & \bar{z}_{(\alpha,j)} \\ \bar{z}_{(\beta,i)} & \bar{z}_{(\beta,j)} \end{vmatrix}, \; etc\ldots$$

As a basis of Fock-Bargmann space but we must normalize this basis then

$$|\Psi_N(z)\rangle = \sum_n \frac{A^{+n}}{n!\sqrt{n!}}|\Phi\rangle = |\Phi\rangle + \left[\sum_{i\alpha} \bar{z}_{(\alpha,i)}\, b_\alpha^+ a_i |\Phi\rangle\right] +$$
$$+\left[\sum_{ijkl} \frac{1}{\sqrt{2!}}\begin{vmatrix} \bar{z}_{(\alpha,i)} & \bar{z}_{(\alpha,j)} \\ \bar{z}_{(\beta,i)} & \bar{z}_{(\beta,j)} \end{vmatrix} b_\alpha^+ b_\beta^+ a_j a_i |\Phi\rangle + \ldots\right] \quad (4.3)$$

We write the generating function as:

$$|G(z)\rangle = \bar{P}_m(z)|m\rangle = (\sum_n \frac{A^{+n}}{n!\sqrt{n!}})|\Phi\rangle \quad (4.4)$$

And for the applications we use: $\hat{X}|G(Z)\rangle = (G(Z)|X$

# 5. The quasiboson development of the Hamilton

## 5.1. The Hamiltonian of the nucleus is:

$H_{20} = 0 \quad if \; |\Phi_0\rangle$ is the Hartree-Fock wave Function.

We write

$$H = E_0 + H_{11} + (H_{20} + H'_{20}) + H_{22} + (H_{40} + H'_{40}) + H'_{22} + (H_{31} + H'_{31})$$

$$E_0 = \langle \Phi_0 | H | \Phi_0 \rangle = \sum_\alpha h_{\alpha\alpha} + (\sum_{\alpha\beta} \widetilde{V}_{\alpha\beta,\alpha\beta})/2$$

$$H_{11} = \sum_{ij}(h_{ij} + \sum_\alpha \widetilde{V}_{\alpha i,j\alpha})a_i^+ a_j - \sum_{\alpha\beta}(h_{\alpha\beta} + \sum_\gamma \widetilde{V}_{\alpha\gamma,j\gamma})b_\alpha^+ b_\beta$$

$$H_{20} = \sum_{i\alpha}(h_{i\alpha} + \sum_\beta \widetilde{V}_{i\beta,\alpha\beta})a_i^+ b_\alpha$$

$$H_{22} = \sum_{i\alpha\beta}(\sum_\beta \widetilde{V}_{i\beta,\alpha j})a_i^+ b_\alpha^+ b_\beta a_j$$

$$H_{40} = \sum_{i\alpha\beta}(\widetilde{V}_{i\beta,\alpha\beta}) \times a_i^+ b_\alpha^+ a_j^+ b_\beta^+ / 4$$

$$H'_{22} = \sum_{ijkl}\widetilde{V}_{ij,kl}a_i^+ a_j^+ a_l a_k / 4 + \sum_{\alpha\beta\gamma\delta}\widetilde{V}_{\alpha\beta,\gamma\delta}b_\delta^+ b_\gamma^+ b_\alpha b_\beta / 4$$

$$H_{31} = \sum_{ijk\alpha}\widetilde{V}_{ij,\alpha k}a_i^+ b_\alpha^+ a_j^+ a_k / 2 + \sum_{i\alpha,\beta\gamma}\widetilde{V}_{i\alpha,\beta\gamma}a_i^+ b_\gamma^+ b_\beta^+ b_\alpha / 2$$

(5.1)

## 5.2-The image of the operators: particles-holes

Using the formula $\hat{A}|G(z)\rangle = \langle G(z)|A$ we find the image of particles-holes in terms Of $\{z_{ij}\}$:

$$\widehat{(a_i^+ a_j)} = \sum_\gamma Z_{i\gamma}\frac{\partial}{\partial Z_{j\gamma}}$$

$$\widehat{(b_\alpha^+ b_\beta^+)} = \sum_j Z_{j\alpha}\frac{\partial}{\partial Z_{j\beta}}$$

(5.2)

$$\widehat{(b_\alpha^+ a_i)} = \alpha_0 \hat{K}_0 + \alpha_1 \hat{K}_1 + ...,$$

with $\hat{K}_0 = Z_{i\alpha}$, $\hat{K}_1 = \sum_{\beta j} Z_{j\beta} \frac{\partial}{\partial Z_{j\beta}} \frac{\partial}{\partial Z_{i\beta}} B_{j\beta}^+ B_{j\alpha} B_{i\beta}$, ....

The latest development is rapidly converging with $\alpha_0 = 1$, $\alpha_1 = 1 - \sqrt{2}$,

And $\quad \alpha_2 \approx -0.048, \alpha_3 \approx -o.0076$.

The general term $\{\alpha_i\}$ being defined by the recurrence relation:

$$n\alpha_0 - n(n-1)\alpha_1 + n(n-1)(n-2)\alpha_2 + ... + (-1)^n n!\alpha_{n-1} = n\sqrt{n}, \; \alpha_0 = 1$$

Taking into account the correspondence between the Bargmann-Fock space {F} and the bosons space {B} we write:

$$Z_{i\alpha} \to B^+_{i\alpha}, \quad \frac{\partial}{\partial Z_{i\alpha}} \to B_{i\alpha}$$

We find simply the transformation of the operators with the operators of bosons.

$$(a^+_i a_j)_B = \sum_\gamma B^+_{i\gamma} B_{j\gamma}$$
$$(b^+_\alpha b^+_\beta)_B = \sum_j B^+_{j\alpha} B_{j\beta} \tag{5.3}$$
$$(b^+_\alpha a_i)_B = \alpha_0 K_0 + \alpha_1 K_1 + ...,$$
with $K_0 = B_{i\alpha}, \; K_1 = \sum_{\beta j} B^+_{j\beta} B_{j\alpha} B_{i\beta}, \; ....$

*5.3-The image of the operators: two particles-two holes*

$$1 - (a^+_i b^+_\alpha b_\beta a_j)_B = B^+_{i\alpha} B_{j\beta} - \sum_{m\gamma} B^+_{i\gamma} B^+_{m\alpha} B_{m\gamma} B_{j\beta}$$

$$2 - (b_\beta a_j b_\alpha a_i)_B = \alpha_1 B_{j\beta} B_{i\alpha} + \alpha_2 \sum_{k\gamma} B^+_{k\gamma} B_{k\gamma} B_{j\gamma} B_{i\alpha} + ...$$

$$\alpha_1 = \sqrt{2}, \; \alpha_2 = \sqrt{2} - \sqrt{6} = -1{,}035$$

The third term coefficient is:

$$\alpha_2 = (2\sqrt{3} + \sqrt{2} - 2\sqrt{6}) \approx -0.01$$

This coefficient is negligible.

$$3 - (a^+_k a_j b_\alpha a_i)_B = \sqrt{2} \sum_\beta B^+_{k\beta} B_{j\beta} B_{i\alpha} + ......$$
$$4 - (a^+_i b^+_\alpha b_\beta a_j)_B = -\sum_{\alpha\beta} B^+_{i\alpha} B^+_{j\beta} B_{l\alpha} B_{k\beta}$$
$$5 - (b^+_\delta b^+_\gamma b_\alpha b_\beta)_B = -\sum_{mn} B^+_{m\delta} B_{n\gamma} B_{m\alpha} B_{n\beta} \tag{5.4}$$
$$6 - (b^+_\alpha b_\beta b_\gamma a_i)_B = \sqrt{2} \sum_m B^+_{m\alpha} B_{m\beta} B_{i\gamma} + ...$$

## 5.4 The expression of the Hamiltonian in terms of bosons operators

$$H = E_0 + \sum_{i\alpha}(e_i - e_\alpha)B^+_{i\alpha}B_{i\alpha} + \sum_{i\alpha j\beta}(\widetilde{V}_{i\beta,\alpha j})B^+_{i\alpha}B_{j\beta} +$$

$$+ \frac{1}{2\sqrt{2}} \sum_{ij,\alpha\beta}\left[\widetilde{V}_{ij,\alpha\beta} B^+_{i\alpha}B^+_{j\beta} + hc\right] +$$

$$\frac{\sqrt{2}}{2} \sum_{ijk\alpha\beta}\left[\widetilde{V}_{ij,\alpha k} B^+_{i\alpha}B^+_{j\beta}B_{k\beta} + hc\right] + \frac{\sqrt{2}}{2} \sum_{im\alpha\beta\gamma}\left[\widetilde{V}_{i\alpha,\beta\gamma} B^+_{i\gamma}B^+_{m\beta}B_{m\alpha} + hc\right] \quad (5.5)$$

$$- \sum_{ijm\alpha\beta\gamma}\widetilde{V}_{i\beta,\alpha\gamma} B^+_{i\gamma}B^+_{m\alpha}B_{m\gamma}B_{j\beta} - \frac{1}{4}\sum_{ijkl\alpha\beta}\widetilde{V}_{ij,kl} B^+_{i\alpha}B^+_{j\beta}B^+_{l\alpha}B_{k\beta}$$

$$- \frac{1}{4} \sum_{mn\alpha\beta\gamma\delta}\widetilde{V}_{\alpha\beta,\gamma\delta} B^+_{m\delta}B_{n\gamma}B_{m\alpha}B_{n\beta} + \frac{1-\sqrt{3}}{2\sqrt{2}} \sum_{ijkl\alpha\beta\gamma}\left[\widetilde{V}_{ij,\alpha\beta} B^+_{i\alpha}B^+_{j\gamma}B^+_{k\beta}B_{k\gamma} + hc\right]$$

Note that the developments of Belyaev-Zelevinsky and Marumori et al. different from ours by the presence of surplus operators whose action on all elements of the subspace {B} of bosons is null. Such an operator is for example the following:

$$(\frac{\partial}{\partial Z_{j\beta}}\frac{\partial}{\partial Z_{i\alpha}} + \frac{\partial}{\partial Z_{j\alpha}}\frac{\partial}{\partial Z_{i\beta}})f(Z) = 0$$

Or
$$B_{j\beta}B_{i\alpha} + B_{j\alpha}B_{i\beta} = 0$$

Therefore in the generating function method the development of the operators converges quickly, conserve the commutations relations and the matrix elements ,verify the Pauli principle and the calculus of the image of the Hamiltonian of fermions in terms quasi-bosons is elementary and simple. The development will be very rapidly convergent and therefore very useful [80]. Then the method described here will therefore ultimately to obtain rigorous development in terms of quasibosons, observables of a system of fermions, valid up to an order as high as desired.

After 90, the calculations conducting by many authors and the comparison with experimental results showed the importance of this formula in many body problems [89]

# Chapter VI

## On the Euler angles for the classical groups and The Wigner's Symbols for SU (3) multiplicity free

**1-Introduction**
**2. The classical groups**
   2.1 The special orthogonal group SO(*n*)
   2. 2 special unitary group SU(*n*)
   2.3 The symplectic group Sp(*n*)
   2.4 The infinitesimal group generators
**3. On the Euler angles for the classical groups**
   3.1 Generalization of the Euler parameterization of SO(3)
   3.2- Parameterization of SO(n)
   3.3 Parameterization of SU(n)
   3.4 Parameterization of SO(6)
**4. The invariant measure on the group SU (n)**
   4.1 The invariant measure of the group SO (n) and Euclidean measure
   4.2 The invariant measure of the group SU(n) and Fock-Bargmann spaces
**5- Generating function of the basis SU(2) ⊂SU(3)**
   5.1 The basis of the group SU (2) ⊂ SU (3)
   5.2 The generating function of the basis SU(2)⊂SU(3)
   5.3 The expression of the generating function of SU(3)
**6. Generating function of the 3j symbols for multiplicity-free of SU (3)**
   6.1 The invariant of the 3j symbols for multiplicity-free $\mu_1 = \mu_2 = 0$.
   6.2 The generating function of the 3j symbols for multiplicity-free $\mu_1 = \mu_2 = 0$.
   6.3 Expression of 3j symbols of SU (3).

# Chapter VI

## On the Euler angles for the classical groups and
## The Wigner's Symbols for SU (3) multiplicity free

### 1. Introduction

The applications of the SU (n) group theory have occurred in numerous research areas: nuclear physics, high energy particle theory and experimental nano-scale physics.
Several problems remain under investigation:
 a-The parameterization of these groups [41-43],
 b-The explicit determination of Wigner's D-functions is not found [4-6].
 c-The Wigner's 3j coefficients are very important for applications and are not completely solved despite the extensive efforts made by many authors [44-46].
 In this work we start from the order of the classical groups to determine new recurrences formulas of parameterization and from which we derive the generalization of Euler's angles for these groups.
 We prove the connection of the measure of SU(n) with the measure of product of cylindrical basis of harmonic oscillator or the two dimensions Fock-Bargmann spaces.
 The basis of the representation of SU (3) was constructed by many authors [93-96] and we've built the generating function of this basis using Schwinger's coupling method of angular momentum in Fock- Bargmann space [91-92].
 The invariants for 3j symbols of multiplicity-free are functions of the powers of the elementary invariants of SU (3) and the normalization is feasible in this case. The expression of the isoscalar factor in a compact form is found for the first time.

### 2. The classical groups

We give a quick revision of the properties of classical groups, then we derive from two kinds of recurrences relations the parameterization of the classical groups and then the measures of integration on SO (n), SU (n) and the connection of the measure of unitary groups with the measures of integration in Fock-Bargmann spaces.

*2.1 The special orthogonal group SO(n)*

The special orthogonal group SO(*n*) is the group of *n*×*n* orthogonal matrices ($A_n^0$) with unit determinant. They form real compact lie groups of dimension *n* (*n*-1)/2.
The real special orthogonal matrices leave invariant the real quadratic form:

$$\sum_{i=1}^{n} x_i^2 \qquad (2.1)$$

## 2.2 special unitary group SU(n)

The special unitary group of degree $n$, denoted SU($n$), is the group of $n \times n$ unitary matrices ($A_n^1$) with unit determinant. The special unitary group SU($n$) is a real matrix of dimension f(n)= $n^2 - 1$.
The unitary group leaves invariant the hermitian form:

$$\sum_{i=1}^{n} z_i \bar{z}_i \tag{2.2}$$

## 2.3 The symplectic group Sp(n)

This is the Lie algebra of Sp(n), the group of n x n quaternionic matrices ($A_n^2$) that preserve the standard hermitian form on $\mathbf{Q}^n$:

$$\langle x, y \rangle = x_1 y_1 + x_2 y_2 + ... + x_n y_n \tag{2.3}$$

That is, SP(n) is just the quaternionic unitary group, Sp(n) is a real Lie group of Dimension f(n)=n(2n+1).

## 2.4 The infinitesimal group generators

The elements A of group G are composed of nonsingular matrices of degree n and can be expressed in terms of r continuous parameters

$$A = A(\alpha_1,...,\alpha_r).$$

Such that the infinitesimal group generators are:

$$X_k = \left( \partial A / \partial \alpha_k \right)_{\alpha=0} \tag{2.4}$$

We have the important group theory formula:

$$Of(x) = f({}^t o(x)), \ x \in E_n \tag{2.5}$$

We have also the important group theory formula:

$$Uf(z) = f({}^t u(z)), \ z \in C_n \tag{2.6}$$

The generators of unitary group may be written in terms of creations and destruction of n-dimensional harmonic oscillators as:

$$E_{ij} = \sum_{ij1}^{n} a_i^+ a_j, (i,j = 1,...,n) \tag{2.7}$$

Using theses formulas we derive the generators of SU(3) in part five.

# 3. On the Euler angles for the classical groups

We establish recurrences formulas of the order of the classical groups that allow us to find a generalization of Euler's angles for classical groups and the invariant measures of these groups.

### 3.1 Generalization of the Euler parameterization of SO(3)

In Quantum mechanic we write the matrix elements of rotation by

$$\langle lm'|R(\psi\theta\varphi)|lm\rangle = \langle lm'|e^{-i\psi L_z}e^{-i\theta L_y}e^{-i\varphi L_z}|lm\rangle$$

$$\phantom{xx}\uparrow\phantom{xxxxxx}\uparrow$$
$$\phantom{xx}(2)\phantom{xxxx}(1)$$

(*1*) *is a rotation in the space*

(*2*) *is a rotation in the dual – space*, (3.1)

We observe that for every rotation in the space $\{|lm\rangle\}$ there is a rotation in the dual space $\{\langle lm'|\}$. In this interpretation, we can write the finite transformation of classical groups in the form:

$$A_n^m = A_{n-1}^m B_n^m A_{n-1}^m \tag{3.2}$$

With m = 0, 1 and 2 for orthogonal, unitary and symplectic groups.
In the following we derive two kinds of recurrences formulas

### 3.1.1 First recurrences relations for the number of parameters

It's simple to verify the recurrences relations

$$\begin{aligned}
a)\quad & \frac{n(n-1)}{2} = \frac{(n-1)(n-2)}{2} + n - 1 \\
b)\quad & n^2 - 1 = [(n-1)^2 - 1] + 2n - 1 \\
c)\quad & n(2n+1) = [(n-1)[2(n-1)+1]] + 4n - 1
\end{aligned} \tag{3.3}$$

We obtain the relation

$$N(n) = N(n-1) + 2^m n - 1, \; m = 0,1,2. \tag{3.4}$$

So the order of the matrix n has parameters more than the matrix of order n-1.
Since the point $(0,...,0,1)$ is invariant by the group of order n-1 this means that the last column and the last row are the components of the unit vectors of points on the unit sphere $S^{m,n-1}$ of the Euclidian space $E_n(K), K = R, C, H = Q$.

$$A_n^m = \begin{pmatrix} \vdots & \cdots & Last \\ \vdots & \cdots & Col. \\ Last & Row & a_{nn} \end{pmatrix}$$

### 3.1.2 Second recurrences relations for the number of parameters

It's also simple to verify the recurrences relations

$$\begin{aligned}
a)\quad & \frac{n(n-1)}{2} = 2[\frac{(n-1)(n-2)}{2}] - [\frac{(n-2)(n-3)}{2}] + 1 \\
b)\quad & n^2 - 1 = 2[(n-1)^2 - 1] - [(n-2)^2 - 1] + 2 \\
c)\quad & n(2n+1) = 2[(n-1)[2(n-1)+1]] - [(n-2)[2(n-2)+1]] + 4
\end{aligned}$$

We can write these expressions in the form
$$N(n) = 2N(n-1) - N(n-2) + 2^m, \quad m = 0,1,2.$$
In the expression $\quad A_n^m = A_{n-1}^m B_n^m A_{n-1}^m$

It is quite evident that the parameters of left and right are different.

But
$$A_n^m = A_{n-2}^m B_{n-1}^m A_{n-2}^m B_n^m A_{n-2}^m B_{n-1}^m A_{n-2}^m$$
$$= A_{n-2}^m B_{n-1}^m [A_{n-2}^m B_n^m A_{n-2}^m] B_{n-1}^m A_{n-2}^m$$

We choose
$$[B_n^m, A_{n-2}^m] = 0 \qquad (3.5)$$

Then number of parameters $A_n^m$ of becomes $2(N-1, m) - N(n-2, m) + 2^m$ and the number of $B_n^m$ parameters of is $2^m$. Therefore we find the same result of the recurrence relation.

Therefore we write $\quad A_n^m = A_{n-2}^m B_{n-1}^m A_{n-2}^m B_n^m A_{n-2}^m B_{n-1}^m A_{n-2}^m$

To find $A_n^m$ we must choose the parameters such that the last line, or the last column, are the components of the vector $\vec{r} = (x_1, x_2, ..., x_n)$, $\vec{r} \bullet \vec{r} = 1$ and $[B_n^m, A_{n-2}^m] = 0$. In this case the range of parameters is imposed by the range of the variation of the angles of the vector $\vec{r}$.

It is important to note that every parameterization components of the vector $\vec{r} = (x_1, x_2, ..., x_n)$ corresponds to a parameterization of classical groups and therefore the parameterization is not unique.

## 3.2- Parameterization of SO(n)

In this case $A_n^m = A_{n-1}^m B_n^m A_{n-1}^m$, m = 0 the matrix $B_n^0$ is function of one variable and

The expression $[B_n^0, A_{n-2}^1] = 0$ means that $B_n^0$ leave invariant $A_{n-2}^0$.

Then we write
$$B_n^0 = \begin{pmatrix} I_{n-2} & 0 & 0 \\ 0 & \cos\theta_{n-1}^{n-1} & \sin\theta_{n-1}^{n-1} \\ 0 & -\sin\theta_{n-1}^{n-1} & \cos\theta_{n-1}^{n-1} \end{pmatrix} \qquad (3.6)$$

If we choose in $E_n$ the spherical coordinates $\theta_1, \theta_2, ... \theta_{n-1}$ we write
$$\xi_1 = \sin\theta_{n-1} ... \sin\theta_2 \sin\theta_1$$
$$\xi_2 = \sin\theta_{n-1} ... \sin\theta_2 \cos\theta_1$$
$$\xi_3 = \sin\theta_{n-1} ... \cos\theta_2$$
$$\vdots$$
$$\xi_{n-1} = \sin\theta_{n-1} \cos\theta_{n-2}$$
$$\xi_n = \cos\theta_{n-1}$$

With $\quad 0 \leq \theta_1 \leq 2\pi, \quad 0 \leq \theta_j \leq \pi, j = 2, ..., n-1$

The position vector $\vec{r} = \vec{r}^{\,n} = (x_1, x_2, ..., x_n)$

By use of the polar coordinates $(r, \theta_1, \theta_2, ..., \theta_{n-1})$ defined by $x_i = r\xi_i$.
We find the Vilenkin's parameterization [121] for SO(n) and therefore we shall use the same notations. Any rotation g of the group SO (n) can be set as follows

$$g = g^{(n-1)}...g^{(1)}$$

Where
$$g^{(k)} = g_1(\theta_1^k)...g_k(\theta_k^k) \tag{3.7}$$

And $g_{(n-1)}(\theta_{(n-1)}^{(n-1)}) = B_n^0$ is the transformation

$$\begin{aligned}x'_{n-1} &= x_{n-1}\cos\theta_{n-1}^{n-1} + x_n \sin\theta_{n-1}^{n-1} \\ x'_n &= x_{n-1}\sin\theta_{n-1}^{n-1} + x_n \cos\theta_{n-1}^{n-1}\end{aligned} \tag{3.8}$$

## *3.3 Parameterization of SU(n)*

In the case of m = 1 the matrix $B_n^1$ is function of two variables and $\det(B_n^1) = 1$. The expression $[B_n^1, A_{n-2}^1] = 0$ means that $B_n^1$ leave invariant $A_{n-2}^1$ and the solution is not unique for n>2. If we parameterize like above the last column by the spherical Coordinates:

$$z_i = re^{i\psi_i}\xi_i, r = 1.$$

We write
$$u_1^k = \begin{pmatrix} e^{-i\psi_1^k} & 0 \\ 0 & e^{+i\psi_1^k} \end{pmatrix}, \quad u_i^k(\theta_i^k, \psi_i^k) = B_n^2 = g_i(\theta_i^k)d_i(\psi_i^k)$$

$$d_i(\psi_i^k) = \begin{pmatrix} I_{n-2} & 0 & 0 \\ 0 & e^{-i\psi_i^k} & 0 \\ 0 & 0 & e^{+i\psi_i^k} \end{pmatrix} \tag{3.9}$$

Where
$$u = u^{(n-1)}...u^{(1)}$$

And
$$u^{(k)} = u_0^k(\psi_1^k)u_1(\theta_1^k, \psi_2^k)...u_k(\theta_k^k, \psi_{k+1}^k) \tag{3.10}$$

We can also consider other useful options (22), for example

$$u_1^k = \begin{pmatrix} e^{-i\psi_1^k} & 0 \\ 0 & e^{+i\psi_1^k} \end{pmatrix}, \quad u_i^k(\theta_i^k, \psi_i^k) = B_n^2 = g_i(\theta_i^k)d_i(\psi_i^k)$$

$$d_i(\psi_i^k) = \begin{pmatrix} e^{-i\psi_i^k}I_{n-1} & 0 \\ 0 & e^{-i(n-1)\psi_i^k} \end{pmatrix}$$

$$u = u^{(n-1)}...u^{(1)} \tag{3.11}$$

Where
$$u^{(k)} = u_0^k(\psi_1^k)u_1(\theta_1^k, \psi_2^k)...u_k(\theta_k^k, \psi_{k+1}^k) \tag{3.12}$$

$$SU(2) \qquad U_2 = A_2^2 = \begin{pmatrix} a_1 & a_2 \\ -\bar{a}_2 & \bar{a}_1 \end{pmatrix} \qquad (3.13)$$

$$SU(3) \qquad U_3 = A_3^2 = A_2^2 [B_3^2] A_2^2$$

$$\begin{pmatrix} a_1 & a_2 & 0 \\ -\bar{a}_2 & \bar{a}_1 & 0 \\ 0 & 0 & 1 \end{pmatrix} \left[ \begin{pmatrix} 1 & 0 & 0 \\ 0 & \cos\frac{v_3}{2} & \sin\frac{v_3}{2} \\ 0 & -\sin\frac{v_3}{2} & \cos\frac{v_3}{2} \end{pmatrix} \begin{pmatrix} d_3 & 0 & 0 \\ 0 & d_3 & 0 \\ 0 & 0 & \bar{d}_3^2 \end{pmatrix} \right] \begin{pmatrix} b_1 & b_2 & 0 \\ -\bar{b}_2 & \bar{b}_1 & 0 \\ 0 & 0 & 1 \end{pmatrix} \qquad (3.14)$$

with $0 \le v_3 \le \pi, \qquad d_3 = e^{i\beta_3}, 0 \le \beta_3 \le \pi.$

### *3.4 Parameterization of SO(6)*

It's known that the Lie algebra of SO(6) and the Lie algebra SU(4) are isomorphic. Therefore, there are a non-singular mapping between the generators of SO(6) and SU(4). Since such mapping preserves the Lie bracket structure, we can deduce a parameterization of SO(3), SO(4), SO(5) and SO(6) using the expressions of the generators (2.5) and the harmonic oscillator basis .

## 4. The invariant measure on the group SU (n)

the invariant measure is the result of the product of invariants measure on the sphere $S^{2n-1}$ With n = 1... n. We determine first the invariant measure of the group SO(n) and then for the group SU (n).

### *4.1 The invariant measure of the group SO (n) and Euclidean measure*

The metric on the sphere $S^n$ is:
$$ds^2 = dr^2 + r^2 [\sum_{i=1}^{n} d\xi_i^2] \qquad (4.1)$$

By use of the polar coordinates

$$dS^2 = dr^2 + r^2 d\theta_{n-1}^2 + r^2 \sin^2\theta_{n-1} d\theta_{n-2}^2 + ..... + r^2 \sin^2\theta_{n-1} .. \sin^2\theta_2 d\theta_1^2$$

Hence
$$dV_s = r^{(n-1)} d\xi = A r^{(n-1)} \sin^{n-2}\theta_{n-1} ... \sin\theta_2 d\theta_1 ... d\theta_{n-1}$$

We choose the constant A so that the measure on the sphere is equal to one.

Since $\qquad \int_{-\infty}^{\infty} e^{-x^2} dx = 1/(\pi)^{\frac{1}{2}},$

And in the Cartesian n-dimensional harmonic oscillator we have

$$\int e^{-r^2} \prod_{i=1}^{n} dx_i = \int e^{-r^2} d\vec{r} = \int_0^\infty e^{-r^2} r^{n-1} dr d\xi$$

$$= \frac{1}{2}\Gamma(n/2)\frac{1}{(\pi)^{\frac{n}{2}}}\int_0^\infty d\xi = A$$

So
$$A = \Gamma(n/2)/(2\pi^{n/2}).$$

Finally
$$dV_s = \frac{\Gamma(n/2)}{2\pi^{n/2}}\sin^{n-2}\theta_{n-1}...\sin\theta_2 d\theta_1...d\theta_{n-1} \quad (4.2)$$

The invariant measure on the group SO(n) is:
$$dg = A_n \prod_{k=1}^{n-1}\prod_{j=1}^{k}\sin^{j-1}\theta_j^k d\theta_j^k \quad (4.3)$$

With
$$A_n = \prod_{k=1}^{n}\frac{\Gamma(k/2)}{2\pi^{k/2}}$$

The invariant measure of SO (n) is the angular part of product measure of Cartesian harmonic oscillator.

$$\prod_{i=1}^{n} e^{-(r^i)^2} d\vec{r}^i = (\prod_{i=1}^{n} e^{-(r^i)^2}(r^i)^{i-1} dr^i) d\xi,$$
$$d\xi = (\prod_{i=1}^{n} d\xi^i) \quad (4.4)$$

The number of parameters of SO(n) is $\frac{n(n-1)}{2} = 2^0(\prod_{i=1}^{n} i) - n$ with n is the number of parameters $r_i$.

### *4.2 The invariant measure of the group SU(n) and Fock-Bargmann spaces*

By use of the polar coordinates $z = (z_1, z_2,...z_n)$, $z_i = r\hat{z} = re^{-i\psi_i}\xi_i$

The metric on the sphere $S^{2n-1}$ is:
$$ds^2 = dr^2 + r^2[\sum_{i=1}^{n} d(e^{-i\psi_i}\xi_i)d(e^{+i\psi_i}\xi_i))$$
$$ds^2 = dzd\bar{z} = dr^2 + r^2[\sum_{i=1}^{n} d\psi_i^2 \xi_i^2 + d\xi_i d\xi_i]$$

Therefore
$$dV_s = d\bar{z} = A(\prod_{i=1}^{n}\xi_i)\sin^{n-2}\theta_{n-1}...\sin\theta_2 d\theta_1...d\theta_{n-1} d\psi_1...d\psi_n$$
$$= A(\prod_{i=1}^{n}\xi_i)(\prod_{i=1}^{n}d\xi_i)\prod_{i=1}^{n}d\psi_i \quad (4.5)$$

We deduce the connection between the 2n-dimensional cylindrical bases of harmonic oscillator, the measure of integration of Bargmann spaces of dimension 2n and the measure on the sphere $S^{2n-1}$.

We obtain

$$e^{-r^2}\prod_{i=1}^{n}dz_i d\bar{z}_i = e^{-r^2}\prod_{i=1}^{n}\xi_i d\xi_i d\psi_i = e^{-r^2}r^{2n-1}(\prod_{i=1}^{n}\xi_i)(\prod_{i=1}^{n}d\xi_i)\prod_{i=1}^{n}d\psi_i$$

And therefore we note for the following of this work

$$d\mu(U_n) = \prod_{i=2}^{n}d\mu(z^n),\ d\mu(z^n) = e^{-z^n\bar{z}^n}\prod_{i=1}^{n}dz_i^n d\bar{z}_i^n$$

We determine A by observing that

$$\int d\mu(U_n) = \int_0^\infty e^{-r^2} r^{2n-1} dr d(\widehat{U}_n) = \frac{1}{2}\Gamma(n)\frac{1}{\pi^n}\int d(\widehat{U}_n) = A$$

And $A = \dfrac{\Gamma(n)}{2\pi^n}$. Finally:

$$dV_s = \frac{\Gamma(n)}{2\pi^n}\prod_{i=1}^{n}\xi_i \sin^{n-2}\theta_{n-1}...\sin\theta_2 d\theta_1...d\theta_{n-1}d\psi_1...d\psi_n \qquad (4.6)$$

The same arguments for the derivation of the measure of integration of SO (n) remain valid in the case of SU (n). It follows that the measure of integration of the group SU (n) must be taken as the angular part of the measure of product of basis of the cylindrical harmonic oscillators.

$$d\mu(U_n^g) = \prod_{i=1}^{n}e^{-(r^i)^2}dz^i d\bar{z}^i = (\prod_{i=1}^{n}e^{-(r^i)^2}(r^i)^{2i-1}dr^i)d(U_n^g),$$

With
$$z^i = (z_1^i, z_2^i,...,z_i^i), (r^i)^2 = z^i\bar{z}^i$$

And
$$d(U_n^g) = (\prod_{i=2}^{n}d(U_i)) \qquad (4.7)$$

The number of parameters SU (n) is $n^2 - 1 = 2^1(\sum_{i=2}^{n}i) - (n-1)$ with (n-1) is the number of parameters $\{r_i\}$ and the sum is the dimension of the space. We obtain then the relationship between the measure of Fock- Bargmann space and the measure on the This property is very useful for the calculation of the isoscalar factors of unitary groups, using the Fock spaces, after the introduction of the additional parameters $\{r_i\}$.

## 5- Generating function of the basis SU(2) ⊂SU(3)

### 5.1 The basis of the group SU (2) ⊂ SU (3)

Let $D_{[\lambda,\mu]}$ the space of homogeneous polynomials and $V_{(t,t0,y)}^{\lambda\mu}(z^1,z^2)$ is the orthogonal basis with:

$$z^1 = (z_1^1, z_2^1, z_3^1) = (\xi_1, \eta_1, \sigma_1)$$
$$z^2 = (z_1^2, z_2^2, z_3^2) = (\xi_2, \eta_2, \sigma_2) \qquad (5.1)$$

The space is homogeneous then

$$T_{11}V_{(\alpha)}^{\lambda\mu} = (\lambda + \mu)V_{(\alpha)}^{\lambda\mu}, \qquad T_{22}V_{(\alpha)}^{\lambda\mu} = (\mu)V_{(\alpha)}^{\lambda\mu} \tag{5.2}$$

With
$$T_{ij} = \sum_k z_k^i (\partial/\partial z_k^j) \tag{5.3}$$

The vectors $V_{(t,t0,y)}^{\lambda\mu}(z^1,z^2)$ are eigenfunctions of the Casimir operator of the second order $\vec{T}^2$, the projection of $\vec{T}$ on the z axis and the hypercharge Y. The eigenvalue of these operators are respectively t (t + 1), $t_0$ and the triple of the hypercharge quantum number y. The numbers $t, t_0$ are the isospin and the component of isospin on the z axis.
We have:
$$YV_{(\alpha)}^{\lambda\mu} = yV_{(\alpha)}^{\lambda\mu}, \qquad T_z V_{(\alpha)}^{\lambda\mu} = t_0 V_{(\alpha)}^{\lambda\mu}$$

And
$$\vec{T}^2 V_{(\alpha)}^{\lambda\mu} = t(t+1)V_{(\alpha)}^{\lambda\mu}. \tag{5.4}$$

the condition of Young tableau on $V_{(\alpha)}^{\lambda\mu}$ imposes the further condition:

$$T_{12}V_{(\alpha)}^{\lambda\mu}(z^1,z^2) = 0 \tag{5.5}$$

$$T_+ = C_{12} = \xi\frac{\partial}{\partial \eta}, \qquad T_- = C_{21} = \eta\frac{\partial}{\partial \xi},$$

With
$$T_0 = \frac{1}{2}(C_{11} - C_{22}) = \frac{1}{2}(\xi\frac{\partial}{\partial \xi} - \eta\frac{\partial}{\partial \eta})$$

$$Y = \sum_{i=1}^{3} C_{ii} - 3C_{33} = \xi\frac{\partial}{\partial \xi} + \eta\frac{\partial}{\partial \eta} - 2\sigma\frac{\partial}{\partial \sigma}$$

The Casimir operator of second order is:
$$\vec{T}^2 = T_0(T_0 + 1) + T_+T_-$$

According to the relation (5.3) we have
$$T_{ij} = \xi_i\frac{\partial}{\partial \xi_j} + \eta_i\frac{\partial}{\partial \eta_j} + \sigma_i\frac{\partial}{\partial \sigma_j}$$

The expression of $V_{(t,t_0,y)}^{\lambda\mu}(z^1,z^2)$ is given by many other s [94-97]:

$$V_{(t,t_0,y)}^{\lambda\mu}(z^1,z^2) = N(\lambda\mu;\alpha)(-1)^q \times \sum_k \binom{r}{k} \frac{(\mu-q)!p!}{(\mu-q-k)![p-(r-k)]!}$$
$$\times \xi_1^{p-(r-k)}\eta_1^{r-k}\sigma_1^{\lambda-p}(\Delta_1^{(1,2)})^k(-\Delta_2^{(1,2)})^{\mu-q-k}(\Delta_3^{(1,2)})^q$$

and
$$N(\lambda\mu;\alpha) = \{\frac{(\lambda+1)!(\mu+p-q+1)!}{p!q!(\mu-q)!(\lambda-p)!(\mu+p+1)!(\lambda+\mu-q+1)!} \times \frac{(2t-r)!}{(2t)!r!}\}^{\frac{1}{2}}$$

$$y = -(2\lambda + \mu) + 3(p+q), \quad 0 \le p \le \lambda,$$
$$t = \frac{\mu}{2} + \frac{p-q}{2}, \quad\quad 0 \le q \le \mu,$$
$$t_0 = t - r, \quad\quad r = 0,1,\ldots,2t.$$

## *5.2 Generating function of the basis SU(2)⊂SU(3)*

The vectors $V_{(\alpha)}^{\lambda\mu}(z^1, z^2)$ belong to the space $D_{t_1} \otimes D_{t_2} \otimes D_{t_3}$ which has the basic Elements $\varphi_{j_1 m_1}(\xi_1,\xi_2)\varphi_{j_2 m_2}(\eta_1,\eta_2)\varphi_{j_3 m_3}(\sigma_1,\sigma_2)$ and has the generating:

$$\exp[(x^1\xi^1) + (x^2\eta^1) + (x^3\sigma)] .$$

The generating functions of the eigenfunctions of $\vec{T}^2$ and $T_0$ can be deduced by applying the Schwinger's coupling method.
We get the first coupling:

$$\exp\{[t_2[\frac{\partial}{\partial x_1^1}\frac{\partial}{\partial x_2^2} - \frac{\partial}{\partial x_2^1}\frac{\partial}{\partial x_1^2}] + \tau_1(Z\frac{\partial}{\partial x^1}) + \tau_2(Z\frac{\partial}{\partial x^2})\}\exp[(x^1\xi^1) + (x^2\eta^1)$$
$$= \exp\{[t_2\delta_3^{(1,2)}]\} + \tau_1(Z\xi^1) + \tau_2(Z\eta^1)]\}$$

We obtain the second coupling by the application of Schwinger's coupling method

$$\exp\{[t_2[\frac{\partial}{\partial\tau_1}\frac{\partial}{\partial x_2^3} - \frac{\partial}{\partial\tau_2}\frac{\partial}{\partial x_1^3}] + \tau_1^{'}(Z^{'}\frac{\partial}{\partial\tau}) + \tau_2^{'}(Z^{'}\frac{\partial}{\partial x^3})\}\exp\{\tau_1(Z\xi^1) + \tau_2(Z\eta^1)]\}\exp[(x^1\sigma)]$$
$$= \exp[t_2\delta_3^{(1,2)} + t_2^{'}[Z_2\delta_1^{(1,2)} - Z_1\delta_2^{(1,2)}] + \tau_1^{'}Z_1^{'}[(Z_1\xi_1 + Z_2\eta_1)] + \tau_1^{'}Z_1^{'}\sigma_1]$$
$$+ \tau_2^{'}Z_1^{'}[(Z_1\xi_2 + Z_2\eta_2)] + \tau_2^{'}Z_1^{'}\sigma_2]$$

Nous avons comme puissance de $Z'_1$ et $Z'_2$ :
$$Z'_1 \to j' + m_3 + j_1 - j_2, \quad Z'_2 \to j' - m_3 - j_1 + j_2$$

We note in the following the minors by $\vec{\delta}^{(1,2)}$

$$\vec{\delta}^{(1,2)} = (\delta_1^{(1,2)}\vec{i} + \delta_2^{(1,2)}\vec{j} + \delta_3^{(1,2)}\vec{k}) = \vec{z}^{(i)} \times \vec{z}^{(j)}$$

## *5.3 The expression of the generating function of SU(3)*

So that the relation (5.5) is satisfied we put $Z'_2 = 0$ in the result of the second coupling then we get the generating function of the basis vectors $V_{(t,t_0,y)}^{\lambda\mu}(z^1, z^2)$.

Put $\quad x_1 = \tau_1^{'}, \ x_2 = \tau_2^{'}, \ y_1 = t_2^{'}, \ y_2 = t_2, \ Z_1 = \xi, \ Z_2 = \eta$

we write the generating function in a compact form

$$G((x,y,u),z) = \exp[f.\vec{z}^{(1)} + \vec{g}.\vec{z}^{(12)}] =$$
$$\sum_{\lambda\mu t t_0 y} \varphi_{(t,t_0,y)}^{(\lambda\mu)}(\vec{f},\vec{g})\, V_{(t,t_0,y)}^{(\lambda\mu)}(z^{(1)}, z^{(12)}) \tag{5.6}$$

With
$$\vec{f} = (x_1\xi, x_1\eta, x_2), \quad \vec{g} = (y_1\eta, -y_1\xi, y_2) \qquad (5.7)$$
and
$$\vec{z}^{(i)} = (z_1^i, z_2^i, z_3^i), \quad \vec{z}^{(ij)} = \vec{z}^{(i)} \times \vec{z}^{(j)} \qquad (5.8)$$

We have
$$\varphi_{(t,t_0,y)}^{(\lambda\mu)}(\vec{f},\vec{g}) = N[(\lambda\mu),(\alpha)](x_1^p x_2^{\lambda-p} y_1^{(\mu-q)} y_2^q)(\xi^{(t+t_0)} \eta^{(t-t_0)}) \qquad (5.9)$$

And
$$N[(\lambda\mu),(\alpha)] = (-1)^q \sqrt{\frac{(\mu+p+1)!(\mu+\lambda-q+1)!}{(\lambda+1)!(2t+1)\lambda![p!(\lambda-p)!q!(\mu-q)!(t+t_0)!(t-t_0)!}}$$

We have also:
$$\begin{aligned} y &= -(2\lambda+\mu)+3(p+q), & 0 \leq p \leq \lambda, \\ t &= \mu/2 + (p-q)/2, & 0 \leq q \leq \mu, \\ t_0 &= t - r, & r = 0,1,...,2t. \end{aligned} \qquad (5.10)$$

## 6. Generating function of the 3j symbols for Multiplicity-free of SU (3)

### 6.1 The invariant of the 3j symbols for multiplicity-free $\mu_1 = \mu_2 = 0$.

The invariant of 3j symbols of SU (3) is given by:

$$h = \sum_{\alpha_i} \begin{pmatrix} \lambda_1 0 & \lambda_2 0 & \lambda_3 \mu_3 \\ (\alpha_1) & (\alpha_2) & (\alpha_3) \end{pmatrix} V_{(\alpha_1)}^{(\lambda_1 0)}(z^1, 0) V_{(\alpha_2)}^{(\lambda_2 0)}(z^3, 0) V_{(\alpha_3)}^{(\lambda_3 \mu_3)}(z^{(5)}, z^{(56)})^c \qquad (6.1)$$

The conjugate vector $(V_{(\alpha)}^{\lambda\mu})^c$ is deduced from $V_{(\alpha)}^{\lambda\mu}$ by R- Conjugation [6, 7]:
$$\lambda \leftrightarrow \mu, \; p \to \mu - q, q \to \lambda - p$$
And
$$(V_{(\alpha)}^{\lambda\mu})^c = (-1)^{y/2-t_0} V_{(-\alpha)}^{\lambda\mu}, \quad (-\alpha) = (-y, t, -t_0)$$

the invariant are functions of the elementary invariants:
$$\vec{z}^{(1)} \cdot (\vec{z}^{(3)} \times \vec{z}^{(5)}), \quad \vec{z}^{(1)} \cdot \vec{z}^{(56)}, \quad \vec{z}^{(3)} \cdot \vec{z}^{(56)} \qquad (6.2)$$
consequently we write:

$$h = N(k_i) \frac{[\vec{z}^{(1)} \cdot (\vec{z}^{(3)} \times \vec{z}^{(5)}))]^{k_1} (\vec{z}^{(3)} \cdot \vec{z}^{(56)})^{k_2} (\vec{z}^{(1)} \cdot \vec{z}^{(56)})^{k_3}}{k_1! \quad k_2! \quad k_3!} \qquad (6.3)$$

We can determine the constant of normalization by our method [108-109] but it is simpler in this case to do the direct calculation [50], we write:

$$k_2 + k_3 = \lambda_3, \ k_1 + k_3 = \lambda_1, \ k_i \geq 0,$$
$$k_1 = \mu_3, \ k_1 + k_2 = \lambda_2, \tag{6.4}$$

So we have the decomposition:

$$(\lambda_1,0) \otimes (\lambda_2,0) \to \sum_{\mu_3}(\lambda_3 = \lambda_1 + \lambda_2 - 2\mu_3, \mu_3) \tag{6.5}$$

the normalization is:

$$N(k_i) = \sqrt{\frac{2\mu_3!(\lambda_2 - \mu_3)!(\lambda_1 - \mu_3)!}{(\lambda_1 + \lambda_2 - \mu_3 + 2)!(\lambda_1 + \lambda_2 - \mu_3 + 1)!}} \tag{6.6}$$

### 6.2 The generating function of the 3j symbols for multiplicity-free $\mu_1 = \mu_2 = 0$.

We find the 3j symbols from the expression (9.1) as:

$$\begin{pmatrix} \lambda_1 0 & \lambda_2 0 & \lambda_3 \mu_3 \\ (\alpha_1) & (\alpha_2) & (\alpha_3) \end{pmatrix} = \left\langle h \middle\| V_{(\alpha_1)}^{(\lambda_1 0)}(z^1,0) V_{(\alpha_2)}^{(\lambda_2 0)}(z^3,0) V_{(\alpha_3)}^{(\lambda_3\mu_3)}(z^{(5)}, z^{(56)})^c \right\rangle \tag{6.7}$$

Multiplying this expression by ($\prod_i t_i^{k_i}$) and using (6.7) we write:

$$G((f,g),t) = \int \Big\{\exp[\vec{f}_1 \cdot \vec{z}^{(1)}]\exp[+\vec{f}_3 \cdot \vec{z}^{(3)}]\exp[\vec{f}_5 \cdot \vec{z}^{(5)} + \vec{g} \cdot \vec{z}^{(56)}] +$$

$$[t_1 \vec{z}^{(1)} \cdot (\vec{z}^{(3)} \times \vec{z}^{(5)}) + t_2 \vec{z}^{(1)} \cdot \vec{z}^{(56)} + t_3 \vec{z}^{(3)} \cdot \vec{z}^{(56)}]\Big\} d\mu(z^{(1)}, z^{(3)}, z^5, z^6) =$$

$$\sum_{\alpha_i} \begin{pmatrix} \lambda_1 0 & \lambda_2 0 & \lambda_3 \mu_3 \\ (\alpha_1) & (\alpha_2) & (\alpha_3) \end{pmatrix} (N(k_i))^{-1} \varphi_{(\alpha_1)}^{(\lambda_1 0)}(f_1) \varphi_{(\alpha_2)}^{(\lambda_2 0)}(f_3) \varphi_{(\alpha_3)}^{(\lambda_3\mu_3)}(f_5,g)(\prod_i t_i^{k_i}) \tag{6.8}$$

In carrying out the integration over $\vec{z}^{(1)}, \vec{z}^{(3)}, \vec{z}^{(6)}$ we find that the quantity in brackets is written as:

$$\exp[\vec{f}_5 \cdot \vec{z}^{(5)} + t_1 \vec{z}^{(5)} \cdot (\vec{f}_1 \times \vec{f}_3) + (\vec{g} \times \vec{z}^{(5)}) \cdot ((t_2 \vec{f}_3 + t_3 \vec{f}_1) \times \vec{z}^{(5)})] \tag{6.9}$$

Let $\vec{h} = (t_2 \vec{f}_3 + t_3 \vec{f}_1)$ and using the expression

$$\vec{z}^{(5)} \times (\vec{g} \times \vec{z}^{(5)}) = \vec{g}(\vec{z}^{(5)} \cdot \vec{z}^{(5)}) - \vec{z}^{(5)}(\vec{z}^{(5)} \cdot \vec{g}) \tag{6.10}$$

we find for the third term the result:

$$\exp[(\vec{z}^{(5)})^t \begin{pmatrix} h_1 g_1 + h_2 g_2 & -h_1 g_2 & -h_1 g_3 \\ -h_2 g_1 & h_1 g_1 + h_3 g_3 & -h_2 g_3 \\ -h_3 g_1 & -h_3 g_2 & h_2 g_2 + h_3 g_3 \end{pmatrix} (z^{(5)})] \tag{6.11}$$

Using the formula the Gaussian integral we find the generating function:

$$G((f,g),t) = \frac{1}{(1-\vec{h}\cdot\vec{g})^2} \exp[\frac{t_1}{1-\vec{h}\cdot\vec{g}}(\vec{f}_5 \cdot (\vec{f}_1 \times \vec{f}_3))] \qquad (6.12)$$

## *6.3 Expression of 3j symbols of SU (3).*

We observe that $\vec{h}\cdot\vec{g} = t_2 \vec{f}_3 \cdot \vec{g} + t_3 \vec{f}_1 \cdot \vec{g}$.

And

$$\vec{f}_1 \cdot \vec{g} = [-x_1^1 y_1^5 [u^5 u^1] + x_2^1 y_2^5], \quad \vec{f}_3 \cdot \vec{g} = [x_1^3 y_1^5 [u^3 u^5] + x_2^3 y_2^5], \qquad (6.13)$$

$$\vec{f}_5 \cdot (\vec{f}_1 \times \vec{f}_3) = [x_2^1 x_1^5 x_1^3 [u^3 u^5] + x_1^3 x_2^5 x_1^1 [u^1 u^3] + x_1^5 x_1^1 x_2^3 [u^5 u^1]] \qquad (6.14)$$

developing (6.12) first and after that we use (6.13), (6.14) and the generating function of the 3j symbols of SU(2) we find the expression of 3j symbols of SU (3):

$$\begin{pmatrix} \lambda_1 0 & \lambda_2 0 & \lambda_3 \mu_3 \\ (\alpha_1) & (\alpha_2) & (\alpha_3) \end{pmatrix} = \frac{(-1)^{y_3/2-(t_0)_3}(-1)^{(2t_1-\mu_3)} N(k_i)}{N[(\lambda_1 0),(y_1 00)]N[(\lambda_2 0),(y_2 00)]N[(\lambda_3 \mu_3),(\alpha_3)]} \times$$

$$\left[\sum_{k_1}(-1)^{k_1} \frac{1}{k_1!(\mu_3-T+2t_3-k_1)!(T-2t_1-k_1)!(2t_1-\mu_3+k_1)!} \times \right.$$

$$\left. \frac{1}{(\lambda_2-\mu_3-T+2t_1+k_1)!(\lambda_1-2t_1-k_1)!}\right] \times \begin{pmatrix} t_1 & t_2 & t_3 \\ (t_1)_0 & (t_2)_0 & (t_3)_0 \end{pmatrix}. \qquad (6.15)$$

We write the quantity between brackets:

$$[] = \frac{1}{(\mu_3-T+2t_3)!(T-2t_1)!(2t_1-\mu_3)!(\lambda_2-\mu_3-T+2t_1)!(\lambda_1-2t_1)!} \times$$

$$_3F_2(-\mu_3+T-2t_3,-T+2t_1,-\lambda_1+2t_1;2t_1-\mu_3+1,\lambda_2-\mu_3-T+2t_1+1;1) \qquad (6.16)$$

Thus the isoscalar factor expression is found for the first time in this compact form
And which shows the great interest of the generating function method.

# Chapter *VII*

## The Gel'fand basis of SU(n) and the Wigner's Coefficients with multiplicity for the canonical basis

# Chapter *VII*

## The Gel'fand basis of SU(n) and the Wigner's Coefficients with multiplicity for the canonical basis

### 1. Introduction

The theory of unitary groups is of great interest in quantum physics, nuclear and elementary particle. The study of these groups was started in mathematics and several methods have been proposed: the infinitesimal method developed by Shur, Cartan, Killing, Weyl, etc. .., and the Weyl global method [108-116] whose starting point the matrix elements of SU(n). Weyl find the connection between the representation of the symmetric group and the unitary group. Weyl also find the basis vectors of the irreducible representation labeled by the highest weights $[h]_n = [h_{1n}, h_{n-1},..., h_{nn}]$ and the dimension formula. The reduction of the representation with highest weight $[h]_n$ of U(n) to U(n-1) with highest weight $[h]_{n-1}$ is given in terms of Weyl branching law.

$$[h_{1,n} \geq h_{1,n-1} \geq h_{2,n} \geq h_{2,n-1} ... \geq h_{n-1,n-1} \geq h_{nn}]$$

Using the "Weyl's branching law" Gelfand-Zeitlin introduce the basis of representation of U (n), function of $n(n+1)/2$ indices, and later proved the orthogonality of this basis. Moreover, Cartan has already found that these irreducible representations are polynomials of the fundamental representations $[1,..,0],..,[1,..,1]$, whose number is $2^n - 1$.

In physics the Schwinger's method [23] of bosons calculus, has been extended to study the homogenous polynomials basis for the irreducible representation of U(n) by Bargmann and Moshinsky and other [113]. Biedenharn et al. [44-45] used the Weyl tableau techniques of construction of some vectors [45] of the Gelfand-Zeitlin basis in terms of the bosons operators. The maximal and semi-maximal states of SU(n) are defined by Biedenharn et al.[45], and their importance for the study of the space of representation was observed by Moshinsky [90] and their extension to kernel and the branching operators was find in the papers of Louck [114] and Henrich [115] .
Furthermore, Nagel et Moshinsky[46] derive the Gel'fand basis polynomials in terms of the raising and lowering operators but the calculus[115-116] was very complexes and difficult to find the number of summations N of these polynomials for n>3, $N = (2^n - 1) - n(n+1)/2$ [22-23]. After that, Heinrich use the kernel and the branching operators to determine the polynomials and he is unable to find it for n> 3.

In other side the Wigner coefficients of SU(3) for the canonical basis were discussed by many authors[99-107]: Moshinsky observed that the Kronecker product of k representations of SU(n) could be analyzed in terms of certain representation of SU(N), where N=k(n-1). Furthermore, a large class considered of theses coefficients, for example Biedenharn et al. using the canonical unit tensor operator method and Le blanc and Rowe use the vector coherent state theory [100-105]. The method of invariants applied by Van der Wearden and finds the generating function of 3-j symbols of SU (2). This method was generalized by Resnikoff to SU (3) and derives only the results for multiplicity free. Parakash et al. [107] uses the latest methods and the expression obtained contains 33 summations and the normalization factor is difficult to calculate.

All theses methods are very complex and the Gel'fand basis of homogenous polynomials is not found for n>3 and the Wigner coefficients with multiplicity in the canonical basis are very difficult to calculate.

To solve these important and difficult problems we proposed a simple method [22,118-120], the generating function method [117], for the calculation of Gel'fand basis polynomials, the Wigner coefficients and isoscalar factors for SU(n).

Recently the author has returned to these problems [43] and we applied our method to calculate the Wigner coefficients for multiplicity free. However, in this work we will do a review of this method and we focus our attention to the practical sides to do the calculations of Gel'fand basis polynomials, the Wigner coefficients and isoscalar factors with multiplicity for SU(n). .

The generalization of the generating functions of SU (2) and SU (3) to SU(n) is easy after our introduction of the binary representations of the vectors of the fundamental representations. We observe also that there is a connection between the generating function, the kernel and the branching operators expressed as functions of complex variables of SU (n). We use these functions and a recurrence method for the determination of the vectors basis of representation of SU (4). We also use the space of parameters of the generating function and the invariants method to find an algebraic expression of Wigner's coefficient in the general case, multiplicity free or not, and the isoscalar of SU (3).

This chapter is organized as follows: Part two and three are a simple revision of Gel'fand basis, The fundamental representations, Matrix elements, Bosons polynomials and kernel function of SU(n). The next section is devoted to the derivation of the Generating function of SU(n). We outline the method for calculating the bosons polynomials of Gel'fand basis vector and we apply it to the case of SU(3) and SU (4) in part 6. In part 7 we present the invariant method for the calculation of Wigner's coefficients of SU(n) and we apply it to SU (2). The parts eight are devoted to the derivation of the analytic function of the 3-j symbols and the Isoscalar factors with multiplicity of SU(3). In the appendix we give a maple program very useful for the derivation of the generating function of U(n) and the normalization of Gel'fand basis.

## 2. Gel'fand basis and the fundamental representations

We summarize in this part the results of the determination of Gel'fand basis of the irreducible representation and the properties of this basis. By analogy with the theory of angular momentum, the maximal and the semi maximal of this basis are derived.

We define also the vectors of the fundamental representations.
Nagel and Moshinsky have found that the states of SU(n) may be written in terms of raising and lowering as in the SU(2) theory and we also summarize this work.

### *2.1 The Weyl generators and the Weyl branching law of U(n)*

The $n^2$ Weyl infinitesimal generators $E_{ij}$, $(i, j = 1....n)$ of the unitary group U(n) obey the commutation relations

$$[E_{ij}, E_{kl}] = \delta_{jk} E_{il} - \delta_{il} E_{ki}, \qquad (2.1)$$

These generators may be written in terms of creations and destruction of n-dimensional harmonic oscillators as:

$$E_{ij} = \sum_{ij} a_i^+ a_j \qquad (2.2)$$

The irreducible representations of U(n) are labeled by n-integer numbers

$$[h_{1n}, h_{2n}, ..., h_{nn}]. \qquad (2.3)$$

When the group U(n) is restricted to the subgroup U(n-1) we find the Weyl branching law:

$$h_{1,n} \geq h_{1,n-1} \geq h_{2,n} \geq h_{2,n-1} \geq .... \geq h_{n-1,n-1} \geq h_{n,n}.$$

### *2.2 Gel'fand basis for SU(n)*

Gel'fand and al. [5] extend the Weyl branching law to U(n) and derived the individual orthogonal states of the representation, called Gel'fand basis $|(h)_n\rangle$:

$$|(h)_n\rangle = \begin{vmatrix} h_{1n} & h_{2n} & \cdots & h_{nn} \\ & h_{1n-1} & \cdots & h_{n-1n-1} \\ & & \cdots\cdots\cdots\cdots & \\ & h_{12} & h_{22} & \\ & & h_{11} & \end{vmatrix} = \begin{vmatrix} [h]_n \\ (h)_{n-1} \end{vmatrix} = \begin{vmatrix} [h]_n \\ [h]_{n-1} \\ (h)_{n-2} \end{vmatrix} \cdots \qquad (2.4)$$

With $\qquad [h]_n = [h_{1n}\, h_{2n}\, ...\, h_{nn}]$

And
$$h_{i,n} \geq h_{i,n-1} \geq h_{i+1,n}$$

$$h_{i,n} \quad h_{i+1,n} \qquad SU(n)$$
$$\quad h_{i,n-1} \qquad SU(n-1)$$

In angular momentum and in particles physics [20] we have the notations:

For SU(2) $\qquad h_{12} = j + m, \ h_{11} = j - m$

For SU(3)

$$h_{12} = I + \frac{Y}{2} + B, \ h_{22} = -I + \frac{Y}{2} + B, h_{11} = I_3 + \frac{Y}{2} + B \qquad (2.5)$$

## 2.3 The Weyl dimension formula

The dimension of subspaces $[h_{\mu\nu}]$ is given by the Weyl formula:

$$d_{[h_{\mu\nu}]} = \left[\prod_{i<j}(p_{in} - p_{jn})\right] / [1!2!\cdots(n-1)!] \tag{2.6}$$

With $p_{in} = h_{in} + n - i$

## 2.4 The maximal and the semi-maximal states

The eigenvalue of the diagonal generators $E_{ii}$ is:

$$E_{ii}|(h)_n\rangle = \omega_{in}|(h)_n\rangle, \text{ with } \omega_{in} = \left(\sum_{j=1}^{i} h_{j,i} - \sum_{j=1}^{i-1} h_{j,i-1}\right) \tag{2.7}$$

We associate to each state $|h_{\mu\nu}\rangle$ a vector or weight vector which has components

$$\omega(h) = (\omega_{1n}(h), \omega_{2n}(h),\ldots \omega_{nn}(h)).$$

A weight $\omega(h')$ is higher than a weight $\omega(h)$ if the first nonzero component in the difference $\omega(h') - \omega(h)$ is positive.

We note respectively $\left|\begin{array}{c}[h]_n\\(\max)_{n-1}\end{array}\right\rangle$ and $\left|\begin{array}{c}[h]_n\\(\min)_{n-1}\end{array}\right\rangle$ are the states that have the maximum and minimum of weight.

The vector $\left|\begin{array}{c}[h]_n\\ [h]_{n-1}\\(\max)_{n-2}\end{array}\right\rangle$ is the semi-maximal vector.

## 2.5 The fundamental representations

We can express an arbitrary irreducible representation of U(n) in terms of a set of subspace called the fundamental representations [20].
The fundamental representations of U(n) are the irreducible subspaces:

$$[1,0,\cdots,0], [1,1,\cdots,0], \cdots,[1,1,\cdots,1] \tag{2.8}$$

The dimension of the subspace $[\overbrace{1,1,1,..,1}^{p},0,...,0,n]$ is $C_n^p$. Then we deduce that the total number of vector bases of the fundamental representations is $2^n - 1$. And we observe that the weight vectors of these bases were expressed in terms of the binary number and it is easy to establish a correspondence between these weight vectors and the fundamentals Gel'fand basis.

We denote these fundamentals basis vectors by $\left|\Delta_{n,[i]}^p\right\rangle$, $i = 1,2\cdots,2^n - 1$.

Using the binomials formula $C_n^p = C_{n-1}^p + C_{n-1}^{p-1}$ and a symbolic program (Maple 8 see appendix1) we derive by recurrence all Gel'fand fundamental representations for n > 2 and the binary representation of the fundamental representations (B.F.R).

*2.6 Explicit expression of Gel'fand basis vectors*

Nagel and Moshinsky have found that each vector $|h_{\mu\nu}\rangle$ of the basis $[h_{\mu n}]$ may be deducted from the vector $\left|\begin{matrix}[h]_n\\(min)_{n-1}\end{matrix}\right\rangle$ or the vector $\left|\begin{matrix}[h]_n\\(max)_{n-1}\end{matrix}\right\rangle$ by applying the raising operators $R_\lambda^\mu$ or the lowering operators $L_\lambda^\mu$ and derived the explicit expressions of these operators. We write:

$$|(h)_n\rangle = N\prod_{\lambda=2}^{n}\prod_{\mu=1}^{k-1}(L_\lambda^\mu)^{L_\lambda^\mu}\left|\begin{matrix}[h]_n\\(max)_{n-1}\end{matrix}\right\rangle$$
$$= N'\prod_{\lambda=2}^{n}\prod_{\mu=1}^{k-1}(R_\lambda^\mu)^{R_\lambda^\mu}\left|\begin{matrix}[h]_n\\(min)_{n-1}\end{matrix}\right\rangle \quad (2.9)$$

With $\quad L_\lambda^\mu = h_{\mu,\lambda} - h_{\mu,\lambda-1} \searrow \quad R_\lambda^\mu = h_{\mu,\lambda-1} - h_{\mu+1,\lambda} \quad \nearrow$

N and N' are the constants of normalization.

It is quite clear that this result is the generalization of the well-known result of angular momentum [8]. And it is very important to mention that the computation of Gel'fand basis vectors with this formula is very difficult and complicate for n > 3 [115-116].

# 3. Matrix elements, Bosons polynomials and Kernel function of SU(n)

After the classification of elementary particles a great effort has been made to study the matrix elements of unitary groups using the Gel'fand basis and the maximal and semi-maximal cases of the D-Wigner matrix elements of SU(n) are found. The maximal and semi-maximal polynomials basis in terms of bosons operators introduced by Biedenharn et al. [44] or in term of complexes variables are used by many authors [90,113]. Theses polynomials are functions of minors determinants as variables and it's extension to the derivation of the kernel and the branching kernel function is found [115-116]. We also give in term of bosons operators the basis of U (2) and SU (3) which are very useful later in this work.

*3.1 The D-Wigner matrix elements of SU(n)*

The application of the unitary transformation to the basis $\left|\begin{matrix}[h]_n\\(h)_n\end{matrix}\right\rangle$ is:

$$T_{U_n}\left|\begin{matrix}[h]_n\\(h)_n\end{matrix}\right\rangle = \sum_{(h')}(D_{(h'),(h)}^{[h]_n}(U_n))\left|\begin{matrix}[h]_n\\(h)_n\end{matrix}\right\rangle \quad (3.1)$$

$D_{(h'),(h)}^{[h]_n}(U_n)$ Are the elements of the matrix of SU(n).

The Gel'fand states for which $h_{rs} = h_{rn}$, $1 \leq r \leq s \leq n$, is the state of highest weight.

$$D^{[h]_n}_{(\max),(\max)}(U_n) = (\det(U_n))^{h_{n,n}} \prod_{k=1}^{n-1} (u^{(12..k)}_{(12..k)})^{h_{k,n-1}-h_{k+1,n}} \qquad (3.2)$$

A special result which is immediately available from tableau techniques [18] is the so called semi-maximal case:

$$D^{[h]_n}_{\substack{[h]_{n-1}\\(\max)},(\max)}(U_n) = \frac{1}{\sqrt{N}} \prod_{k=1}^{n-1} (u^{(12..k)}_{(12..k)})^{h_{k,n-1}-h_{k+1,n}} \prod_{k=1}^{n} (u^{(12..k)}_{(12..k-1,n)})^{h_{k,n}-h_{k,n-1}} \qquad (3.3)$$

$u^{(12..k)}_{(12..k)}(U_n)$ Is the minors constructed from the matrix of (Un).
The normalization is:

$$N = \prod_{\substack{i<j\\1}}^{n} \frac{(p_{i,n-1}-p_{jn})!}{(p_{i,n}-p_{j,n}-1)!} \prod_{\substack{i<j\\1}}^{n-1} \frac{(p_{i,n}-p_{j,n-1}-1)!}{(p_{i,n-1}-p_{j,n-1})!} \qquad (3.4)$$

***The conjugate representation***
Define the transformation

$$T_{U_n} \left| \begin{matrix} [h]_n \\ (h)_n \end{matrix} \right\rangle_c = \sum_{(h')} (D^{[h]_n}_{(h'),(h)}(U_n))^* \left| \begin{matrix} [h]_n \\ (h)_n \end{matrix} \right\rangle_c \qquad (3.5)$$

The conjugate of the basis states is

$$\left| \begin{matrix} [h]_n \\ (h)_n \end{matrix} \right\rangle_c \quad \text{With} \quad \left( \left| \begin{matrix} [h]_n \\ (h)_n \end{matrix} \right\rangle_c \right)_c = \left| \begin{matrix} [h]_n \\ (h)_n \end{matrix} \right\rangle \qquad (3.6)$$

***3.2 The bosons polynomials basis of U(n)***
The well known isomorphism between the spaces of Fock - Bargmann with the harmonic oscillator [17] implies that we can use one or the other of these spaces.
In this work we give the expressions of kernel and branching kernel functions in the Fock-Bargmann space because the computation in this space is very convenient.
We also give the expressions of known expressions of the bases of SU(2) and SU(3).

*3.2.1 The Fock space*
We consider the analytic Hilbert $(z_1, z_2, \cdots, z_n)$, $z_i \in C_n$ with the Gaussian measure and the scalar product is:

$$(f, g) = \int \overline{f(z)} g(z) d\mu(z) \qquad (3.7)$$

With $d\mu(z) = \pi^{-n} \exp(-(z,z)) \prod_{i=1}^{n} d\operatorname{Re}(z_i) d\operatorname{Im}(z_i)$

*3.2.2 The polynomials basis of U(n)*
We consider transformation

$$\left| \begin{pmatrix} [h]_n \\ (h)_n \end{pmatrix} \right\rangle \to \Gamma\left( \begin{pmatrix} [h]_n \\ (h)_n \end{pmatrix} \right)(\Delta z) \qquad (3.8)$$

In this representation the Gel'fand basis will be noted by $\Gamma\binom{[h]_n}{(h)_n}(\Delta(z))$.

$\{\Gamma\left(\binom{[h]_n}{(h)_n}\right)(\Delta z)\}$ is a homogenous polynomials basis of the space $B([h]_n)$ with coordinates:

$$\Delta_i^1(z) = z_i^1, \Delta_{i_1 i_2}^{12}(z) = \begin{vmatrix} z_{i_1}^1 & z_{i_1}^2 \\ z_{i_2}^1 & z_{i_2}^2 \end{vmatrix}, \cdots, \Delta_{i_1..i_k}^{1...k}(z) = \begin{vmatrix} z_{i_1}^1 & \cdots & z_{i_1}^k \\ \vdots & \vdots & \vdots \\ z_{i_k}^1 & \cdots & z_{i_k}^k \end{vmatrix}, \Delta_{12...n}^{12...n}(z) = \det(z) \qquad (3.9)$$

$\{\Delta(z)\} = \{\Delta_{i_1..i_l}^{12..l}(\Delta z), i,j = 1,...,n\}$ are the minors constructed from the matrix $(z_j^i), i,j = 1,\cdots,n$ by the selection of rows $1,2,...,l$ and columns $i_1, i_2, ..., i_l$.

These coordinates are independent vectors [23-24],

And if $\delta = diag(\delta_1, \delta_2, \cdots, \delta_n)$

We have $\quad \Gamma\left(\binom{[h]_n}{(h)_n}\right)(\Delta(\delta z)) = \delta^{\omega_1} \delta^{\omega_2 - \omega_1} \cdots \delta^{\omega_n - \omega_{n-1}} \Gamma\left(\binom{[h]_n}{(h)_n}\right)(\Delta z)$

And $\quad \omega_j = h_{1,j} + ... + h_{j,j}$, $\quad \langle z \| \Delta_{n,[i]}^k \rangle = \Delta_{i_1..i_k}^{1...k}(z)$ $\qquad (3.10)$

### 3.3 The kernel and the branching kernel function of SU(n)

We give only the analytical expressions of kernel function and the branching kernel functions of unitary groups [115].

### 3.3.1 The kernel function is:

$$K^n(\Delta(z), \Delta(u)) = (A_n)^{-1} \Delta^e(zu^*) = \sum_{(h)_n} \Gamma\left(\binom{[h]_n}{(h)_n}\right)(\Delta z) \overline{\Gamma}\left(\binom{[h]_n}{(h)_n}\right)(\Delta u) \qquad (3.11)$$

$$\Delta^e(z) = (\Delta_1^1(z))^{e_1} (\Delta_{12}^{12}(z))^{e_2} \cdots (\Delta_{12..n}^{12...n}(z))^{e_n}$$

$$e_i = h_{i,n} - h_{i+1,n}, i \leq n-1, \text{ and } e_n = h_{nn}$$

And $\quad A_n = (\prod_{j=1}^n (p_{jn})!) \times (\prod_{j=1}^{n-1} \prod_{k=j+1}^n (p_{j,n} - p_{k,n}))^{-1}$

### 3.3.2 The branching kernel function is:

$$R_{n-1}^n(\Delta(z), \Delta(u)) = \left[\frac{A_n}{A_{n-1}^n}\right]^{1/2} \prod_{k=1}^{n-1} (\Delta_{12..(k-1),k}^{12....k}(z,u))^{R_n^k} \prod_{k=1}^n (\Delta_{12..(k-1),n}^{12....n}(z,u))^{L_n^k}$$

$$= \sum_{(h)_{n-2}} \Gamma_n \begin{pmatrix} [h]_n \\ [h]_{n-1} \\ (h)_{n-2} \end{pmatrix}(\Delta z) \overline{\Gamma}_{n-1} \begin{pmatrix} [h]_{n-1} \\ (h)_{n-2} \end{pmatrix}(\Delta u) \qquad (3.12)$$

With $\quad L_n^m = h_{j,n} - h_{j,n-1}, \ L_n^n = h_{n,n}, \ R_n^m = h_{j,n-1} - h_{j+1,n}, \ 1 \leq j \leq n-1$

And $A_{n-1}^n = A_n \left( \prod_{i<j}(p_{in-1} - p_{jn})! \prod_{i \leq j}(p_{in} - p_{jn-1} + 1)! \right) / \left( \prod_{i<j}(p_{in} - p_{jn})!(p_{in-1} - p_{jn-1})! \right)^{-1}$

### *3.4 The SU(2) and SU(3) basis in terms of bosons expansion*

The expressions of U(2) and SU(3) in the base of the harmonic oscillator are well known [114].

### *3.4.1 The bosons expansion of U(2)*

$$\begin{pmatrix} h_{12} & h_{22} \\ & h_{11} \end{pmatrix} |0\rangle = N_2 (\Delta_{12}^{12})^{h_{22}} (\Delta_{13}^{12})^{h_{23}-h_{23}} (\Delta_1^1)^{h_{11}-h_{22}} (\Delta_2^1)^{h_{12}-h_{11}} |0\rangle \quad (3.13)$$

With $\quad N_2 = \left[ \dfrac{(h_{12} - h_{22} + 1)!}{(h_{11} - h_{22})!(h_{12} - h_{11})!(h_{12} + 1)!(h_{22})!} \right]^{1/2}$

### *3.4.2 The bosons expansion of U(3)*

$$\begin{pmatrix} h_{13} & h_{23} & 0 \\ h_{12} & h_{22} & \\ & h_{11} & \end{pmatrix} |0\rangle = (N_3)^{-\tfrac{1}{2}} (\Delta_{12}^{12})^{h_{22}} (\Delta_{13}^{12})^{h_{23}-h_{23}} (\Delta_1^1)^{h_{11}-h_{23}} (\Delta_2^1)^{h_{12}-h_{11}} (\Delta_3^1)^{h_{13}-h_{12}}$$

$$\times {}_2F_1 \left( h_{22} - h_{23}, h_{11} - h_{12} \middle| h_{11} - h_{23} + 1 \middle| \dfrac{\Delta_1^1 \Delta_{23}^{12}}{\Delta_2^1 \Delta_{13}^{12}} \right) \quad (3.14)$$

With

$$(N_3)^{-\tfrac{1}{2}} = \left[ \dfrac{(h_{11} - h_{22})!(h_{12} - h_{23})!(h_{12} - h_{22} + 1)!(h_{13} - h_{23} + 1)!}{(h_{11} - h_{23})!(h_{12} - h_{22})!(h_{12} + 1)!(h_{22})!} \right.$$

$$\left. \times \dfrac{(h_{12} - h_{11})!(h_{11} - h_{23})!(h_{23} - h_{22})!}{(h_{13} - h_{22} + 1)!(h_{13} - h_{12})!} \right]^{\tfrac{1}{2}} \quad (3.15)$$

## 4. Generating function of SU(n)

We observe that the parameters and their powers in the generating function of the basis of SU(2) and SU(3) are linked to the raising and lowering operators and their powers, then we generalized it by an empirical way [39] to SU(n) basis. And we derive it also using the kernel function.

Our introduction of the binary fundamental representation basis (B.F.R) is very useful for the calculations of the generating function and the invariance, which is connected with the complement of binary numbers [118-120].

This generating function is practical for the derivation of the invariant polynomials of SU(n) from the Gel'fand basis of unitary group SU(3(n-1)).

## 4.1 The generating function of SU(2) and SU (3)

We write only the generating functions of SU (2) and SU (3) then, we deduce simply the generating function of SU (n).

### 4.1.1 The generating function of SU(2)

$$\sum_{h_{\mu\nu}} g_2 \varphi^2(h_{\mu\nu},(x,y)) \Gamma_2 \begin{pmatrix} [h]_2 \\ (h)_2 \end{pmatrix} (\Delta(z)) = \exp[(\Delta_1^I y_2^I + \Delta_2^I x_2^I)] \qquad (4.1)$$

With $\quad g_2 = \dfrac{1}{\sqrt{(h_{12}-h_{11})!(h_{11})!}}\quad$ and $\quad \varphi^2(h_{\mu\nu},(x,y)) = x^{h_{12}-h_{11}} y^{h_{11}}$

### 4.1.2 The generating function of SU(3)

The generating function of SU (3) may be written in Fock-Bargmann basis in the form:

$$\sum_{h_{\mu\nu}} g_3 \varphi^3(h_{\mu\nu},(x,y)) \Gamma_3 \begin{pmatrix} [h]_3 \\ (h)_3 \end{pmatrix} (\Delta(z)) =$$
$$\exp[\Delta_{12}^{(12)} y_3^2 + (\Delta_{23}^{(12)} x_2^I z_3^2 + \Delta_{13}^{(12)} y_2^I z_3^2) + (\Delta_1^I y_2^I y_3^I + \Delta_2^I x_2^I y_3^I) + \Delta_3^I x_3^I] \qquad (4.2)$$

We write $\qquad \phi^3(h,(x,y)) = \displaystyle\prod_{\ell=2}^{3}\prod_{m=1}^{\ell}\left[(x_\ell^m)^{L_\ell^m}(y_\ell^m)^{R_\ell^m}\right] \qquad (4.3)$

We find this generating function using Schwinger's approach of angular momentum.

## 4.2 The generating function of SU(n)

The generalization of (4.2) to the generating functions of SU (n) is immediate and in the representation of Fock-Bargmann [6-7] we write

$$\sum_{h_{lm}} g_n \phi^n(h,\varphi^n(x,y)) \Gamma_n \begin{bmatrix} [h]_n \\ (h)_n \end{bmatrix} (\Delta z) = \exp\left[\sum_k \phi_k^{n,[i]}(x,y) \Delta_{n,[i]}^k(z)\right] \qquad (4.4)$$

And $\qquad \phi^n(h,(x,y)) = \displaystyle\prod_{\ell=2}^{n}\prod_{m=1}^{\ell}\left[(x_\ell^m)^{L_\ell^m}(y_\ell^m)^{R_\ell^m}\right] \qquad (4.5)$

We will calculate $\Delta_{n,[i]}^k(z)$ by the introduction of the binary fundamental representation and then we use two simple rules for the calculation of $\phi_k^{n,[i]}(x,y)$, the constant will be calculated later.

### 4.2.1 The binary fundamental representation (B.F.R) of $\Delta_{n,[i]}^k$

We associate to each miner $\Delta_{i_1...i_l}^{12...l}$ a table of n-boxes numbered from 1 to n.
We put "one" in the boxes $i_1, i_2, ..., i_l$ and zeros elsewhere.

$$\Delta_{n,[i]}^k = \Delta_{i_1...i_k}^{12...k} = \begin{array}{cccccccc} 1 & 2 & ... & i_1 & ... & i_k & ... & n \\ |\,0 & 0 & ... & 1 & ... & 1 & ... & 0\,\rangle \end{array} \qquad (4.6)$$

It' is very important to mention from the fact that the B.F.R. $\Delta^k_{n,[i]}$ is anti-symmetric then there are a connection between this basis and the Fock space of the second quantization hence the theory of unitary group plays an important role in physics.

### *4.2.2 Calculus of coefficients $\phi_k^{n,[i]}(x,y)$*

The coefficients $\phi_k^{n,[i]}(x,y)$ may be written as product of parameters $y_\lambda^\mu = y(\lambda,\mu)$ and $x_\lambda^\mu = x(\lambda,\mu)$. We determine the indices of these parameters by using the following rules:

a- We associate to each "one" which appeared after the first zero a parameter $y(\lambda,\mu)$ whose index $\lambda$ are the number of boxes and $\mu$ the number of "one" before him, plus one.

b- We associate to each zero after the first "one" a parameter $x(\lambda,\mu)$ whose index $\lambda$ is the number of boxes and $\mu$ the number of "one" before him.

### *4.3 The generating function and the kernel function of SU(n)*

We have $K^n(\Delta(z),\Delta(u)) = A_n^{-1}\Delta^e(zu^*)$

In multiply by $\dfrac{A_n(h_{11})!}{e_1!e_2!\cdots e_n!}$ and by summing we find

$$\sum_{e_i}\frac{A_n(h_{11})!}{e_1!e_2!\cdots e_n!}\Delta^e(zu^*) = (\sum_k \Delta^k_{n,[i]}(z)\Delta^k_{n,[i]}(u^*))^{h_{11}}$$

Replace $\Delta^k_{n,[i]}(u^*)$ by $\phi_k^{n,[i]}(x,y)$ and summing with respect to $h_{\mu\nu}$ we find:

$$\sum_{h_{\mu\nu}}\frac{A_n(h_{11})!}{e_1!e_2!\cdots e_n!}\Delta^e(z\varphi) = \exp\left[\sum_k \phi_k^{n,[i]}(x,y)\Delta^k_{n,[i]}(z)\right] \qquad (4.7)$$

### *4.4 Invariance by complementary of binary numbers (R-reflexion).*

We know that each binary number has a complement then we deduce that $\Delta^k_{n,[i]}(z)$ has a complement $\overline{\Delta}^k_{n,[i]}(z)$, Therefore the B.F.R. is invariant by the transformation

$$\Delta^k_{n,[i]}(z) \longrightarrow \overline{\Delta}^k_{n,[i]}(z). \qquad (4.8)$$

- For SU (2) we have the transformation $\varphi_{jm} \to (-1)^{j+m}\varphi_{j-m}$ taken into account that the complement of [0  1] is [1  0] and conversely.

- For SU (3) we also deduce the R-Conjugation of Gell-Mann (Resnikoff)

$$V^{\lambda\mu}_{(t,t_0,y)} \to (-1)^{y/2-t_0} V^{\lambda\mu}_{(t,-t_0,-y)} \qquad (4.9)$$

The expression of complement $\overline{\phi}_k^{n,[i]}$ may be deduced from $\phi_k^{n,[i]}$ by changing $y(\ell,m)$ by $z(\ell,-m+\ell)$ and $z(\ell,m)$ by $z(\ell,-m+\ell)$, and then the expression (4.5) is invariant by this transformation. We call this property of invariance by reflection or complementarily invariance. We also note that in the basis of U(n) the complement of $[1,1,\cdots,1]$ is $|0\rangle$ in the oscillator basis and 1 in the Fock-Bargmann space.

## 4.5 The generating functions of SU(3), U(4) and U(5)

We find simply by a direct calculation of rules a and b or using the results of the symbolic program (appendix1) the generating functions of U(4) and U(5) which are very useful for later.

### 4.5.1 The generating function of SU(3)
We write the generating function in a manner useful for computations

$$\sum_{h_{\mu\nu}} g_3 \varphi^3(h_{\mu\nu},(x,y)) \Gamma_3 \begin{pmatrix} [h]_3 \\ (h)_3 \end{pmatrix}(\Delta(z)) =$$

$$\exp[\Delta_{12}^{(12)} y_3^2 + (\Delta_{23}^{(12)} x_2^1 + \Delta_{13}^{(12)} y_2^1)z_3^2 + (\Delta_1^1 y_2^1 + \Delta_2^1 x_2^1)y_3^1 + \Delta_3^1 x_3^1]. \qquad (4.10)$$

### 4.5.2 The generating function of U(4)

$$\sum_{h_4} g_4 \phi_4(h, \varphi^4(x,y)) \Gamma_4 \begin{bmatrix} [h]_4 \\ (h)_4 \end{bmatrix}(\Delta z)$$

$$= \exp \begin{bmatrix} ((\Delta_1^1 y_2^1 + \Delta_2^1 x_2^1)y_3^1 + \Delta_3^1 x_3^1)y_4^1 + \Delta_4^1 x_4^1 + \\ ((\Delta_{13}^{12} y_2^1 + \Delta_{23}^{12} x_2^1)x_3^2 + \Delta_{12}^{12} y_3^2)y_4^2 + ((\Delta_{14}^{12} y_2^1 + \Delta_{24}^{12} x_2^1)y_3^1 + \Delta_{34}^{12} x_3^1)x_4^2 + \\ ((\Delta_{134}^{123} y_2^1 + \Delta_{234}^{123} x_2^1)x_3^2 + \Delta_{124}^{123} y_3^2)x_4^3 + \Delta_{123}^{123} y_4^3 + \Delta_{1234}^{1234} y_4^4 \end{bmatrix} \qquad (4.11)$$

### 4.5.3 The generating function of U(5)

$$\sum_{h_4} g_5 \phi_5(h, \varphi^5(x,y)) \Gamma_5 \begin{bmatrix} [h]_5 \\ (h)_5 \end{bmatrix}(\Delta z)$$

$$= \exp \begin{bmatrix} (((\Delta_1 y_2^1 + \Delta_2 x_2^1)y_3^1 + \Delta_3 x_3^1)y_4^1 + \Delta_4 x_4^1)y_5^1 + \Delta_5 x_5^1 + \\ ((\Delta_{15} y_2^1 + \Delta_{25} x_2^1)y_3^1 + \Delta_{35} x_3^1)y_4^1 + \Delta_{45} x_4^1)x_5^2 \\ (((\Delta_{13} y_2^1 + \Delta_{23} x_2^1)x_3^2 + \Delta_{12} y_3^2)y_4^2 + ((\Delta_{14} y_2^1 + \Delta_{24} x_2^1)y_3^1 + \Delta_{34} x_3^1)x_4^2)y_5^2 + \\ (((\Delta_{135} y_2^1 + \Delta_{235} x_2^1)x_3^2 + \Delta_{125} y_3^2)y_4^2 + ((\Delta_{145} y_2^1 + \Delta_{245} x_2^1)y_3^1 + \Delta_{345} x_3^1)x_4^2))x_5^3 + \\ ((\Delta_{1345} y_2^1 + \Delta_{2345} x_2^1)x_3^2 + \Delta_{1245} y_3^2)x_4^3 + \Delta_{1235} y_4^3)x_5^4 + \\ ((\Delta_{134} y_2^1 + \Delta_{234} x_2^1)x_3^2 + \Delta_{124} y_3^2)x_4^3 + \Delta_{123} y_4^3)y_5^3 + \Delta_{1234} y_5^4 \end{bmatrix} \qquad (4.12)$$

# 5. The Gel'fand basis vectors of U(n)

We will calculate by recurrence the polynomials of the irreducible representations of SU (n) using the branching kernel function. We consider the base of U (2) as a starting point, then we presents the recurrence method and we determine the bases of the groups U (3) and U (4).

### 5.1 The Gel'fand basis of U(2).
We have
$$\Gamma(h_{11}) = \Delta_1^{h_{11}} / \sqrt{(h_{11})!}$$

And
$$\Gamma\begin{pmatrix} h_{12} & h_{22} \\ & h_{11} \end{pmatrix} = M_2^{1/2}\ \Delta_1^{h_{12}-h_{22}} \Delta_2^{h_{12}-h_{11}} \Delta_{12}^{h_{22}} \qquad (5.1)$$

In the notation of angular momentum [20] we write:

$$j+m = h_{12} - h_{22},\ j-m = h_{12} - h_{11}.$$

## 5.2 The recurrence method for the calculation of U(n) Polynomials

By considering the product of coefficients of $y_n^i = y(n,i)$ and $x_n^i = x(n,i)$, $i = 1,\cdots,n$ appearing in the generating function of SU(n) we find the branching kernel.
We have

$$R_{n-1}^n(\Delta(z), \varphi^{n-1}(x,y)) = \left[\frac{A_n}{A_{n-1}^n}\right]^{1/2} \times$$
$$\prod_{k=1}^{n-1}(\Delta_{12..(k-1),k}^{12.....k}(z,\varphi^{n-1}))^{R_n^k} \prod_{k=1}^{n}(\Delta_{12..(k-1),n}^{12.....n}(z,\varphi^{n-1}))^{L_n^k} \qquad (5.2)$$

$$= \sum_{(h)_{n-2}} \Gamma_n\begin{pmatrix} [h]_n \\ [h]_{n-1} \\ (h)_{n-2} \end{pmatrix}(\Delta z) \overline{\Gamma}_{n-1}\begin{pmatrix} [h]_{n-1} \\ (h)_{n-2} \end{pmatrix}(\varphi^{n-1}(x,y)) \qquad (5.3)$$

But
$$\overline{\Gamma}_{n-1}\begin{pmatrix} [h]_{n-1} \\ (h)_{n-2} \end{pmatrix}(\varphi^{n-1}(x,y)) = N_{n-1}\phi^{n-1}(h,(x,y))P_{n-1}(1) \qquad (5.4)$$

And $P_2(1) = 1$.
After identification of the two sides of (5.4) we find the polynomial representations of the irreducible of U(n)

$$\Gamma_n\begin{pmatrix} [h]_n \\ (h)_{n-1} \end{pmatrix}(\Delta z) = N_n P_n(\Delta(z)),\ N_n = \frac{\sqrt{A_n}}{N_{n-1}P_{n-1}(1)\sqrt{A_{n-1}^n}} \qquad (5.5)$$

## 5.3 Calculation of $P_n(1)$

By replacing (5.4) in (5.5) we identify the results and then we do the summation for the convenience of calculations, we find the expression:

$$\prod_{k=1}^{n-1}(\Delta_{12..(k-1),k}^{12.....k}(\varphi^n(a,b),\varphi^{n-1}(x,y)))^{R_n^k} \prod_{k=1}^{n}(\Delta_{12..(k-1),n}^{12.....n}(\varphi^n(a,b),\varphi^{n-1}(x,y)))^{L_n^k} = \qquad (5.6)$$

$$\sum_{(h)_{n-1}} \varphi^n(h_{\mu\nu},(a,b)) P_n(1) \varphi^{n-1}(h_{\mu\nu},(x,y))$$

But $\phi^n(h,(x,y)) = \prod_{\ell=2}^{n}\prod_{m=1}^{\ell}\left[(x_\ell^m)^{L_\ell^m}(y_\ell^m)^{R_\ell^m}\right] = \phi^{n-1}(h,(x,y))\prod_{m=1}^{n}\left[(x_\ell^m)^{L_\ell^m}(y_\ell^m)^{R_\ell^m}\right]$

And $\phi^{n-1}(h,(a,b))\phi^{n-1}(h,(x,y)) = \phi^{n-1}(h,(ax,by))$

If we put u = ax and v = by we find after identification of the two sides of
The expression (5.6) :

$$\prod_{k=1}^{n-1}(\Delta_{12..(k-1),k}^{12.....k}(1,\varphi^{n-1}(u,v))^{R_n^k} \prod_{k=1}^{n}(\Delta_{12..(k-1),n}^{12.....n}(1,\varphi^{n-1}(u,v))^{L_n^k} =$$
$$\sum_{(h)_{n-1}} P_n(1)\varphi^{n-1}(h_{\mu\nu},(u,v)) \quad (5.7)$$

The constants $N_n$ and $P_n(1)$ are functions of Gel'fand indices of U(n).
The expression (5.7) is very important for the computing of $P_n(1)$.

### 5.4 Calculation of $P_n(1)$ for n=3, 4, 5.

We will compute P3 (1), P4 (1) using the formula (5.7).

#### 1- Calculation of $P_3(1)$

Using (5.7) we find:
$$(y_2^1 + x_2^1)^{h_{12}-h_{23}}(y_2^1 + x_2^1)^{h_{23}-h_{22}} = (y_2^1 + x_2^1)^{h_{12}-h_{22}} \quad (5.8)$$

We deduce from the above expression $P_3(1) = C_{h_{12}-h_{22}}^{h_{12}-h_{11}}$

#### 2- Calculation of $P_4(1)$

We will compute P4 (1) using the formula (5.7).
$$((v_2^1 + u_2^1)v_3^1 + u_3^1)^{L(4,1)+R(4,2)} \times ((v_2^1 + u_2^1)u_3^2 + v_3^2)^{L(4,2)+R(4,3)}$$
$$= \sum_{(h)_{n-1}} P_4(1)\varphi^3(h_{\mu\nu},(u,v)) \quad (5.9)$$

After development of the first member and the identification with the second member we find $P_4(1)$

$$P_4(1) = \frac{(L(4,1)+R(4,2))!}{(L(3,1))!(R(3,1))!} \frac{(L(4,2)+R(4,3))!}{(R(3,2))!(L(3,2))!} \frac{(L(3,1)+R(3,2))!}{(L(2,1))!(R(2,1))!} \quad (5.10)$$

#### 3- Calculation of $P_5(1)$

We will compute P5 (1) using the formula (5.7).
$$(((v_2^1 + u_2^1)v_3^1 + u_3^1)v_4^1 + u_4^1)^{L(5,1)+R(5,1)} \times$$
$$(((v_2^1 + u_2^1)u_3^2 + v_3^2)v_4^2 + ((v_2^1 + u_2^1)v_3^1 + u_3^1)u_4^2)^{L(5,2)+R(5,2)} \times \quad (5.11)$$
$$((v_2^1 + u_2^1)u_3^2 + v_3^2)u_4^3 + v_4^3)^{R(5,4)+L(5,3)} = \sum_{(h)_{n-1}} P_5(1)\varphi^4(h_{\mu\nu},(u,v))$$

After development of the first member and the identification with the second member we find $P_5(1)$

$$P_5(1) = \frac{(L(5,1)+R(5,1))!}{(R(4,1))!(L(4,1))!} \frac{(L(5,2)+R(5,2))!}{(R(4,2))!(L(4,2))!} \frac{(L(5,4)+R(5,3))!}{(R(4,3))!(L(4,3))!} \times$$
$$\frac{(L(4,2)+R(4,3))!}{(R(3,2))!(L(3,2))!} \frac{(L(4,1)+R(4,2))!}{(R(3,1))!(L(3,1))!} \frac{(L(3,1)+R(3,2))!}{(R(2,1))!(L(2,1))!} \quad (5.12)$$

# 6. The Gel'fand basis of U(3) and U(4)

We will determine the polynomials basis of SU(3) and SU(4).

## *6.1 The Gel'fand basis of U(3)*
We know that $P_2(I) = I$ so we can do the calculations with the aid of (5.5) and (5.6).
In this case, we write

$$R_2^3(\Delta(z), \varphi^2)) = \sqrt{\frac{A_2}{A_2^3}}[(\Delta_1(z)y_2^1 + \Delta_2(z)x_2^1)y_3^1]^{h_{12}-h_{23}} \Delta_3(z)^{h_{13}-h_{12}} \times$$
$$\Delta_{12}(z)^{h_{22}-h_{23}}(y_3^2)^{m_{22}}[(\Delta_{13}(z)y_2^1 + \Delta_{23}(z)x_2^1)x_3^2]^{h_{23}-h_{22}} \Delta_{123}(z)^{h_{33}} \quad (6.1)$$

Using (5.5) we find:

$$R_2^3(\Delta(z), \varphi^2)) = \sum_{h_{11}} \Gamma\binom{[h]_3}{(h)_2}(\Delta(z)) \times (N_2 \times (y_2^1 y_3^1)^{h_{12}-h_{22}}(x_2^1 y_3^1)^{h_{12}-h_{11}}(y_3^2)^{h_{22}}) \quad (6.2)$$

After identification we find the expression of the vector basis of U(3):

$$\Gamma\binom{[h]_3}{(h)_2}(\Delta(z)) = \sqrt{\frac{A_2}{A_2^3}} \sum_{i+j=h_{11}-h_{22}} \left(C_i^{h_{12}-h_{23}} \times C_j^{h_{23}-h_{22}}\right)$$
$$\times \Delta_1^i \Delta_2^{h_{12}-h_{23}-i} \Delta_3^{h_{13}-h_{12}} \Delta_{12}^{h_{22}-h_{33}} \Delta_{13}^j \Delta_{23}^{h_{23}-h_{22}-j} \Delta_{123}^{h_{33}} \quad (6.3)$$

We find the same expression already found in paper [19, 23].

## *6.2 The Gel'fand basis of U(4)*
We have

$$R_3^4(\Delta(z), \varphi) = \sum_{(h)_2} \Gamma\binom{[h]_4}{(h)_3}(\Delta(z)) \Gamma\binom{[h]_3}{(h)_2}(\varphi) \quad (6.4)$$

This is also written in the form

$$R_3^4(\Delta(z), \varphi) = \sqrt{\frac{A_3}{A_3^4}}((\Delta_1^1 y_2^1 + \Delta_2^1 x_2^1)y_3^1 + \Delta_3^1 x_3^1)^{L_4^1} \times (\Delta_4^1)^{R_4^1} \times$$
$$\times ((\Delta_{13}^{12} y_2^1 + \Delta_{23}^{12} x_2^1)x_3^2 + \Delta_{12}^{12} y_3^2)^{L_4^2} \times ((\Delta_{14}^{12} y_2^1 + \Delta_{24}^{12} x_2^1)y_3^1 + \Delta_{34}^{12} x_3^1)^{R_4^2}$$
$$\times ((\Delta_{134}^{123} y_2^1 + \Delta_{234}^{123} x_2^1)x_3^2 + \Delta_{124}^{123} y_3^2)^{R_4^3} \times (\Delta_{123}^{123})^{L_4^3} \times (\Delta_{1234}^{1234})^{L_4^4} \quad (6.5)$$

*b-the "bosons" polynomial of the irreducible representations of U(4)*
by the development of (6.5) and using (5.5) we find the relation between the indices:

$$i + i_1 = R_3^1, \quad j + j_1 = L_3^2, \quad L_4^1 + R_4^2 - i - i_1 = L_3^1, \quad L_4^2 + R_4^3 - j - j_1 = R_3^2,$$
$$k + k_1 = L_4^1 - i, \quad \ell + \ell_1 = L_4^2 - j, \quad m + m_1 = R_4^2 - i_1, \quad n + n_1 = R_4^3 - j_1 \quad (6.6)$$
$$k + \ell + m + n = R_2^1$$

We find that the number of indices five which is the exact number.
Finally the bosons polynomial is:

$$\Gamma_4 \begin{pmatrix} [h]_4 \\ [h]_3 \end{pmatrix}(\Delta(z)) = N_4 \sum_{ijklm} \frac{(L_4^1)!(L_4^2)!(R_4^2)!(R_4^3)!}{i!i_1!j!j_1!k!k_1!\ell!\ell_1!m!m_1!n!n_1!} \times \quad (6.7)$$

$$(\Delta_1^1)^{k_1}(\Delta_2^1)^k(\Delta_3^1)^i(\Delta_4^1)^{R_4^1} \times (\Delta_{13}^{12})^{\ell_1}(\Delta_{23}^{12})^{\ell}(\Delta_{12}^{12})^{i_1} \times$$

$$(\Delta_{14}^{12})^{m_1}(\Delta_{24}^{12})^m(\Delta_{34}^{12})^j \times (\Delta_{134}^{123})^{n_1}(\Delta_{234}^{123})^n(\Delta_{124}^{123})^{j_1} \times (\Delta_{123}^{123})^{L_4^3} \times (\Delta_{1234}^{1234})^{L_4^4}$$

With $N_4$ is the normalization constant.
It is clear that our method is the only one who can solve this problem from the practical point of view.

## 7. The Wigner's symbols and the invariants of SU(n)

In this section we give the definition of invariant and its connection with the Wigner coefficients. By using the binary representation of invariants and the parameter space we show that our method gives the Van der Wearden's result of SU(2).

### 7.1 The Wigner's symbols
The direct product of two representations may be reduced according to the formula
$$[h^1] \otimes [h^2] = \sum (\rho)[h^3]_{(\rho)} \quad (7.1)$$
Where (ρ) is the multiplicity or the number of time the representation is contained in $[h^1] \otimes [h^2]$.

With 
$$\left| \begin{matrix} [h^3] \\ (h^3) \end{matrix} \right\rangle_\rho = \sum_{h^1 h^2} \left\langle \begin{matrix} [h^1] & [h^2] \\ (h^1) & (h^2) \end{matrix} \middle\| \begin{matrix} [h^3] \\ (h^3) \end{matrix} \right\rangle_\rho \times \left| \begin{matrix} [h^1] \\ (h^1) \end{matrix} \right\rangle \left| \begin{matrix} [h^2] \\ (h^2) \end{matrix} \right\rangle \quad (7.2)$$

The coefficients in this expression are the Clebsh-Gordan coefficients.

The vector 
$$\frac{1}{\sqrt{d_{h^3}}} \sum_{(h^3)} \left| \begin{matrix} [h^3] \\ (h^3) \end{matrix} \right\rangle_\rho \left| \begin{matrix} [h^3] \\ (h^3) \end{matrix} \right\rangle_c \quad (7.3)$$

Is an invariant by unitary transformation with unity norm in the product of trois spaces.
When we replace it with the above mentioned:

$$H_{(\rho)} = \sum_{h^1 h^2} \begin{pmatrix} [h^1] & [h^2] & [h^3] \\ (h^1) & (h^2) & (h^3) \end{pmatrix}_\rho \prod_{i=1}^3 \left| \begin{matrix} [h^i] \\ (h^i) \end{matrix} \right\rangle \quad (7.4)$$

The coefficients 
$$\begin{pmatrix} [h^1] & [h^2] & [h^3] \\ (h^1) & (h^2) & (h^3) \end{pmatrix}_\rho = \frac{1}{\sqrt{d_{h^3}}} \left\langle \begin{matrix} [h^1] & [h^2] \\ (h^1) & (h^2) \end{matrix} \middle\| \begin{matrix} [h^3] \\ (h^3) \end{matrix} \right\rangle_{c\rho} \quad (7.5)$$

Are Wigner's 3j symbols of SU(n) and ρ is the indices of multiplicity.
$H_{(\rho)}$ is the generalization of the Van der Wearden's invariant of the group SU(2). These invariants has the following

$$T_U^{(1,2,3)} H_{(\rho)} = H_{(\rho)}, \quad \left\langle H_{(\rho)} \middle| H_{(\rho')} \right\rangle = \delta_{(\rho),(\rho')} \quad (7.6)$$

These properties mean that the invariant polynomial is function of elementary invariants. We choose $H_{(\rho)}$ as subspace of SU(3 (n-1)) which are function of the compatible elementary invariants.

$$H_{(\rho)}(\phi^1,\phi^2,\phi^3) = \sum_{h^1 h^2} \begin{pmatrix} [h^1]_n & [h^2]_n & [h^3]_n \\ (h^1)_n & (h^2)_n & (h^3)_n \end{pmatrix}_\rho \prod_{i=1}^{3} \Gamma_n \begin{pmatrix} [h^i]_n \\ (h^i)_n \end{pmatrix} (^S\phi^i) =$$

$$\Gamma_{3(n-1)} \begin{pmatrix} [h]_{3(n-1)} \\ (h)_{3(n-1)} \end{pmatrix} (^S\phi) \tag{7.7}$$

We note for the remainder of the variables by xi (λ ,μ) ,yi (λ ,μ), $N^i_{(3,0)}$, $P^i_{(3,0)}$ (1) Li(λ ,μ), Ri(λ, μ).

## 7.2 The elementary invariants $^s\Delta^i_n(z)$ and $^s\phi^i_n$

We determine the elementary scalars $^s\Delta^i_n(z)$ which are the basic elements of the Gel'fand basis of the SU (3 (n-1)). These scalars are formed of three rows of tables, Where each row of (n-1) boxes and $\alpha_i$ "one" and zero elsewhere.

$\alpha_i$ Satisfies the following conditions

$$0 \leq \alpha_i \leq n-1, \quad \sum_{i=1}^{3} \alpha_i = n \tag{7.8}$$

## 7.3 The Wigner's coefficients of SU(2)

We will apply the formula (7.7) for the determination of 3-j symbols.

### 7.3.1 The Invariants in the Gel'fand basis

We find for SU (2) the three elementary scalars

$$\boxed{1 \quad 1 \quad 0}, \quad \boxed{1 \quad 0 \quad 1}, \quad \boxed{0 \quad 1 \quad 1} \tag{7.9}$$

The parameters {x, y} that are not in the $\{\phi_k^{3,[i]}(x,y)\}$ of elementary scalars must have the power null. We put $y_3^1 = x_3^1 = 0$ then $h_{13} = h_{23} = h_{12}$ and the invariants $H_{(\rho)}$ are the Gel'fand bases:

$$\begin{pmatrix} h_{12} & h_{12} & 0 \\ & h_{12} & h_{22} \\ & & h_{11} \end{pmatrix} \tag{7.10}$$

We can write this expression in term of well known quantum numbers of angular momentum:

$$h_{22} = J_3, \; h_{12} - h_{11} = J_1, \; h_{11} - h_{22} = J_2$$

### 7.3.2 The elementary invariants in the space of parameters $\{^s\phi_3^i\}$

The elementary invariants in the space of parameters are:

$$\overline{|1\ 1\ 0|} \Rightarrow \begin{vmatrix} z_1^1 & z_1^2 \\ z_2^1 & z_2^2 \end{vmatrix} = \Delta_1^1\Delta_2^2 - \Delta_2^1\Delta_1^2 \Rightarrow \Xi(1,2) = \begin{vmatrix} y1(2,1) & y2(2,1) \\ x1(2,1) & x2(2,1) \end{vmatrix} \quad (7.11)$$

$$\overline{|1\ 0\ 1|} \Rightarrow \Delta_1^1\Delta_2^3 - \Delta_2^1\Delta_1^3 \Rightarrow \Xi(1,3) = \begin{vmatrix} y1(2,1) & y3(2,1) \\ x1(2,1) & x3(2,1) \end{vmatrix},$$

$$\overline{|0\ 1\ 1|} \Rightarrow \Delta_1^2\Delta_2^3 - \Delta_2^2\Delta_1^3 \Rightarrow \Xi(2,3) = \begin{vmatrix} y2(2,1) & y3(2,1) \\ x2(2,1) & x3(2,1) \end{vmatrix} \quad (7.12)$$

### 7.3.3 The generating function of 3-j symbols of SU(2)

The expression (7.7) in the case of SU(2) becomes:

$$\sum_{(h^i)_2} \begin{pmatrix} h_{12}^1 & h_{12}^2 & h_{12}^3 \\ h_{11}^1 & h_{11}^2 & h_{11}^3 \end{pmatrix} \prod_{i=1}^{3} \Gamma_2\begin{pmatrix} h_{12}^i & 0 \\ & h_{11}^i \end{pmatrix}(^s\phi^i) = \Gamma_3\begin{pmatrix} h_{12} & h_{12} & 0 \\ & h_{12} & h_{22} \\ & & h_{11} \end{pmatrix}(^s\phi^3) \quad (7.13)$$

We obtain the well known expression of Van der Wearden with ρ=1.

$$\sum_{(h^i)_2} \begin{pmatrix} h_{12}^1 & h_{12}^2 & h_{12}^3 \\ h_{11}^1 & h_{11}^2 & h_{11}^3 \end{pmatrix} \prod_{i=1}^{3} \frac{(xi(2,1))^{h_{12}^i - h_{11}^i}(yi(2,1))^{h_{11}^i}}{\sqrt{(h_{12}^i - h_{11}^i)!(h_{11}^i)!}} = \quad (7.14)$$

$$\frac{\sqrt{A_2}}{N_2\sqrt{A_2^3}} C_{h_{11}-h_{22}}^{h_{23}-h_{22}} \times (\Xi(1,2))^{h_{22}} (\Xi(1,3))^{h_{11}-h_{22}} (\Xi(2,3))^{h_{12}-h_{11}}$$

To simplify the notations we write: $u^i = (xi(2,1), yi(2,1))$.
Then we find the generating function of SU(2) or the well known Van der Wearden invariant of SU(2):

$$\sum_{m_i} \left[\prod_{i=1}^{3} \varphi_{j_i,m_i}(u^i)\right] \begin{pmatrix} j_1 & j_2 & j_3 \\ m_1 & m_2 & m_3 \end{pmatrix} = \frac{[u^2 u^3]^{(J-2j_1)}[u^3 u^1]^{(J-2j_2)}[u^1 u^2]^{(J-2j_3)}}{\sqrt{(J+1)!(J-2j_1)!(J-2j_2)!(J-2j_3)!}} \quad (7.15)$$

We have: J=j1+j2+j3 and P1=J-2j1, P2=J-2j2, P3=J-2j3.

## 8. The 3-j symbols and the Isoscalar factors of SU(3)

We deduce that the Gel'fand pattern is reduced to 7 indices variables:
The invariants polynomials are formed from one term or monomials and function of compatible product of elementary invariant scalars.

### 8.1 The Invariants of the Gel'fand basis

We find for SU (3) seven scalar elementary compatible, which are represented by the following tables:

$$|1\ 0\ 1\ 1\ 0\ 0|, \quad |1\ 1\ 1\ 0\ 0\ 0|, \quad |1\ 0\ 0\ 0\ 1\ 1|$$

$$|1\ 1\ 0\ 0\ 1\ 0|, \quad |0\ 0\ 1\ 0\ 1\ 1|, \quad |0\ 0\ 1\ 1\ 1\ 0|$$

$$|1\ 0\ 1\ 0\ 1\ 0| \tag{8.1}$$

The parameters $\{x, y\}$ that are not present in the elementary scalars $\phi_k^{n,[i]}(x, y)$ must have the power null.

We find:
$$k_1 = h_{34} - h_{33}, \ k_2 = h_{33}, \ k_3 = h_{12} - h_{23}, \ k_4 = h_{22} - h_{33},$$
$$k_5 = (h_{13} - h_{24}) - (h_{12} - h_{23}), \ k_6 = h_{35} - h_{23}, \ k_7 = (h_{23} - h_{34}) - (h_{22} - h_{33}) \tag{8.2}$$

The basis of Gel'fand for the invariants is:

$$\Gamma_6 \begin{pmatrix} h_{13} & h_{13} & h_{13} & 0 & 0 & 0 \\ & h_{13} & h_{13} & h_{24} & 0 & 0 \\ & & h_{13} & h_{24} & h_{34} & 0 \\ & & & h_{13} & h_{23} & h_{33} \\ & & & & h_{12} & h_{22} \\ & & & & & h_{12} \end{pmatrix} =^s \Gamma_6 \begin{pmatrix} [h]_6 \\ (h)_6 \end{pmatrix} \tag{8.3}$$

## 8.2 Calculus of the invariants in the space of parameters $^s\phi_6^i$

To determine the images of invariants in the space of parameters we write

a- $|1\ 0\ 1\ 1\ 0\ 0| \Longrightarrow$

$$\begin{vmatrix} z_1^1 & z_1^3 & z_1^4 \\ z_2^1 & z_2^3 & z_2^4 \\ z_3^1 & z_3^3 & z_3^4 \end{vmatrix} = \Delta_1^1 \Delta_{23}^{56} - \Delta_2^1 \Delta_{13}^{56} + \Delta_3^1 \Delta_{12}^{56} \Rightarrow W^1 = y1(3,1)x2(3,2)\Xi(1,2) + x1(3,1)y2(3,2)$$

We apply the same method for the calculation of the image of the invariants.

b- $|1\ 1\ 1\ 0\ 0\ 0| \Rightarrow W^2 = -y2(3,1)x1(3,2)\Xi(1,2) + x2(3,1)y1(3,2)$

c- $|1\ 0\ 0\ 0\ 1\ 1| \Rightarrow W^3 = y1(3,1)x3(3,2)\Xi(1,3) + x1(3,1)y3(3,2)$

d- $|1\ 1\ 0\ 0\ 1\ 0| \Rightarrow W^4 = -y3(3,1)x1(3,2)\Xi(1,3) + x3(3,1)y1(3,2)$

e- $|0\ 0\ 1\ 1\ 1\ 0| \Rightarrow W^5 = y2(3,1)x3(3,2)\Xi(2,3) + x2(3,1)y3(3,2)$

f- $|1\ 0\ 1\ 0\ 1\ 0| \Rightarrow W^6 = -y3(3,1)x2(3,2)\Xi(2,3) + x3(3,1)y2(3,2)$

g- $|1\ 0\ 1\ 0\ 1\ 0| \Rightarrow$

$$W^7 = x3(3,1)y1(3,1)y2(3,1)\Xi(1,2) - x2(3,1)y1(3,1)y3(3,1)\Xi(1,3)$$
$$+ x1(3,1)y2(3,1)y3(3,1)\Xi(2,3) \tag{8.4}$$

## 8.3 The generating function of 3-j symbols of SU(3)
The expression (7.7) is written in this case as:

$$\sum_{(h^i)_3} \begin{pmatrix} [h^1]_3 & [h^2]_3 & [h^3]_3 \\ (h^1)_3 & (h^2)_3 & (h^3)_3 \end{pmatrix}_\rho \prod_{i=1}^{3} \Gamma_3 \begin{pmatrix} [h^i]_3 \\ (h^i)_3 \end{pmatrix} (^S\phi^i) = N_6 \prod_{i=1}^{7} [W^i]^{k_i} \quad (8.5)$$

The development of the second side is

$$N_6 \left( \prod_{i=1}^{7} k_i! \right) \sum_{(h^i)_3} \left( \prod_{i=1}^{} (i_{15}!)^{-1} \right) \left[ \Xi(1,2)^{P3} \Xi(1,3)^{P2} \Xi(2,3)^{P3} \right] \times \prod_{\ell=2}^{3} \prod_{m=2}^{\ell} \left[ (xi(\ell,m))^{L_\ell^m} (yi(\ell,m))^{R_\ell^m} \right] \quad (8.6)$$

**a-** We have

$$\prod_{i=1}^{3} \Gamma_3 \begin{pmatrix} [h^i]_3 \\ (h^i)_3 \end{pmatrix} (^S\phi^i) = \left( \prod_{i=1}^{3} (N_3^i) P_3^i(I) \right) \prod_{\ell=2}^{3} \prod_{m=1}^{\ell} \left[ (xi(\ell,m))^{L_\ell^m} (yi(\ell,m))^{R_\ell^m} \right] \quad (8.7)$$

**b-** The development of the second side of (8.5) and the identification with the first member lead to a system of equations (Appendix2). The number of indices is fifteen so we have a system of fifteen equations which has the solution:

$$i_1 = R3(3,1) - L3(3,2) - P3 + i_9 + i_{11} - i_6; \quad i_2 = R2(3,2) - i_{11},$$
$$i_3 = R1(3,2) - i_7, \quad i_4 = R2(3,1) - P2 + i_7 - i_9 + i_6,$$
$$i_5 = L3(3,2) - i_9,$$
$$i_8 = -R2(3,1) + P2 + L1(3,2) - i_6 - i_7 + i_9, \quad (8.8)$$
$$i_{10} = R3(3,2) - i_6;$$
$$i_{12} = L2(3,2) - i_1, \quad i_{13} = P3 - R3(3,2) - i_{11} + i_6,$$
$$i_{14} = P2 - i_6 - i_7, \quad i_{15} = P1 - R1(3,2) - R2(3,2) + i_7 + i_{11}.$$

We have also the system
$$k_j = i_j + i_{(j+1)}, \quad j = 1..6, \quad k_7 = i_{13} + i_{14} + i_{15}. \quad (8.9)$$

It is simple to verify that these variables are function of i7-i9 then we choose for simplicity the multiplicity ρ by: ρ =k3.
We write i6 in terms of i9:   i6= k3-L3 (3.2) + i9.
We deduce that the number of summations is three indices: i7, i9, i11.

## 8.4 The algebraic expression of Wigner's coefficients and isoscalors of SU(4)
By replacing (8.10) and (8.6) in (8.5) and by comparison we find the algebraic expression of Wigner's coefficients, and isoscalors factors of SU(3).

$$\begin{pmatrix} [h^1]_3 & [h^2]_3 & [h^3]_3 \\ (h^1)_3 & (h^2)_3 & (h^3)_3 \end{pmatrix}_\rho = N_6 \left( \prod_{i=1}^{3} (N_3^i) P_3^i(I) \right)^{-1} \prod_{i=1}^{7} (k_i!) \times$$

$$\sum_{i_7 i_9 i_{11}} \frac{\sqrt{(P+1)! \prod_{i=1}^{3} (P-2Pi)!}}{\prod_{j=1}^{15} i_j!} \begin{pmatrix} [h^1]_2 & [h^2]_2 & [h^3]_2 \\ (h^1)_1 & (h^2)_1 & (h^3)_1 \end{pmatrix} \qquad (8.10)$$

As in (7.15) we write in this case P=J and Pi=Ji.

We use the well known notations of Wigner's coefficients in terms of isoscalar {},

And 3-j symbols of SU(2). We have:

$$\begin{pmatrix} [h^1]_3 & [h^2]_3 & [h^3]_3 \\ (h^1)_3 & (h^2)_3 & (h^3)_3 \end{pmatrix}_\rho = \begin{Bmatrix} [h^1]_3 & [h^2]_3 & [h^3]_3 \\ [h^1]_2 & [h^2]_2 & [h^3]_2 \end{Bmatrix}_\rho \begin{pmatrix} [h^1]_2 & [h^2]_2 & [h^3]_2 \\ (h^1)_1 & (h^2)_1 & (h^3)_1 \end{pmatrix} \qquad (8.11)$$

We find the analytic expression of the isoscalar for the canonical basis of SU(3):

$$\begin{Bmatrix} [h^1]_3 & [h^2]_3 & [h^3]_3 \\ [h^1]_2 & [h^2]_2 & [h^3]_2 \end{Bmatrix}_\rho = N_6 \frac{\prod_{i=1}^{7} (k_i!)}{\prod_{i=1}^{3} (N_3^i) P_3^i(I)} \left( \sum_{i_7 i_9 i_{11}} \frac{\sqrt{(P+1)! \prod_{i=1}^{3} (P-2Pi)!}}{\prod_{j=1}^{15} i_j!} \right) \qquad (8.12)$$

## 9. Appendixes

*Appendix1*

The maple program for the derivation of the binary representation and it is parameters representation in the generating function and the normalization coefficients of Gel'fand polynomials basis of U(n).

```
> restart:
with(linalg):
geyz:=proc(n,m)
local lam,mu,p,z,y,dlm,dplm;
y:= array(1..n,1..n);   z:= array(1..n,1..n);
dlm:= array(1..n,1..n);  dplm:= array(1..n,1..n);
for lam from 1 to n do
for mu from 1 to n do
dlm[lam,mu]:=0;dplm[lam,mu]:=0;
od;od;
p:=1;
for lam from 1 to n-1 do
for mu from 1 to (n-lam) do
dlm[lam,mu]:=m[mu,lam]-m[mu+1,lam];
dplm[lam,mu]:=m[mu+1,lam]-m[mu,lam+1];
```

```
p:=p*((z[lam,n-mu+1]**dlm[lam,mu])*(y[lam,n-mu+1]**dplm[lam,mu]));
od;od;print("Phi of BFR" ,p);    end;
ibn:=proc(n,m)
local i,i1,j,s,bn,del;
bn:= array(1..n);w:= array(1..n);del:= array(1..n);
for j from 1 to n do
del[j]:=0;od;
bn[1]:=m[n,1];
for j from 1 to n do
s:=0;
for i from 1 to j do
s:=s+ m[n-j+1,i];
od;w[j]:=s;od;
for j from 2 to n do
bn[j]:=w[j]-w[j-1]; od;
print(" BFR", bn);
i:=0;
for j from 1 to n do
if bn[j]=1 then
i:=i+1;
del[i]:=j;fi;od;
i1:=i;print(i1,   "delta", del);      end;
#  la base de Gel'fand et la formule des binomes#
    # (n!/p!(n-p)!)=(((n-1)!/(p-1)!(n-p)!)+(n-1)!/p!(n-p-1)!)#
#SU(2)  SU(3)  SU(4)   SU(5)   SU(6)#
#==========================================#
n1:=1+3+7+15+31+63; n:=6;
nt:= array(1..n);m:= array(1..n,1..n);a:= array(1..n1,1..n,1..n);
i1:=0;
for j from 1 to n do
i1:=i1+2**(j)-1;
nt[j]:=i1; od;
n1:=nt[n];
for j from 1 to n do
for k from 1 to n do
m[j,k]:=0;   od;od;
for i from 1 to n do
for j from 1 to n do
m[i,j]:=0;od;od;
for i from 1 to n1 do
for j from 1 to n do
for k from 1 to n do
a[i,j,k]:=0;
od;od;od;
      a[2,1,1]:=1;  a[2,1,2]:=0;     a[2,2,1]:=0;  a[2,2,2]:=0;
      a[3,1,1]:=1;   a[3,1,2]:=0;    a[3,2,1]:=1;   a[3,2,2]:=0;
```

```
        a[4,1,1]:=1;   a[4,1,2]:=1;   a[4,2,1]:=1;   a[4,2,2]:=0;
                # le programme#
for i from 3 to 5 do
print("=======================================");
print("----------------","the group SU(",i,") ---------------------");
print("=======================================");
i3:=nt[i-1];i4:=nt[i-2];id:=i;       print(" i3= ",i3," i4= ",i4);
          # la formule des elements ai1,1=k<=i#
for j from 1 to n do
for k from 1 to n do
m[j,k]:=0;   od;od;
for k from 1 to i do
i3:=i3+1;
for j from 1 to k do
a[i3,j,1]:=1;   od;
for k1 from 1 to k do
m[k1,1]:= a[i3,k1,1];
od;print("n=",i3,m);ibn(i,m);
geyz(i,m); od;i5:=1:
              # la formule des reccurences #
    # (i!/(j!*(i-j)!))=((i-1)!/j!(i-j-1)!)+((i-1)!/(j-1)!(i-j)!)#
    # *******************************************#
 # part 1 #    print(".........part 1........");
for j from 2 to (i-1) do
t1:=((i-1)!/((j-1)!*(i-j)!));print("part 1",t1);
for k from 1 to t1 do
i3:=i3+1:i4:=i4+1:
for k1 from 1 to (j) do
a[i3,1,k1]:=1;m[1,k1]:=a[i3,1,k1];    od;
for k2 from 2 to n do
for k3 from 1 to (n) do
a[i3,k2,k3]:= a[i4,k2-1,k3];m[k2,k3]:= a[i3,k2,k3];   od;od;
print("n=",i3,m);ibn(i,m);geyz(i,m);   od;"end k";
 # part 2 #  print(".........part 2........");
t2:=((i-1)!/(j!*(i-j-1)!));print("part 2",t2);
i5:=i4;
for k from 1 to t2 do
i3:=i3+1;i4:=i4+1;
for k1 from 1 to (j) do
a[i3,1,k1]:=1;m[1,k1]:=a[i3,1,k1];    od;
for k2 from (2) to n do
for k3 from 1 to (n) do
a[i3,k2,k3]:= a[i4,k2-1,k3];m[k2,k3]:= a[i3,k2,k3];
od;od;print("n=",i3,m);ibn(i,m);geyz(i,m);"end k";od;
i4:=i5;od;"end j";#++++++++++++++++#
           # la formule des elements aii===#
```

```
print("la formule des elements aii=======");
i3:=i3+1:i4:=i4+1:
for k1 from 1 to (id) do
for j1 from 1 to (id-k1+1) do
a[i3,k1,j1]:=1;od;od;
for k1 from 1 to (id) do
for j1 from 1 to (id-k1+1) do
m[k1,j1]:= a[i3,k1,j1]; od;od;
print("n=",i3,m);ibn(i,m);geyz(i,m);
od;"end i";> restart:
with(linalg):
   #calcul de A(m(1,n),m(1,n),...,m(n,n) de Kernel functions#
n:=3; m:= array(1..n,1..n);
coefr:=proc(n,m)
local a,mu1,mup,i,j,k,p,pp,q,qq,mq,coefn,
        coefap,n1,a1,ap,ap1;
coefn:= array(1..n);
              #part 1 Kernel functions#
ap:=1;ap1:=1;n1:=n-1;
for j from 1 to n1 do
a1:=m[j,n1]; ap:=(a1+n1-j)!*ap;
od;
for j from 1 to (n1-1) do
for k from j+1 to n1 do
 a1:=(m[j,n1]-m[k,n1]+k-j)!; ap1:=a1*ap1;
od;od;coefa:=ap1/ap;print(coefa,1);
print("***************");
    #part 2 The branching operators#
         #calcul de P( mu, mu)#p:=1;
for k from 1 to n do
for j from 1 to (k-1) do
mu1:=m[k,n]+n-k;
mup:=m[j,n]+n-j;
p:=p*((mup-mu1)!); od;od;print(p,2);
          #calcul de P( mup, mup)#
pp:=1;
for k from 1 to (n-1) do
for j from (1) to (k-1) do
 mu1:=m[k,n-1]+n-k-2;  mup:=m[j,n-1]+n-j-2;
pp:=pp*((mup-mu1)!);
od;od;print(pp,3);
         #calcul de Q( mu, mup)#q:=1;
for k from 2 to n do
mu1:=m[k,n]+n-k;
for j from 1 to (k-1) do
mup:=m[j,n-1]+n-j-1; q:=q*((mup-mu1)!);
```

```
od;od;print(q,4);
    #calcul de Q( mup, mu)#
qq:=1;
for k from 1 to n-1 do
mu1:=m[k,n-1]+n-k-1;
for j from 1 to (k) do
mup:=m[j,n]+n-j;
qq:=qq*((mup-mu1-1)!); od;od;
print(qq,5);
        #calcul de A( mup, mup)#
mq:=1;
for j from 1 to (n) do
mu1:=m[j,n]+n-j; mq:=mq*((mu1)!);
od;print(mq,6);
coefap:=(pp*p)/((mq*qq*q)); coefn[n]:=coefa*coefap;
coefb:=[(m[1,2]+1)!*(m[2,2])!*((m[1,1]-m[2,2])!)
*((m[1,2]-m[1,1])!)]/[(m[1,2]-m[2,2]+1)!];
print("coefa=",coefa); print("coefap=",coefap);
print("coefn1[n]=",coefn[n]);
coefn[n]:=coefn[n]*coefb;
print("coefb=",coefb); print("coefn[n]=",coefn[n]);
end;coefr(n,m);
```

## Appendix 2

The linear system of indices (part 8):

$$i_2 + i_6 + i_{14} + i_{15} = L1(3,1), \quad i_4 + i_8 = L1(3,2),$$
$$i_3 + i_{10} + i_{13} + i_{15} = L2(3,1), \quad i_1 + i_{12} = L2(3,2),$$
$$i_7 + i_{11} + i_{13} + i_{14} = L3(3,1), \quad i_5 + i_9 = L3(3,2),$$
$$i_1 + i_5 + i_{13} = R1(3,1), \quad i_3 + i_7 = R1(3,2),$$
$$i_4 + i_9 + i_{14} = R2(3,1), \quad i_2 + i_{11} = R2(3,2),$$
$$i_8 + i_{12} + i_{15} = R3(3,1), \quad i_6 + i_{10} = R3(3,2),$$
$$i_6 + i_7 + i_{14} = P1, \; i_2 + i_3 + i_{15} = P2$$
$$i_{11} + i_{13} + i_{10} = P3.$$

# Annex

## *Derivation of classical relativity and Schrödinger equation using Hamilton and Hamilton-Jacobi formalisms*

**Abstract**

Using Hamilton formalism of classical mechanic we derive in a simple way the equations of motions of classical relativity. Applying the canonical transformation and the Lagrange-Euler equation we find the Schrödinger equation. Our objective is a pedagogical point of view.

### 1. Introduction

The Hamilton and Hamilton-Jacobi formalisms didn't play a central role in classical mechanics or in the subsequent development of quantum mechanics. It is probably fair to say that the Hamilton and Hamilton-Jacobi formalisms, which were once taught as part of an advanced course on classical mechanics, have been seldom if ever used by physicists.

It was customary to derive the equations of motion of classical relativity using Lagrange formalism [A4] but in this note Our purpose is to point out that the derivation is more simple and instructive with Hamilton Formalism. Also, in the books of quantum mechanics [A1-4] we find comparisons between Schrödinger equation and the Hamilton-Jacobi equation, or start from Schrödinger equation and use Ritz variation method to derive the Schrödinger equation. These treatments do not satisfy a pedagogical point of view. In this note we purpose to derive the Schrödinger equation using the canonical transformation of the Hamiltonian (with the generating function as a variable) and then apply the Lagrange-Euler equation.

In part two, we do a revision of analytical dynamics formalism. In part three we derive the equations of motions of classical relativity. The derivation of Schrödinger equation constitutes part four. Finally, part five is devoted to some conclusions.

# 2. Revision of principle of least action

## 2.1 Hamilton principle of least action

Starting with the expression:

$$S = \int_{t_1}^{t_2} L(q,\dot{q},t)dt$$

S is the Hamilton action and $L(q,\dot{q},t)$ is the Lagrange's function.
The least action means:

$\delta S = 0$  With $\delta q(t_1) = \delta q(t_2) = 0$ and $\delta(\dot{q}) = \dfrac{d}{dt}(\delta q)$

Using Taylor development and the integration by part the second term becomes

$$\delta S = \left[\frac{\partial L}{\partial \dot{q}}\delta q\right]_{t_1}^{t_2} + \int_{t_1}^{t_2}(\frac{\partial L}{\partial q} - \frac{d}{dt}\frac{\partial L}{\partial \dot{q}})\delta q\, dt = 0$$

The equations of Lagrange can be derived:

$$\frac{d}{dt}\frac{\partial L}{\partial \dot{q}} = \frac{\partial L}{\partial q}$$

And: $p_i = \dfrac{\partial L}{\partial \dot{q}}$ is the generalized momentum.

The Lagrangian is defined to within an additive total time derivative of any function of coordinates and time

$$L'(q,\dot{q},t) = L(q,\dot{q},t) + \frac{d}{dt}F(q,t)$$

$$S' = \int_{t_1}^{t_2} L(q,\dot{q},t) + \int_{t_1}^{t_2}\frac{d}{dt}F(q,t) = S + F(q(t_2),t_2) - F(q(t_1),t_1) \Rightarrow \delta S = \delta S'$$

## 2.2 Hamilton Mechanics

### 2.2.1 The Hamiltonian

Starting with the Lagrangian function $L = L(q,\dot{q})$

$$dL = \sum_i \frac{\partial L}{\partial q_i}dq_i + \frac{\partial L}{\partial \dot{q}}d\dot{q}$$

and Lagrange's equations, we obtain

$$d(\sum_i \dot{q}_i p_i - L) = -\sum_i \dot{p}_i dq_i + \sum_i \dot{q}_i dp_i .$$

By definition the Legendre transformation is:

$$H = (\sum_i \dot{q}_i p_i - L)$$

The function $H = H(q,p,t)$ is Hamiltonian.
For a conservative system:

$$L = T - V$$

And $\qquad H = H(q,p,t) = T + V$.

### 2.2.2 Hamilton's equations of motion

$$dH = \sum_i \frac{\partial H}{\partial p_i} dq_i + \frac{\partial H}{\partial q_i} dp_i = -\sum_i \dot{p}_i dq_i + \dot{q}_i dp_i$$

by comparison we find the canonical equations

$$\frac{dq_i}{dt} = \frac{\partial H}{\partial p_i}, \quad \frac{dp_i}{dt} = -\frac{\partial H}{\partial q_i}.$$

## 2.3 Hamilton-Jacobi equation

### 2.3.1 Derivation of the equation

If $t_2$ is variable in the expression of $S = S(q,t)$ then

$$\frac{dS}{dt} = L$$

and

$$\frac{dS}{dt} = \frac{\partial S}{\partial t} + \sum_i \frac{\partial S}{\partial q_i} \dot{q}_i = \frac{\partial S}{\partial t} + \sum_i p_i \dot{q}_i$$

Therefore

$$\frac{\partial S}{\partial t} = L - \sum_i p_i \dot{q}_i = -H$$

finally:

$$\frac{\partial S}{\partial q_i} = p_i$$

And

$$\frac{\partial S}{\partial t} + H(q, \frac{\partial S}{\partial q}, t) = 0 .$$

### 2.3.2 Canonical transformation

Starting with the coordinate's transformation:

$$q, p, t \xrightarrow{T} Q(q,p,t), Q(q,p,t), t$$

$$S, L, H \xrightarrow{T} S', L', H'$$

T is a canonical transformation if the form of Hamilton's equations is conserved

$$\frac{dQ_i}{dt} = \frac{\partial H'}{\partial P_i}, \quad \frac{dP_i}{dt} = -\frac{\partial H'}{\partial Q_i}$$

And $L = L' + \frac{dF}{dt}$ or $\sum_i p_i \dot{q}_i - H = \sum_i P_i \dot{Q}_i - H' + \frac{dF}{dt}$,

Put $\varphi = F + \sum_i P_i Q_i$, we obtain $d\varphi(q, P, t) = \sum_i p_i dq_i + \sum_i Q_i dP_i + (H - H')dt$

We derive easily

$$p_i = \frac{\partial \varphi}{\partial q_i}, \quad Q_i = \frac{\partial \varphi}{\partial P_i}, \quad H' = H + \frac{\partial \varphi}{\partial t}$$

The function $\varphi$ is called generating function.

# 3. Derivation of classical Relativity using Hamilton formalism

## 3.1 Metric of classical relativity
We start from the definition of the action
$$S = \int L dt = \int p_x dx + p_y dy + p_z dz - H dt$$
We observe that $p_x, p_y, p_z$ are momentum and $H$ is the energy. Then we must divide by a velocity, a, to make some coherence in the expression of $S$.

So we write:
$$S = \int L dt = \int p_x dx + p_y dy + p_z dz + \frac{iH}{a}(iadt)$$
$$= \int p_x dx + p_y dy + p_z dz + p_u (du)$$

With a being a constant velocity then $p_u = \frac{iH}{a}$ is a momentum and
$$S = \int \vec{P} \cdot d\vec{R} = \int (p_x \vec{i} + p_y \vec{j} + p_z \vec{k} + p_u \vec{n}) \cdot (dx\vec{i} + dy\vec{j} + dz\vec{k} + du\vec{n})$$

$\vec{P}$ And $\vec{R}$ are vectors in the four Euclidean space.
The choice of the velocity implies that the element of length in the four dimensional Euclidean space is:
$$d\ell^2 = (du)^2 + (dy)^2 + (dz)^2 + (du)^2$$
$d\ell^2$ Is an invariant by Lorentz transformation.

## 3.2 Choice of the velocity a
The invariant operator in the four dimensions space is\ the Laplace-Beltrami Operator:
$$\Delta_S = \frac{\partial^2}{\partial x^2} + \frac{\partial^2}{\partial y^2} + \frac{\partial^2}{\partial z^2} + \frac{\partial^2}{\partial u^2} = \frac{\partial^2}{\partial x^2} + \frac{\partial^2}{\partial y^2} + \frac{\partial^2}{\partial z^2} - \frac{1}{a^2}\frac{\partial^2}{\partial t^2}$$

Now if we choose $a = c$ where $c$ the velocity of light is, the equation $\Delta_S = 0$ is the Maxwell equation. We deduce the momentum

$$p_x = \frac{\partial S}{\partial x}, p_y = \frac{\partial S}{\partial y}, p_z = \frac{\partial S}{\partial z}, p_u = \frac{\partial S}{\partial (ict)} = i\frac{H}{c}$$

The last equation is the Hamilton-Jacobi.

## 3.3 Expression of the Hamiltonian
The invariant in 4-dimension Euclidean space is
$$\vec{P}^2 = \vec{p}^2 + (i\frac{H}{c})^2 = -(\frac{E_0}{c})^2$$

In a system at rest $\vec{p}^2 = 0$, $E_0$ is a constant of motion or the energy at rest:

We deduce the Hamiltonian

$$H = \sqrt{(\frac{E_0}{c}) + \vec{p}^2}$$

for $(\frac{c\vec{p}}{E_0})^2 \ll 1$, $H = (\frac{E_0}{c})\sqrt{1+(\frac{c\vec{p}}{E_0})^2} \approx (\frac{E_0}{c})(1+(\frac{c\vec{p}}{E_0})^2) = const + \frac{p^2}{2m}$

We deduce the expression $E_0 = mc^2$ and $(\frac{c\vec{p}}{E_0})^2 \ll 1$ which means that $v \leq c$.

### 3.4 Equations of motions of classical relativity

Using the canonical equations we simply derive the equations of motions of classical relativity

$$\vec{p} = \frac{m\vec{v}}{\sqrt{1-\frac{v^2}{c^2}}}, \qquad E = \frac{mc^2}{\sqrt{1-\frac{v^2}{c^2}}}$$

It is important to point out that we can use the quaternion field instead of Euclidean space.

## 4. Derivation of Schrödinger equation using Hamilton-Jacobi formalism

We adopt the Hartree-Fock method used in the theory of many body problems. That is, we consider a transformation and we search to find the minimum of energy. In our case we take the canonical transformation of the Hamiltonian $H$ Given by:

$$H' = H + \frac{\partial \varphi}{\partial t}$$

Where $\varphi$ is the generating function and $p_i = \frac{\partial \varphi}{\partial q_i}$.

In this case $q_1 = x, q_2 = y, q_3 = z$,

And $H = \frac{1}{2m}(p_x^2 + p_y^2 + p_z^2) + V$,

We obtain

$$H = \frac{1}{2m}((\frac{\partial \varphi}{\partial x})^2 + (\frac{\partial \varphi}{\partial x})^2 + (\frac{\partial \varphi}{\partial t})^2) + V$$

Put $\Psi = e^{i\varphi/\hbar}$ then $\Psi^* = e^{-i\varphi/\hbar}$

Therefore $\varphi = \frac{\hbar}{i}\ln\Psi$ and $\varphi = -\frac{\hbar}{i}\ln\Psi^*$

Then

$$(\frac{\partial \varphi}{\partial x})^2 = \hbar^2 \frac{\partial \Psi}{\partial x}\frac{\partial \Psi^*}{\partial x}/\Psi\Psi^*, \quad (\frac{\partial \varphi}{\partial y})^2 = \hbar^2 \frac{\partial \Psi}{\partial y}\frac{\partial \Psi^*}{\partial y}/\Psi\Psi^*,$$

$$(\frac{\partial \varphi}{\partial z})^2 = \hbar^2 \frac{\partial \Psi}{\partial z}\frac{\partial \Psi^*}{\partial z}/\Psi\Psi^*,$$

But $\quad \frac{\partial \varphi}{\partial t} = \frac{\hbar}{i}\frac{\partial \Psi}{\partial t}/\Psi, \quad \frac{\partial \varphi^*}{\partial t} = -\frac{\hbar}{i}\frac{\partial \Psi^*}{\partial t}/\Psi^*$

We can write $\quad \frac{\partial \varphi}{\partial t} = \frac{1}{2}(\frac{\hbar}{i})[\Psi^* \frac{\partial \Psi}{\partial t} - \Psi \frac{\partial \Psi^*}{\partial t}]/\Psi\Psi^*$

Using these expressions in the formula

$$H' = H + \frac{\partial \varphi}{\partial t}$$

We obtain

$$\Psi^* H' \Psi = \frac{\hbar^2}{2m}[\frac{\partial \Psi^*}{\partial x}\frac{\partial \Psi}{\partial x} + \frac{\partial \Psi^*}{\partial y}\frac{\partial \Psi}{\partial y} + \frac{\partial \Psi^*}{\partial z}\frac{\partial \Psi}{\partial z}] + V\Psi^*\Psi + \frac{1}{2}(\frac{\hbar}{i})[\Psi^* \frac{\partial \Psi}{\partial t} - \Psi \frac{\partial \Psi^*}{\partial t}]$$

Using the Euler-Lagrange differential equation we will show that the variation of this expression yields the Schrödinger equation. $\Psi$ and $\Psi^*$ have to be varied Independent.

$$J(\Psi) = \int \Psi^* H \Psi = \int \frac{\hbar^2}{2m}[\frac{\partial \Psi^*}{\partial x}\frac{\partial \Psi}{\partial x} + \frac{\partial \Psi^*}{\partial y}\frac{\partial \Psi}{\partial y} + \frac{\partial \Psi^*}{\partial z}\frac{\partial \Psi}{\partial z}]dxdydz$$

$$+ \int V\Psi^*\Psi + \frac{1}{2}(\frac{\hbar}{i})[\Psi^* \frac{\partial \Psi}{\partial t} - \Psi \frac{\partial \Psi^*}{\partial t}]dxdydz = \int H(\Psi, \frac{\partial \Psi}{\partial q})dxdydz$$

By choosing a Lagrange density £($\Psi,\partial\Psi/\partial q$)=-H($\Psi,\partial\Psi/\partial q$) we find the expression
Find by Greiner [A56] without any indication of its origin.
The Euler-Lagrange equation, split up into space and time components, reads:

$$\frac{\partial £}{\partial \Psi_\sigma} - \frac{\partial}{\partial x_i}\frac{\partial £}{\partial(\partial \Psi_\sigma/\partial x_i)} - \frac{d}{dt}\frac{\partial £}{\partial \dot{\Psi}_\sigma} = 0$$

Where the summation over *i* runs through 1, 2, and 3 or
First we vary with respect to $\Psi$ and obtain

$$[-\frac{\hbar^2}{2m}\Delta + V]\Psi = -i\hbar \frac{\partial \Psi}{\partial t} = H\Psi$$

Analogously, variation with respect to $\Psi$ yields

$$[-\frac{\hbar^2}{2m}\Delta + V]\Psi = -i\hbar \frac{\partial \Psi^*}{\partial t} = H^*\Psi^*$$

$H$ Is the familiar Schrödinger equation

$$H = -\frac{\hbar^2}{2m}\Delta + V$$

Thus we derived the Schrödinger equation using the Hamilton-Jacobi formalism.